\documentclass{article}

\usepackage{PRIMEarxiv}

\usepackage[utf8]{inputenc} 
\usepackage[T1]{fontenc}    
\usepackage{url}            
\usepackage{booktabs}       
\usepackage{amsfonts}       
\usepackage{nicefrac}       
\usepackage{microtype}      
\usepackage{lipsum}
\usepackage{fancyhdr}       
\usepackage{graphicx}       
\graphicspath{{media/}}     
\usepackage{amssymb}
\usepackage{algorithmic}
\usepackage{graphicx}
\usepackage{textcomp}
\usepackage{caption}
\usepackage{tcolorbox}
\usepackage{makecell}
\usepackage{pifont}
\usepackage{enumitem}
\usepackage[document]{ragged2e}

\title{The FormAI Dataset: Generative AI in Software Security Through the Lens of Formal Verification
\thanks{\textit{\underline{Citation}}: https://doi.org/10.1145/3617555.3617874} 
}

\author{
  Norbert Tihanyi \\
  Technology Innovation Institute \\
  Abu Dhabi \\
  UAE \\
  \texttt{tihanyi.pgp@gmail.com} \\
  \And
  Tamas Bisztray \\
  The University of Oslo \\
  Oslo \\
  Norway \\
  \texttt{tamasbi@ifi.uio.no} \\
  \And
  Ridhi Jain \\
  Technology Innovation Institute \\
  Abu Dhabi \\
  UAE \\
  \texttt{ridhij@iiitd.ac.in} \\
  \And
  Mohamed Amine Ferrag \\
  Technology Innovation Institute \\
  Abu Dhabi \\
  UAE \\
  \texttt{mohamed.amine.ferrag@gmail.com} \\
  \And
  Lucas C. Cordeiro \\
  University of Manchester \\
  Manchester \\
  UK \\
  \texttt{lucas.cordeiro@manchester.ac.uk} \\
  \And
  Vasileios Mavroeidis \\
  The University of Oslo \\
  Oslo \\
  Norway \\
  \texttt{vasileim@ifi.uio.no} \\
}

\begin{document}
\maketitle
\begin{center}
\textbf{This paper has been published at PROMISE 2023: Proceedings of the 19th International Conference on Predictive Models and Data Analytics in Software. 
https://doi.org/10.1145/3617555.3617874} 
\end{center}

\vspace{2cm}

\begin{abstract}
This paper presents the FormAI dataset, a large collection of $112\, 000$ AI-generated compilable and independent C programs with vulnerability classification. We introduce a dynamic zero-shot prompting technique constructed to spawn a diverse set of programs utilizing Large Language Models (LLMs). The dataset is generated by GPT-3.5-turbo and comprises programs with varying levels of complexity. Some programs handle complicated tasks like network management, table games, or encryption, while others deal with simpler tasks like string manipulation.
Every program is labeled with the vulnerabilities found within the source code, indicating the type, line number, and vulnerable function name. This is accomplished by employing a formal verification method using the Efficient SMT-based Bounded Model Checker (ESBMC), which exploits model checking, abstract interpretation, constraint programming, and satisfiability modulo theories, to reason over safety/security properties in programs. This approach definitively detects vulnerabilities and offers a formal model known as a counterexample, thus eliminating the possibility of generating false positive reports. This property of the dataset makes it suitable for evaluating the effectiveness of various static and dynamic analysis tools. Furthermore, we have associated the identified vulnerabilities with relevant Common Weakness Enumeration (CWE) numbers. We make the source code available for the $112,000$ programs, accompanied by a comprehensive list detailing the vulnerabilities detected in each program, making the dataset ideal for training LLMs and machine learning algorithms.
\end{abstract}

\section{Introduction}
\label{sec:introduction}

The advent of Large Language Models (LLMs) is revolutionizing the field of computer science, heavily impacting software development and programming as developers and computer scientists enthusiastically use AI tools for code completion, generation, translation, and documentation \cite{bui_codetf_2023,ross_programmers_2023}. 
Research related to program synthesis using Generative Pre-trained Transformers (GPT)~\cite{chavez_chat_2023} is gaining significant traction, where initial studies indicate that the GPT models can generate syntactically correct yet vulnerable code~\cite{ ma2023scope, charalambous_new_2023}. A recent study conducted at Stanford University suggests that software engineers assisted by \textit{OpenAI’s codex-davinci-002 model} during development were at a higher risk of introducing security flaws into their code~\cite{perry_users_2022}. As the usage of AI-based tools for code generation continues to expand, understanding their potential to introduce software vulnerabilities becomes increasingly important. Considering that GPT models are trained on freely available data from the internet, which can include vulnerable code, AI tools can potentially recreate the same patterns that facilitated those vulnerabilities.

Our primary objective is to explore how proficiently LLMs can produce secure code for different coding objectives without requiring subsequent adjustments or human intervention. Additionally, we aim to uncover the most frequent vulnerabilities that LLMs tend to introduce in the code they generate, identifying common patterns in realistic examples to comprehend their behavior better. This brings forward the following research questions: 

\begin{itemize}
    \item {\textbf{RQ1}:} How likely is purely LLM-generated code to contain vulnerabilities on the first output when using simple zero-shot text-based prompts?
    \item {\textbf{RQ2}:} What are LLMs' most typical coding errors?
\end{itemize}

\begin{figure*}[ht] 
\centering
\includegraphics[width=0.8\textwidth]{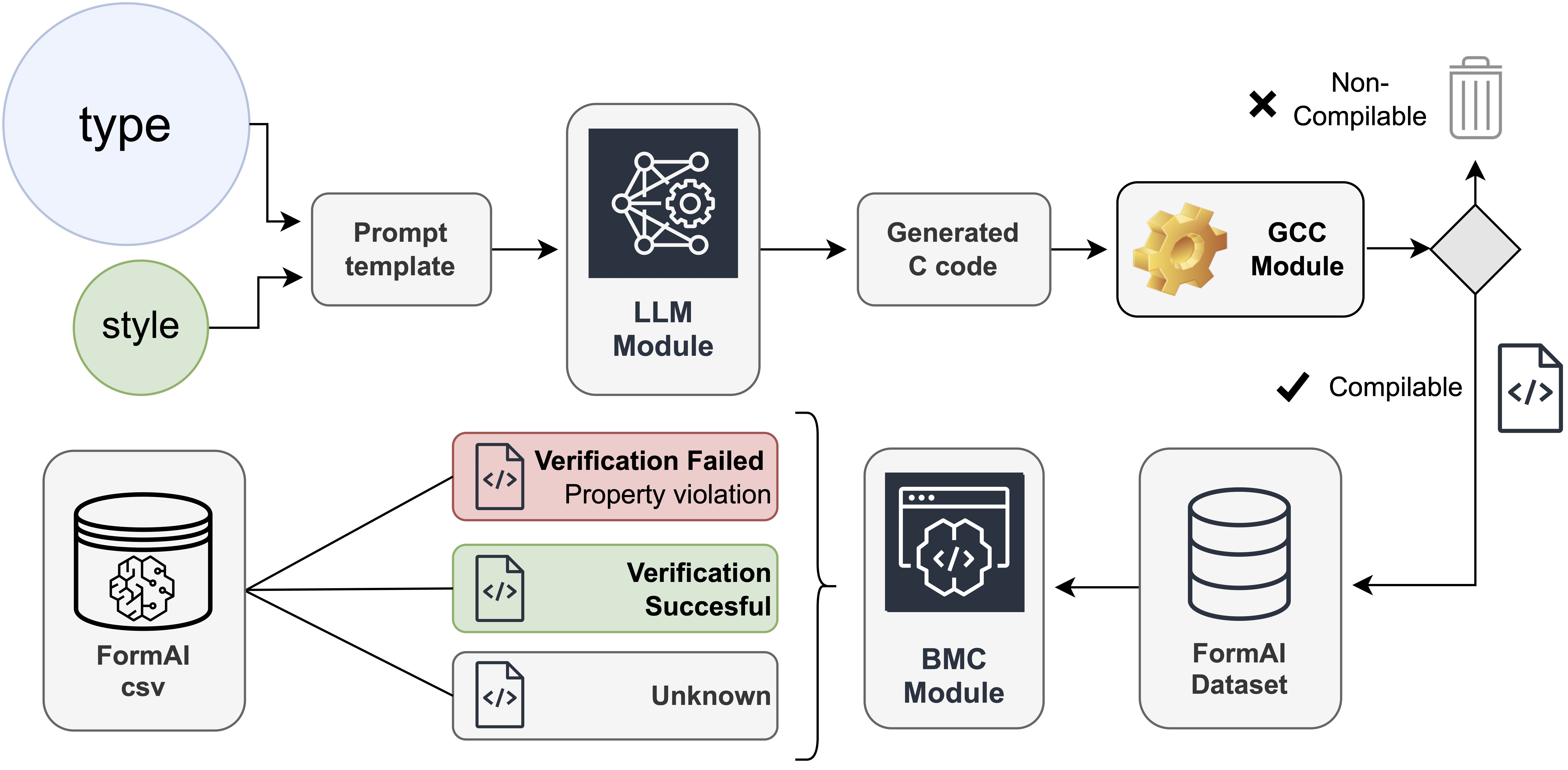}
\caption{AI-driven dataset generation and vulnerability labeling framework. A random type and style combination is selected for each prompt, instructing the LLM module to generate a C program. The compilable programs are fed to the BMC module, which performs the classification based on formal verification techniques.}
\label{pic:method}
\end{figure*}

In particular, we explore these research questions in the context of GPT-3.5 generating C programs. GPT-3.5 is the most widely used LLM available to software developers with a free web interface \cite{somoye_is_2023}. Moreover, C is one of the most popular low-level programming languages for embedded systems, critical security systems, and Internet of Things (IoT) applications \cite{risto_embedded_2022}. 
For our purposes, simply showing through a handful of empirical examples that LLMs can produce vulnerable code is not gratifying and has been demonstrated before for various programming languages \cite{charalambous_new_2023,perry_users_2022,umawing_chatgpt_2023}. 

Two things are required to address the outlined research questions accurately. First, a large database containing a diverse set of C programs. Second, we need to gain insight into the variety and distribution of different vulnerabilities. At the same time, we must determine whether a vulnerability is present in the code. If we label the code as vulnerable, it should not be a false positive. The latter is essential when creating datasets for machine learning purposes \cite{picard_ensuring_2020,hutchinson_towards_2021}.
On that note, deep learning applications also need large datasets of vulnerable source code for training purposes~\cite{chen_diversevul_2023}.

Here, we developed a simple yet effective prompting method to obtain a diverse dataset, prodding the LLM to tackle a mixed bag of tasks. This resulted in a collection of $112,000$ C programs addressing various programming scenarios. Manually labeling the entire dataset is unfeasible for such a large corpus of data. 
Therefore, we use the Efficient SMT-based Bounded Model Checker (ESBMC)~\cite{gadelha2018esbmc}, which can formally falsify the existence of certain vulnerabilities. This state-of-the-art tool showcased exceptional performance in the SV-COMP 2023~\cite{SVCOMP2023} competition by efficiently solving many verification tasks within a limited timeframe~\cite{gadelha2018esbmc}.
Although it can only detect formally verifiable errors through symbolic execution, it does not produce false positives. 

One limitation of this method is that due to its resource-intensive nature, it can only detect vulnerabilities within a predefined search depth bounded by the available computational capacity. Suppose the complexity of the code does not allow the module to check all the nodes in the control-flow graph (CFG)~\cite{Aho:2006:CPT:1177220}  exhaustively under a reasonable time. In that case, we can only know the presence or absence of vulnerabilities within the predefined bound. If we do not find any vulnerabilities up to that depth, the code might still contain some. On the upside, which is why we use this method, we can definitively confirm the presence of the detected vulnerabilities up to a bound, as we can provide a \textit{``counterexample''} as a formal model. Such databases can be useful for various research activities, especially in machine learning, which we remark on in our discussion.

Figure~\ref{pic:method} illustrates the methodology employed in this paper. Initially, we provide instructions to GPT-3.5 to construct a C program for various tasks. This step will be elaborated thoroughly in Section \ref{sec:methodology}.  Next, each output is fed to the GNU C\footnote{\url{https://gcc.gnu.org}} compiler to check if the program is compilable. The compilable source code constitutes the FormAI dataset. 
These programs are used as input for the ESMBC module which performs the labeling process. The labeled data is saved in a \textit{.csv} file, which includes details such as the name of the vulnerable file, the specific line of code containing the vulnerability, the function name, and the type of vulnerability. 

To summarize, this paper holds the following original contributions:
\begin{itemize}

    \item We present FormAI, the first AI-generated large-scale dataset consisting of $112\,000$ independent compilable C programs that perform various computing tasks. Each of these programs is labeled based on the vulnerabilities identified by formal verification, namely, the ESBMC module;
    
    \item A comprehensive analysis on the identification and prevalence of vulnerabilities affecting the safety and security properties of C programs generated by GTP-3.5-turbo. The ESBMC module provides the detection and categorization of vulnerabilities. We connect the identified vulnerability classes with corresponding Common Weakness Enumeration (CWE) numbers.

\end{itemize}

The remaining sections are structured as follows: Section~\ref{sec:motivation} discusses the motivation for our work. Section~\ref{sec:related} overviews the related literature. Section~\ref{sec:preliminaries} presents a short introduction to formal verification and the ESBMC module. Section~\ref{sec:methodology} outlines the approach we employed to create and categorize our dataset, where Section~\ref{sec:experiments} provides an in-depth evaluation of our findings. Section~\ref{sec:limit} overviews limitations related to our work. Finally, Section~\ref{sec:conclusion} concludes the paper with an outlook on possible future research directions.

\section{Motivation}
\label{sec:motivation}

Throughout software development, it is paramount to guarantee the created programs' correctness, safety, and security.
Functionally correct code produces the expected output for each given input.
Safety aims to produce failure tolerant and fail-safe code, resistant against accidental or unexpected inputs that result in correct but undesired outputs, which may cause system failure or erroneous human decisions~\cite{rey1996testing}.
Finally, software security embodies robustness against external hazards and deliberate attacks. 
Our objective in this paper is to examine the safety and security properties of AI-generated source code.

The term \textit{``generated code''} signifies computer code created by an LLM, capable of using multiple forms of data as input.
Textual prompts are segmented into individual units known as tokens. LLMs generate their response one token at a time, where a pre-defined token cap limits the output length.
Due to this, as of today, LLMs cannot be used to spawn large applications on a single prompt.
The main way developers utilize AI tools is by creating small programs or code snippets incorporated into their projects.
The Stanford study mentioned earlier~\cite{perry_users_2022} captured this important aspect. Some assignments given to students were, for example, creating: 
\begin{itemize}
    \item \textit{two functions in Python where one encrypts and the other decrypts a given string using a given symmetric key;} 
    \item \textit{a function in Python that signs a given message using a given ECDSA signing key}. 
\end{itemize}
We aim to prompt the LLM to produce code for tasks with similar complexity levels.
Furthermore, for the ESBMC module, it is beneficial to have smaller independent programs. These allow the module to execute the verification process piece by piece, adhering to a set boundary, thus making the process manageable and more efficient.
If the programs were heavily interdependent, accurately estimating the time required for the module to finish the verification process would be hardly feasible.
The main area of interest in LLM-based code generation has been related to correctness. Datasets such as HumanEval provide programming challenges to assess the performance of models. For example, GPT-4 achieves a $67\%$ success rate in solving tasks compared to $48.1\%$ for GPT-3.5~\cite{openai_gpt-4_2023}. Measuring correctness is not our goal with the FormAI dataset. For example, if the prompt says \textit{``Create a board game using the C programming language in an artistic style''}, correctness would be difficult to verify, especially for a large dataset. The only requirement is that the program should be syntactically correct and it must be possible to compile it.
To restate our research objective, we aim to uncover the proportion and type of frequent coding errors in C source code generated by GPT-3.5 when prompted to perform simple tasks using natural language. The following real-life example demonstrates and underscores the necessity of this research question. 

Imagine a situation where a programmer submits the following prompt to GPT-3.5:
\textit{``Provide a small C program that adds two numbers together.''}.
\begin{figure}[ht] 
\centering
\includegraphics[width=1\textwidth]{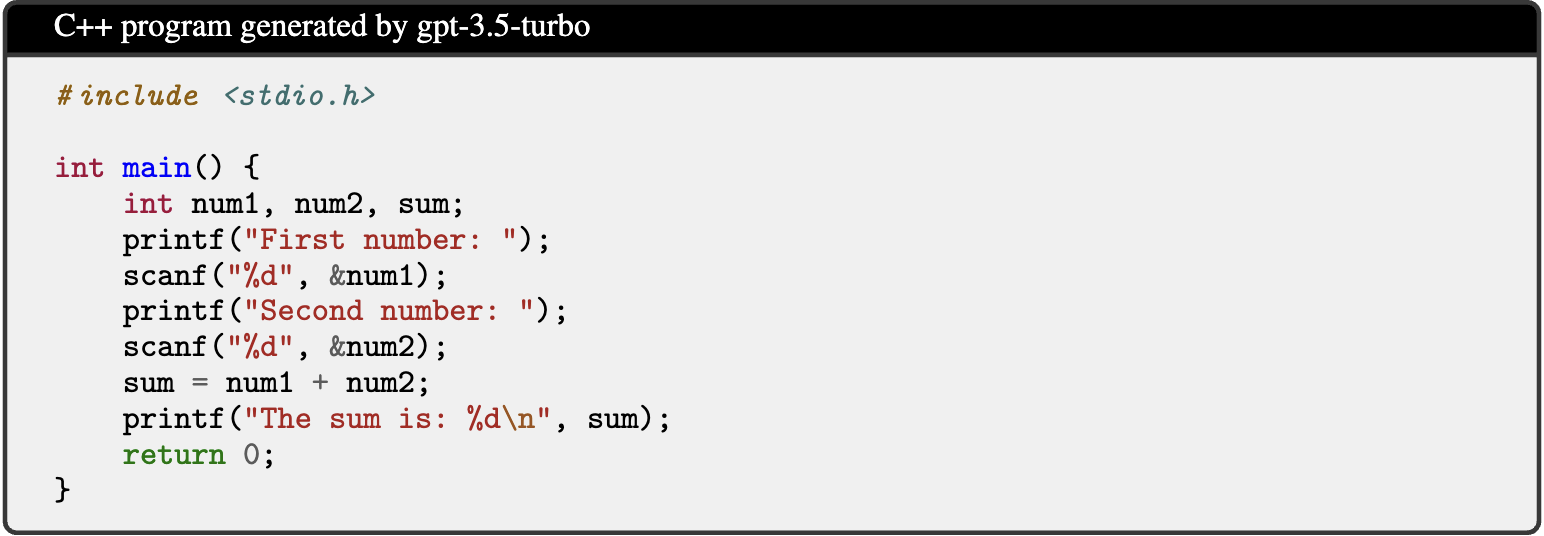}
\caption{Insecure code generated by gpt-3.5-turbo. The program reads two numbers, where the addition can result in a value outside of the range “int” can represent, which may lead to integer overflow.}
\label{list:buffer}
\end{figure}




The resulting code shown in Figure~\ref{list:buffer} is vulnerable, as it contains an integer overflow on the \texttt{scanf()} function. In 32-bit computing architectures, integers are commonly stored as 4 bytes (32 bits), which results in a maximum integer value of $2147483647$, equivalent to $2^{31}-1$. If one attempts to add $2147483647 + 1$ using this small program, the result will be incorrect due to integer overflow. 
The incorrect result will be -2147483648 instead of the expected $2147483648$. The addition exceeds the maximum representable value for a signed 32-bit integer $2^{31} - 1$, causing the integer to wrap around and become negative due to the two's complement representation.

\begin{figure}[ht!] \label{lis:buffer_overflow_example}
\centering
\includegraphics[width=1\textwidth]{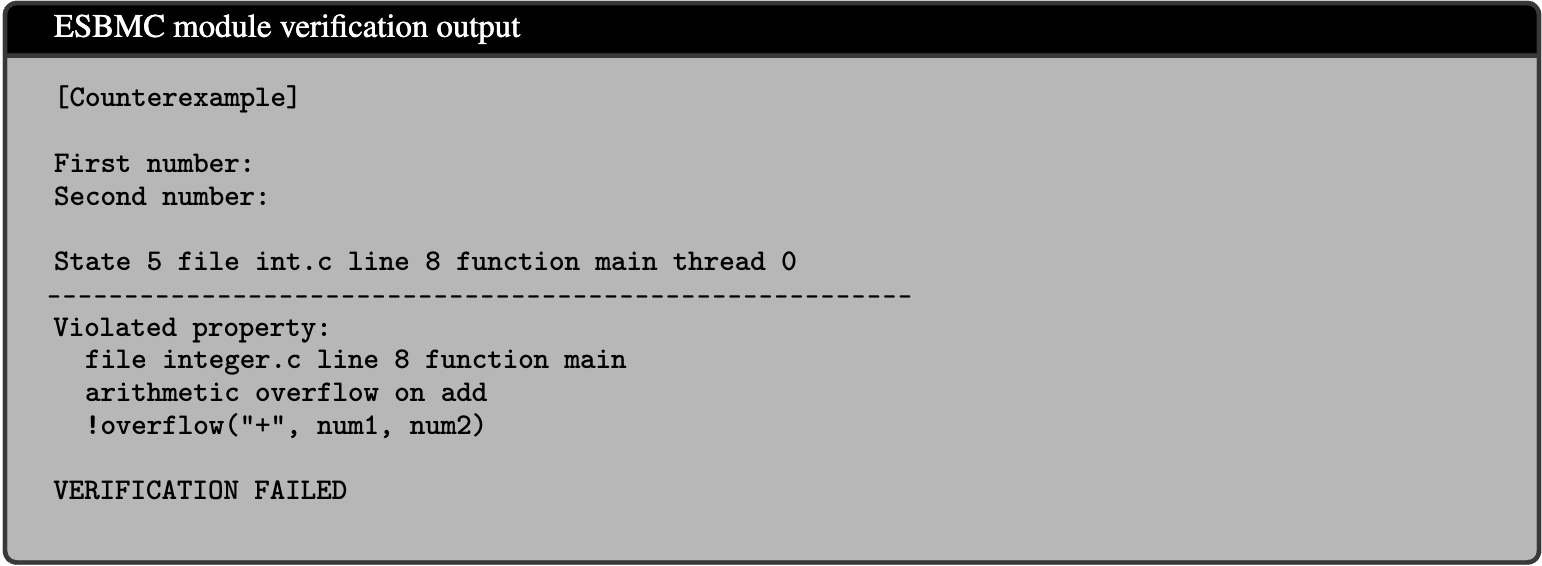}
\caption{The counterexample provided for Figure 2. using ESBMC version 7.2.0. Note this is only part of the output specifically for integer overflow, which we wanted to bring forward for this example.}
\label{fig:esbmc}
\end{figure}

Even when GPT-3.5 is requested to write a secure version of this code --without specifying the vulnerability-- it only attempts to verify against entering non-integer inputs by adding the following code snippet: \texttt{if (scanf("\%d", \&num1) != 1) \{...\}}. 
Clearly, after sanitizing the input, the issue of integer overflow is still present. When prompted to create a C program that adds two numbers, it appears that both GPT-3.5 and GPT-4 generate code with this insecure pattern. 
When asked for a secure version, both models perform input sanitization.
By using the ESBMC module to verify this program, the vulnerability is immediately found through a counterexample, and the following message is created as shown in Figure \ref{fig:esbmc}.





In~\cite{charalambous_new_2023}, the authors demonstrated that GPT-3.5 could efficiently fix errors if the output of the ESBMC module is provided. 
Given only general instruction as \textit{``write secure code''}, or asked to find vulnerabilities, GPT-3.5 struggles to pinpoint the specific vulnerability accurately, let alone if multiple are present.
While advanced models might perform better for certain vulnerabilities, this provides no guarantee that all coding mistakes will be found \cite{pearce_examining_2022}.  
The main challenge is the initial detection without any prior hint or indicator. Doing this efficiently for a large corpus of C code while avoiding false positives and false negatives is still challenging for LLMs \cite{pearce_examining_2022}.
Based on this observation, we want to create an extensive and diverse dataset of properly labeled LLM-generated C programs. Such a dataset can reproduce coding errors often created by LLMs and serve as a valuable resource and starting point in training LLMs for secure code generation.

\section{Related Work}
\label{sec:related}

This section overviews automated vulnerability detection and notable existing datasets containing vulnerable code samples for various training and benchmarking purposes.

\subsection{ChatGPT in Software Engineering}

In \cite{ma_scope_2023} Me et al. assessed the capabilities and limitations of ChatGPT for software engineering (SE), specifically in understanding code syntax and semantic structures like abstract syntax trees (AST), control flow graphs (CFG), and call graphs (CG). ChatGPT exhibits excellent syntax understanding, indicating its potential for static code analysis. They highlighted that the model also hallucinates when interpreting code semantics, creating non-existent facts. This implies a need for methods to verify ChatGPT's outputs to enhance its reliability. This study provides initial insights into why the codes generated by language models are syntactically correct but potentially vulnerable.
Frameworks and techniques for turning prompts into executable code for Software Engineering are rapidly emerging, but the main focus is often functional correctness omitting important security aspects \cite{xing_prompt_2023,white_prompt_2023,yao_tree_2023,wei_chain--thought_2023}.
In \cite{liu_is_2023}, Liu et al. questions the validity of existing code evaluation datasets, suggesting they inadequately assess the correctness of generated code. 

In \cite{khoury_how_2023}  the authors generated 21 small programs in five different languages: C, C++, Python, HTML and Java. Combining manual verification with ChatGPT-based vulnerability detection, the study found that only 5 of the 21 generated programs were initially secure. 
A recent study by Microsoft~\cite{imani2023mathprompter} found that GPT models encounter difficulties when attempting to accurately solve arithmetic operations. This aligns with the findings we presented in the motivation Section.

In a small study involving $50$ students \cite{sandoval_lost_2023}, the authors found that students using an AI coding assistant introduced vulnerabilities at the same rate as their unassisted counterparts. Still, notably, the experiment was limited by focusing only on a single programming scenario.
Contrary to the previous study in \cite{pearce_asleep_2021} Pearce et al. conclude that the control group, which utilized GitHub's Copilot, incorporated more vulnerabilities into their code.
Instead of a single coding scenario like in \cite{sandoval_lost_2023}, the authors expanded the study's comprehensiveness by choosing a diverse set of coding tasks pertinent to high-risk cybersecurity vulnerabilities, such as those featured in MITRE's ``Top 25'' Common Weakness Enumeration (CWE) list.
The study highlights an important lesson: to accurately measure the role of AI tools in code generation or completion, it is essential to choose coding scenarios mirroring a diverse set of relevant real-world settings, thereby facilitating the occurrence of various vulnerabilities. 
This necessitates the creation of code bases replicating a wide range of settings, which is one of the primary goals the FormAI dataset strives to achieve. These studies indicate that AI tools, and in particular ChatGPT, as of today, can produce code containing vulnerabilities.

In a recent study, Shumailov et al. highlighted a phenomenon known as \textit{``model collapse''}~\cite{shumailov_curse_2023}. Their research demonstrated that integrating content generated by LLMs can lead to persistent flaws in subsequent models when using the generated data for training.
This hints that training machine learning models only on purely AI-generated content is insufficient if one aims to prepare these models for detecting vulnerabilities in human-generated code. This is essentially due to using a dataset during the training phase, which is not diverse enough and misrepresents edge cases. We use our dynamic zero-shot prompting method to circumvent the highlighted issue to ensure diversity.
Moreover, our research goal is to find and highlight what coding mistakes AI models can create, which requires a thorough investigation of AI-generated code. On the other hand, AI models themselves were trained on human-generated content; thus, the vulnerabilities produced have roots in incorrect code created by humans.
Yet, as discussed in the next section, existing datasets notoriously include synthetic data (different from AI-generated), which can be useful for benchmarking vulnerability scanners, but has questionable value for training purposes~\cite{chen_diversevul_2023}.

\begin{table}[h]
\caption{Comparisons of various datasets based on their labeling classifications.}
\scriptsize
\center
\renewcommand{\arraystretch}{1.5}
\begin{tabular}{p{0.8cm}cccccccccc}\hline

\textbf{Dataset} & \makecell{\textbf{Only}\\\textbf{C-code}} & \makecell{\textbf{Source}}& \makecell{ \textbf{\#Code}\\ \textbf{ Snippets}} & \makecell{\textbf{\#Vuln.}\\\textbf{Snippets}} & \makecell{\textbf{Multiple}\\ \textbf{Vulns/Snippet}} & \makecell{\textbf{Compiles}/\\ \textbf{Granularity}} & \makecell{\textbf{Vuln.}\\ \textbf{Labelling}} & \makecell{\textbf{\#Avg Line }\\ \textbf{of Code}} & \makecell{\textbf{Labelling}\\\textbf{Method}} \\\hline
\hline
Big-Vul & \ding{56} & Real-World & 188,636  & 100\% & \ding{56} & \ding{56}/Function& CVE/CVW & 30 & PATCH \\ 
\hline

Draper & \ding{56} & Synthetic+Real-World & 1,274,366 & 5.62\% & \ding{52} & \ding{56}/Function & CWE & 29 & STAT \\
\hline

SARD & \ding{56} & Synthetic+Real-World & 100,883 & 100\% & \ding{56} & \ding{52}/Program & CWE & 114 & BDV/STAT/MAN\\
\hline
Juliet & \ding{56} & Synthetic & 106,075 & 100\% & \ding{56} & \ding{52}/Program & CWE & 125 & BDV \\
\hline
Devign & \ding{56} & Real-World & 27,544 & 46.05\% & \ding{56} & \ding{56}/Function & CVE & 112 & ML \\ 
\hline
REVEAL & \ding{56} & Real-World & 22,734  & 9.85\% & \ding{56} & \ding{56}/Function & CVE & 32 & PATCH \\
\hline
DiverseVul & \ding{56} & Real-World & 379,241  & 7.02\% & \ding{56} & \ding{56}/Function & CWE & 37 & PATCH \\
\hline

\textbf{FormAI} & \ding{52} & \textbf{AI-generated} & \textbf{112,000} & 51.24\% & \ding{52} & \ding{52}/Program & CWE & 79 & ESBMC \\
\hline
\end{tabular}
\\
\vspace{5pt}

Legend:\\ \textbf{PATCH}: GitHub Commits Patching a Vuln. \textbf{Man}: Manual Verification, \textbf{Stat}: Static Analyser, \textbf{ML}: Machine Learning Based, \textbf{BDV}: By design vulnerable
\label{tab:compare}
\end{table}

\subsection{Existing databases for Vulnerable C code}

We show how the FormAI dataset compares to seven widely studied datasets containing vulnerable code. The examined datasets are: Big-Vul~\cite{fan_cc_2020}, Draper~\cite{russell_automated_2018,kim_draper_2018}, SARD~\cite{black_software_2018}, Juliet~\cite{jr_juliet_2012}, Devign~\cite{zhou_devign_2019,zhoudataset_devign_2019}, REVEAL \cite{chakraborty_deep_2022}, and DiverseVul\cite{chen_diversevul_2023}. 
Table~\ref{tab:compare} presents a comprehensive comparison of the datasets across various metrics. Some of this data is derived from review papers that evaluate these datasets~\cite{jain2023code,chen_diversevul_2023}.

Big-Vul, Draper, Devign, REVEAL, and DiverseVul comprise vulnerable real-world functions from open-source applications. These five datasets do not include all dependencies of the samples; therefore, they are non-compilable. 
SARD and Juliet contain synthetic, compilable programs. In their general composition, the programs contain a vulnerable function, its equivalent patched function, and a main function calling these functions. All datasets indicate whether a code is vulnerable.
The mentioned datasets use the following vulnerability labeling methodologies:
\begin{itemize}
    \item \texttt{PATCH}: Functions before receiving GitHub commits for detected vulnerabilities are treated as vulnerable. 
    \item \texttt{MAN}: Manual labeling
    \item \texttt{STAT}: Static analyzers
    \item \texttt{ML}: Machine learning-based techniques
    \item \texttt{BDV}: By design vulnerable
\end{itemize}
In the latter case, no vulnerability verification tool is used.
Note that the size of the datasets can be misleading, as many of the datasets contain samples from other languages. For example, SARD contains C, C++, Java, PHP, and C\#.
Moreover, newly released sets often incorporate previous datasets or scrape the same GitHub repositories, making them redundant.

For example, Dreper contains C and C++ code from the SATE IV Juliet Test Suite, Debian Linux distribution, and public Git
repositories. Since the open-source functions from Debian and GitHub were not labeled, the authors used a suite of static analysis tools: Clang, Cppcheck, and Flawfinder~\cite{russell_automated_2018}.
However, the paper does not mention if vulnerabilities were manually verified or if any confirmation has been performed to root out false positives.
In \cite{chen_diversevul_2023}, on top of creating DiverseVul, Chen et al. merged all datasets that were based on GitHub commits and removed duplicates, thus making the most comprehensive collection of GitHub commits containing vulnerable C and C++ code.

\subsection{Vulnerability Scanning and Repair}

Software verification is critical to ensuring correctness, safety, and security. The primary techniques are manual verification, static analysis, and dynamic analysis, where a fourth emerging technique is machine learning-based detection \cite{cordeiro_smt-based_2012,dsilva_survey_2008,wallace_software_1989,ma_scope_2023}. Manual verification techniques such as code review or manual testing rely on human effort and are not scalable. 
Static analysis can test the source code without running it, using techniques such as static symbolic execution, data flow analysis, control flow analysis, and style checking.
On the other hand, dynamic analysis aims at observing software behavior while running the code. It involves fuzzing, automated testing, run-time verification, and profiling. The fourth technique is a promising field where LLMs can be useful in a wide range of tasks, such as code review and bug detection, vulnerability detection, test case generation, and documentation generation; however, as of today, each area has certain limitations. 
Research related to the application of verification tools in analyzing code specifically generated by LLMs remains rather limited.
An earlier work from $2022$ examined the ability of various LLMs to fix vulnerabilities, where the models showed promising results, especially when combined. Still, the authors noted that such tools are not ready to be used in a program repair framework, where further research is necessary to incorporate bug localization. They highlighted challenges in the tool's ability to generate functionally correct code \cite{pearce_examining_2022}.

\section{Formal Verification}
\label{sec:preliminaries}

This section presents the crucial foundational knowledge required to understand the technology employed in this research, specifically Bounded Model Checking (BMC). An intuitive question arises: Could BMC potentially introduce false positives into our dataset? The answer is no, and understanding why is critical to our work. To clarify this theory, we will explain counterexamples and thoroughly discuss the math behind bounded model checking.

Bounded Model Checking (BMC) is a technique used in formal verification to check the correctness of a system within a finite number of steps. It involves modeling the system as a finite state transition system and systematically exploring its state space up to a specified bound or depth.
The latest BMC modules can handle various programming languages \cite{sadowski2014developers, GadelhaSC0N19, white2016deep, zhao2018deepsim, gadelha_esbmc_2023}. This technique first takes the program code, from which a control-flow graph (CFG) is created \cite{Aho:2006:CPT:1177220}. 
In CFG, each node signifies a deterministic or non-deterministic assignment or a conditional statement. Each edge represents a potential shift in the program's control position. Essentially, every node is a block representing a \textit{``set of instructions with a singular entry and exit point''}. Edges indicate possible paths to other blocks to which the program’s control location can transition.
The CFG is first transformed into Static Single Assignment (SSA) and converted into a State Transition System (STS).
This can be interpreted by a Satisfiability Modulo Theories (SMT) solver. This solver can determine if a set of variable assignments makes a given formula true, i.e., this formula is designed to be satisfiable if and only if there's a counterexample to the properties within a specified bound $k$. If there is no error state and the formula is unsatisfiable up to the bound $k$, there is no software vulnerability within that bound. If the solver reaches termination within a bound $\leq k$, we can definitively prove the absence of software errors.

To be more precise, let a given program $\mathcal{P}$ under verification be a finite  state transition system, denoted by a triple $\mathcal{ST}=\left(S, R, I\right)$, where $S$ represents the set of states, $R \subseteq S \times S$ represents the set of transitions and $(s_n, \cdots ,s_m) \in I \subseteq S$ represents the set of initial states.
In a state transition system, a state denoted as $s \in S$ consists of the program counter value, referred to as \textit{pc}, and the values of all program variables. The initial state denoted as $s_1$, assigns the initial program location within the Control Flow Graph (CFG) to \textit{pc}. Each transition $T=(s_i,s_{i+1}) \in R$ between two states, $s_i$ and $s_{i+1}$, is identified with a logical formula $T(s_i,s_{i+1})$. This formula captures the constraints governing the values of the program counter and program variables relevant to the transition.

Within BMC (Bounded Model Checking), properties under verification are defined as follows: $\phi(s)$ represents a logical formula that encodes states satisfying a safety/security property. In contrast, $\psi(s)$ represents a logical formula that encodes states satisfying the completeness threshold, indicating states corresponding to program termination. $\psi(s)$, contains unwindings so that it does not exceed the maximum number of loop iterations in the program. It is worth noting that, in our notation, termination, and error are mutually exclusive: $\phi(s) \wedge \psi(s)$ is by construction unsatisfiable. If $T(s_i, s_{i+1}) \vee \phi(s)$ is unsatisfiable, state $s$ is considered a deadlock state. 
The bounded model checking problem, denoted by $BMC_{\Phi}$ is formulated by constructing a logical formula, and the satisfiability of this formula determines whether $\mathcal{P}$ has a counterexample of length $k$ or less. Specifically, the formula is satisfiable if and only if such a counterexample exists within the given length constraint, i.e.:

\begin{equation}\label{eq:bmc}
 BMC_{\Phi}(k) = I(s_1) \wedge \bigwedge^{k-1}_{i=1} T(s_i, s_{i+1}) \wedge
\bigvee^{k}_{i=1} \neg \phi(s_i).
\end{equation}

In this context, $I$ denotes the set of initial states of $\mathcal{ST}$, and $T(s_i, s_{i+1})$ represents the transition relation of $\mathcal{ST}$, between time steps $i$ and $i+1$. Hence, the logical formula $I(s_1)\wedge\bigwedge^{k-1}_{i=1} T(s_i, s_{i+1})$  represents the executions of $\mathcal{ST}$ with a length of $k$ and $BMC_{\Phi}(k)$ can be satisfied if and only if for some $i \leq k$ there exists a reachable state at time step $i$ in which $\phi$ is violated. If $ BMC_{\Phi}(k) $ is satisfiable, it implies that $\phi$ is violated, and an SMT solver provides a satisfying assignment from which we can extract the values of the program variables to construct a counterexample.

A counterexample, or trace, for a violated property $\phi$, is defined as a finite sequence of states $s_1, \ldots, s_k$, where $s_1, \ldots, s_k \in S$ and $T(s_i, s_{i+1})$ holds for $0 \leq i < k$. If equation (\ref{eq:bmc}) is unsatisfiable, we can conclude that no error state is reachable within $k$ steps or less. This valuable information leads us to conclude that no software vulnerability exists in the program within the specified bound of $k$. 
With this methodology, we aim to classify every generated C program as either vulnerable or not, within a given bound $k$. By searching for counterexamples within this bound, we can establish, based on mathematical proofs, whether a counterexample exists and whether our program $\mathcal{P}$ contains a security vulnerability. This approach allows us to identify security issues such as buffer overflows or access-bound violations.

\subsection{The ESBMC module}
\label{esbmc}

This work uses the Efficient SMT-based Context-Bounded Model Checker (ESBMC)~\cite{gadelha2018esbmc} as our chosen BMC module. ESBMC is a mature, permissively licensed open-source context-bounded model checker for verifying single- and multithreaded C/C++, Kotlin, and Solidity programs. It can automatically verify both predefined safety properties and user-defined program assertions. The safety properties include out-of-bounds array access, illegal pointer dereferences (e.g., dereferencing null, performing an out-of-bounds dereference, double-free of malloced memory, misaligned memory access), integer overflows, undefined behavior on shift operations, floating-point for NaN, divide by zero, and memory leaks. In addition, ESBMC supports the Clang compiler as its C/C++ frontend, the Soot framework via Jimple as its Java/Kotlin frontend, IEEE floating-point arithmetic for various SMT solvers, implements the Solidity grammar production rules as its Solidity frontend. In addition, ESBMC implements state-of-the-art incremental BMC and k-induction proof-rule algorithms based on Satisfiability Modulo Theories (SMT) and Constraint Programming (CP) solvers.

\section{The FormAI dataset}
\label{sec:methodology}

The FormAI dataset consists of two main components: AI-generated C programs and their vulnerability labeling. In the data generation phase, we create a total of $112,000$ samples. In the classification phase, we utilize ESBMC to identify vulnerabilities in the samples. To ensure the reproducibility of the dataset creation process, the exact methodology is thoroughly explained in this section. 

\subsection{Code generation}

To generate the dataset of small C programs, we utilized the GPT-3.5-turbo model~\cite{gpt35}. We employ GPT-3.5 to generate C code instead of GPT-4 as the latter can be up to $60$ times more expensive than the former model. During the creation process, special attention was given to ensuring the diversity of the FormAI dataset, which contains $112,000$ compilable C samples. Requesting the model to generate a unique C program frequently yields similar outcomes, such as adding two numbers or performing basic string manipulation, which does not align with our objectives. Our focus lies in systematically generating a comprehensive and varied collection of small programs that emulates the code creation process undertaken by developers. Therefore, we need a  methodology to circumvent the generation of elementary C programs. To address this issue, we have developed a prompting method consisting of two distinct parts: a dynamic part and a static part. The static component remains consistent and unchanged, while the dynamic portion of the prompt undergoes continuous variation. An example of how a single prompt looks is shown under Listing \ref{fig:prompt}.

\begin{figure}[ht] \label{lis:buffer_overflow_example}
\centering
\includegraphics[width=1\textwidth]{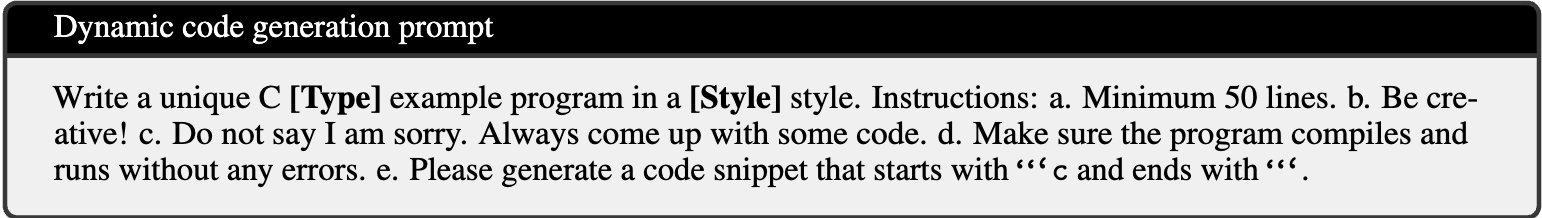}
\caption{Dynamic code generation prompt}
\label{fig:dynamicprompt}
\end{figure}

The dynamic part of the prompt, highlighted as \textbf{[Type]} and \textbf{[Style]}, represent distinct categories within the prompt, each encompassing different elements. In each API call, a different type is selected from a set of $200$ elements for the ``Type'' category. This category contains different topics, such as Wi-Fi Signal Strength Analyzer, QR code reader, Image Steganography, Pixel Art Generator, Scientific Calculator Implementation, etc. In a similar fashion, during each query, a coding style is chosen from a set of $100$ elements within the ``Style'' category. This helps minimize repetition, as specific coding styles such as ``excited'', ``relaxed'', or ``mathematical'' are combined with each Type category. By employing this method, we can generate $200 \times 100 = 20,000$ distinct combinations. This, together with the temperature parameter which regulates the degree of randomness in the model's responses, can enhance diversity among the programs created. The concept of prompt creation can be seen in Figure \ref{fig:dynamicprompt}.

\begin{figure}[ht] \label{pic:dynamicprompt}
\centering
\includegraphics[width=0.5\textwidth]{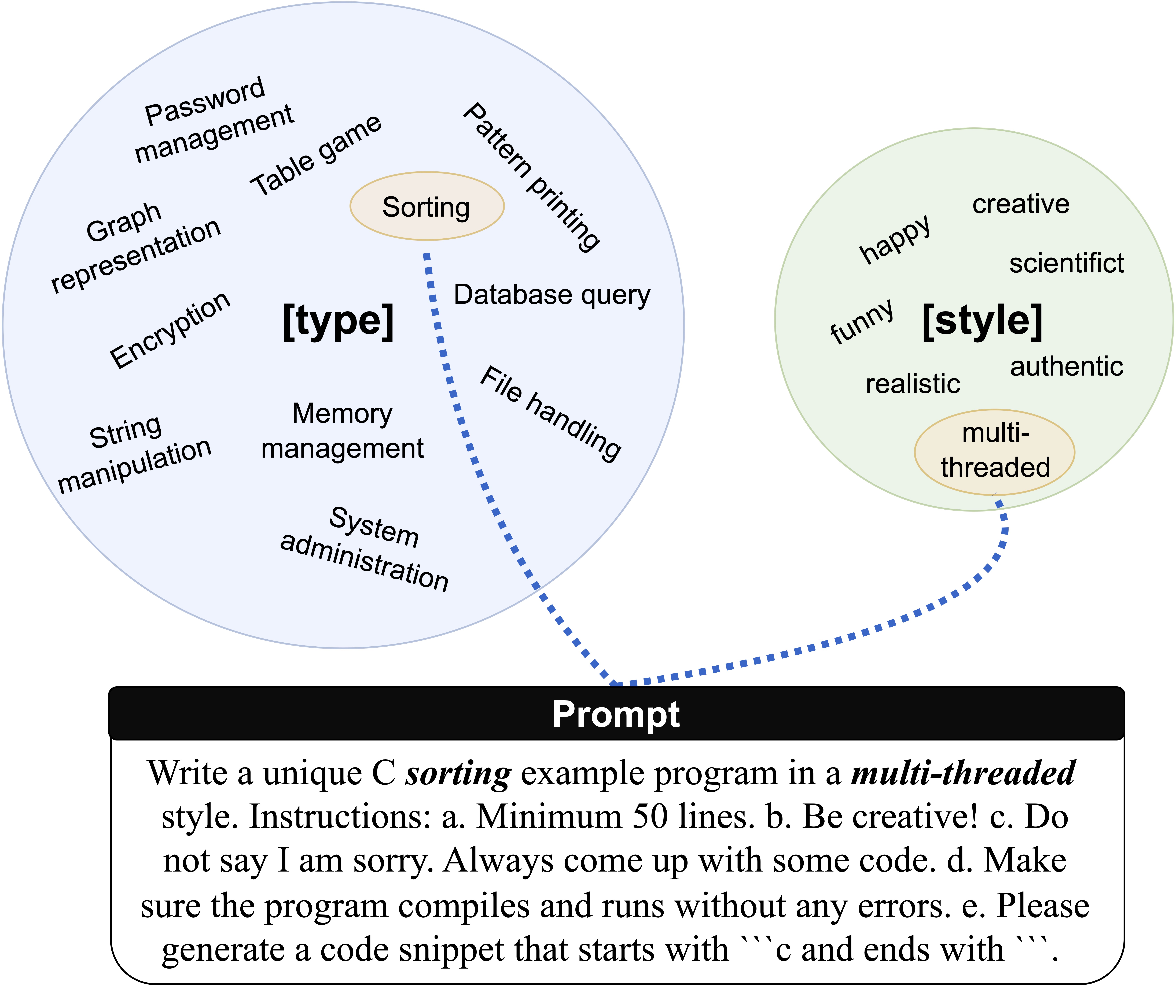}
\caption{Dynamic prompt creation.}
\label{fig:dynamicprompt}
\end{figure}

Decreasing the number of unsuccessful queries is an important factor from an efficiency perspective, as the price for gpt-3.5-turbo is $0.002$ USD\textbackslash 1K token at the time of writing. Hence, refining the prompt to reduce the number of unsuccessful queries holds significant importance. To minimize the error within the generated code, we have established five instructions in each specific prompt:

\begin{enumerate}[label=\alph*.]
    \item \texttt{Minimum 50 lines:} This encourages the LLM to avoid the generation of overly simplistic code with only a few lines (which occasionally still happens);
    \item \texttt{Be creative!:}The purpose of this instruction is to generate a more diverse dataset;
    \item \texttt{Do
not say I am sorry:}The objective of this instruction is to consistently generate code, thereby avoiding responses such as ``As an AI model, I cannot generate code'', and similar statements.
    \item \texttt{Make
sure the program compiles:}This instruction encourages the model to include header files and create a complete and compilable program. 
    \item \texttt{Generate a code snippet that starts with ```c:} Enable easy extraction of the C code from the response.
\end{enumerate}

Once a C code is generated, the GNU C compiler is employed to verify whether the corresponding code is compilable. During the experiment, over $90\%$ of the generated code was compilable. The primary reason for having non-compilable code was due to the absence of necessary headers, such as \texttt{math.h}, \texttt{ctype.h}, or \texttt{stdlib.h}.
During the code generation process, we ensure that the FormAI dataset exclusively consists of compilable code, while excluding any other code that does not meet this criterion.
Code generation does not require high computational power, and for this task, we utilized a standard MacBook Pro $2017$ with $16$ GB of RAM. The generation of $112,000$ code samples was completed within $24$ hours. By leveraging API requests, we ran the creation process in parallel using a single API key. As of the time of writing, the total cost for the creation process was approximately $200$ USD. 

\subsection{Experimental setup for classification}
We ran the classification experiment using an AMD Ryzen Threadripper PRO 3995WX processor with 32 CPU cores. In the worst-case scenario, if we allocate 30 seconds for verification per sample and utilize all 32 threads concurrently, the entire verification process on this machine would take approximately 1.2 days, calculated as $112,000 \times 30 / 3600 / 24 / 32$.

For vulnerability classification, we have selected ESBMC since within a $10$-$30$ second time-limit, this verifier solves the highest amount of verification tasks according to SV-COMP 2023\footnote{\url{https://sv-comp.sosy-lab.org/2023/results/results-verified/quantilePlot-Overall.svg}}. 
Therefore, we use this tool with a verification timeout set to $30$ seconds on each sample. 
\begin{verbatim}
esbmc filename.c  --unwind 1 --overflow --multi-property --timeout 30  
\end{verbatim}
Although it would be possible to collect the vulnerability labels it finds up to that point, we only keep the labels where the verifier finishes within 30 seconds; otherwise, we jump to the next program. 

\subsection{Vulnerability classification}

Let us denote the set of all the generated C samples by $\Sigma$, such that $\Sigma = \{c_1, c_2, \ldots, c_{112000}\}$, where each $c_i$ represents an individual sample. By analyzing the ESBMC verification output, we can classify all the samples into three distinct categories:

\begin{itemize}
    \item $\mathcal{VS} \subseteq \Sigma$: the set of samples for which verification was successful.
    \item $\mathcal{VU} \subseteq \Sigma$: the set of samples for which the verification status is unknown.
    \item $\mathcal{VF} \subseteq \Sigma$: the set of samples for which the verification status failed.
\end{itemize}

These categories are mutually exclusive, meaning a single sample cannot belong to more than one category. Clearly,  $\mathcal{VS} \cap \mathcal{VU} = \mathcal{VS} \cap \mathcal{VF} =\mathcal{VF} \cap \mathcal{VU}  = \emptyset$.
The category labeled as \textit{``verification unknown''} ($\mathcal{VU}$) encompasses all samples where it was not possible to determine a counterexample using ESBMC. This is typically due to the limited search depth, making it uncertain whether a vulnerability exists in the code. It is worth noting that the $\mathcal{VU}$ category is significantly influenced by the runtime duration and the loop unwinding parameter used during ESBMC execution. For instance, if the loop unwinding parameter is set to 30 and the timeout is set to 1 second, many samples are expected to fall into the unknown category. This occurs because only a subset of loops can be unwound up to the specified bound of 30 within the given 1-second timeframe.

Likewise, the category of \textit{``Verification successful''} ($\mathcal{VS}$) indicates that using formal verification methods up to a certain search depth (bound), no counterexample was found in the program. However, it is important to note that increasing the verification time or the unwinding parameter can potentially lead to a change in classification to \textit{``verification failed.''}

On the contrary, \textit{``Verification failed''} represents a completely different scenario and is the main focus of our interest. If a sample is classified as failed by ESBMC, we can be 100\% certain that there is a violation of properties in the program. Additionally, ESBMC provides a counterexample to demonstrate the specific property violation.
We divided \textit{``Verification failed''} into 9 categories, where the first 8 are the most frequently occurring vulnerabilities, while \textit{``Other''} encompasses the remaining results from ESBMC.
\begin{itemize}
 \item $\mathcal{AO} \subseteq \mathcal{VF}$: Arithmetic overflow
  \item $\mathcal{BO} \subseteq \mathcal{VF}$: Buffer overflow on \texttt{scanf()/fscanf()}
\item $\mathcal{ABV} \subseteq \mathcal{VF}$: Array bounds violated 
\item $\mathcal{DFN} \subseteq \mathcal{VF}$ : Dereference failure : NULL pointer
\item $\mathcal{DFF} \subseteq \mathcal{VF}$ : Dereference failure : forgotten memory
\item $\mathcal{DFI} \subseteq \mathcal{VF}$ : Dereference failure : invalid pointer
\item $\mathcal{DFA} \subseteq \mathcal{VF}$ : Dereference failure : array bounds violated
\item $\mathcal{DZ} \subseteq \mathcal{VF}$ :  Division by zero
\item $\mathcal{O} \subseteq \mathcal{VF}$ :  Other vulnerabilities

\end{itemize}

The subsequent subsection will outline the precise distribution of vulnerabilities and the evaluation of the dataset.

\section{Evaluation of the FormAI Dataset}
\label{sec:experiments}

As per our methodology, we verified the compilability of every code piece by utilizing the \texttt{GNU C} compiler. Out of the complete dataset, $109,757$ sample files ($\approx 98$\%) can be compiled without any dependencies solely using the simple command \texttt{gcc -lm -o <filename>}. The remaining $2$\% of samples pose greater complexity, including multithreaded applications, database management applications, and cryptographic applications such as AES encryption. As a result, these samples utilize ten distinct external libraries, including OpenSSL, sqlite3, pthread, and others. Upon successfully installing the following dependencies, all the files can be compiled without any issues:
\textit{libsqlite3-dev, libssl-dev,  libportaudio2, portaudio19-dev,  libpcap-dev, libqrencode-dev, libsdl2-dev,  freeglut3-dev, libcurl4-openssl-dev, libmysqlclient-dev.}

ESBMC is using the clang\footnote{\url{https://clang.llvm.org}} compiler instead of gcc for the verification process. 
Among the $112,000$ samples analyzed, a subset of $786$ samples could not be successfully compiled using clang; therefore, these particular samples were excluded from the classification. Additionally, in a few cases, the ESBMC module crashed when attempting to handle code samples, leading to the exclusion of those samples from the \textit{``.csv''} file containing the vulnerability labels. Despite this, we intentionally chose not to eliminate these samples from the dataset, as they hold value for further research. 


We have examined over 8,848,765 lines of C code, with an average of 79 lines per sample. Programs with 47 lines are the most common, with a total of $1405$ samples. Among the programs in our dataset, only one surpasses a line count of $600$. It is worth mentioning that the FormAI dataset includes all $32$ different C keywords \footnote{\url{https://www.programiz.com/c-programming/list-all-keywords-c-language}}, where for comparison in Juliet, $5$ of the C keywords are not present. The frequency of if-statements, loops, and variables in this context mimics the distribution found in actual real-world projects. We attribute the similarity in patterns exhibited by FormAI to the fact that the training data of GPT models included actual projects from GitHub, which were written by human developers. 

For the classification, in the csv files, we denoted the category \textit{$\mathcal{VS}$:} as \textit{``NOT VULNERABLE up to bound k''}.
The rationale behind this decision is to circumvent potential misinterpretations. We can only confidently assert that a program is devoid of vulnerabilities detectable by ESBMC if we configure both the \texttt{--unwind} and \texttt{timeout} parameters to infinite, and subsequently obtain successful verification. 
This category together with $\mathcal{VU}$ constitutes $48749$ C programs.
In total, from the $112,000$ C programs we performed the verification process on $106139$ files. From this set $57389$ unique programs were found vulnerable, which is over $54\%$ of the examined programs. The overall number of vulnerabilities detected by ESBMC is $197800$.

\subsection{CWE classification} 

Next, we connected the vulnerabilities to Common Weakness Enumeration (CWE) identifiers.
The interconnected and multifaceted nature of software flaws often results in a single vulnerability being associated with multiple CWE identifiers.
Table \ref{tab:vulnerability} showcases a categorization of the most prevalent vulnerabilities and the distribution of the 42 unique CWEs we identified across these categories.
\begin{table}[h!]
\centering
\footnotesize
\renewcommand{\arraystretch}{1.5}
\begin{tabular}{|l|l|p{0.3\textwidth}|}
\hline
\multicolumn{1}{|c|}{\textbf{\#Vulns}} & \textbf{Vuln.} & \textbf{Associated CWE-numbers} \\ \hline
23,312 & $\mathcal{AO}$ & CWE-190, CWE-191, CWE-754,  CWE-680, CWE-681, CWE-682  \\ \hline
11,088 & $\mathcal{ABV}$ &  CWE-119, CWE-125, CWE-129, CWE-131, CWE-193, CWE-787, CWE-788 \\ \hline
88,049 & $\mathcal{BO}$ & CWE-20, CWE-120, CWE-121, CWE-125, CWE-129, CWE-131, CWE-676, CWE-628, CWE-754, CWE-788  \\ \hline
31,829 & $\mathcal{DFN}$ & CWE-391, CWE-476 \\ \hline
24,702 & $\mathcal{DFA}$ & CWE-119, CWE-125, CWE-125, CWE-131, CWE-129, CWE-755, CWE-787 \\ \hline
9823 & $\mathcal{DFI}$ & CWE-416, CWE-476, CWE-690, CWE-822, CWE-824, CWE-825 \\ \hline
5810 & $\mathcal{DFF}$ & CWE-401, CWE-404, CWE-459, CWE-775 \\ \hline
1567 & $\mathcal{DZ}$ & CWE-369 \\ \hline
1620 & $\mathcal{O}$  & CWE-119, CWE-125, CWE-158, CWE-362, CWE-389, CWE-401, CWE-415, CWE-459, CWE-416, CWE-469, CWE-590,  CWE-617, CWE-664, CWE-662, CWE-685, CWE-704, CWE-761, CWE-787, CWE-823, CWE-825, CWE-843  \\ \hline
\end{tabular}
\caption{The vulnerabilities identified by ESBMC, linked to Common Weakness Enumeration identifiers. }
\label{tab:vulnerability}
\end{table}

The \textit{``Other vulnerabilities''} category includes: Assertion failure, Same object violation, Operand of free must have zero pointer offset, function call: not enough arguments, and several types of deference failure issues.

It's vital to emphasize that, in our particular situation, classifying the C programs based on CWE identifiers is not practical, contrary to what is done for other databases like Juliet. As shown in Table \ref{tab:compare}, most datasets contain only one vulnerability per sample. As noted, in the datasets ReVeal, BigVul, Diversevul, a function is vulnerable if the vulnerability-fixing commit changed it. In Juliet, a single vulnerability is introduced for each program.
In our case, a single file often contains not only one but multiple vulnerabilities. Moreover, a single vulnerability can be associated with multiple CWEs. In most cases, multiple CWEs are required as prerequisites for a vulnerability to manifest. 
For example in the case of \textit{``CWE-120: Buffer Copy without Checking Size of Input (Classic Buffer Overflow)''}, there can be other vulnerabilities facilitating the main issue. For example: \textit{``CWE-676: Use of Potentially Dangerous Function''}, which might be the use of \textit{scanf}, and on top of which need \textit{``CWE-20: Improper Input Validation''}.

Labeling the vulnerable function name, line number, and vulnerability type identified by the ESBMC module provides granular information that can be more beneficial to the learning process of the models. This level of detail can allow models to discern patterns and correlations with higher precision, thereby improving vulnerability prediction and detection capabilities.
As our programs contain several vulnerabilities and in some cases multiple instances of the same vulnerability, classifying each into a single CWE group, as done for Juliet, would be less optimal for training purposes.
We also note that while other datasets like DiversVul and Juliet focus more on CWEs related to software security and vulnerabilities that could potentially be exploited, ESBMC detects issues related to software safety as well. 

\subsection{Emerging Research Directions}

The dataset containing all $112,000$ C program files, along with the two .csv files are published on GitHub under the following link: \url{https://github.com/FormAI-Dataset}. 

The diverse structure of the C programs generated in the FormAI dataset made it excellent for an unexpected use case, namely: fuzzing different applications. While running ESBMC on the dataset, we discovered and reported seven bugs in the module. After validating these issues, ESBMC developers managed to resolved them. These included errors in the goto-cc conversion and the creation of invalid SMT solver equations. Additionally, we identified bugs in the CBMC model checker--which is another BMC module--and the Clang compiler which failed to compile several programs that GNU C had no issue with. We promptly communicated these findings to the respective developers.

\section{Limitations and Threats to Validity}
\label{sec:limit}

While ESBMC is a robust tool for detecting many types of errors in C, as of today it is not suited to detect design flaws, semantic errors, or performance issues.
As such, more vulnerabilities might be present in the code besides the detected ones. 
Also, to find all errors detectable by ESBMC, the unwind limit and verification time must be set to infinite. As we provided the original C programs and the instructions on how to run ESBMC, researchers who invest additional computational resources have the potential to enhance our findings.
Our results were reached with: \textit{``--unwind 1 --overflow --multi-property --timeout 30''}. Even with the same settings but using a weaker CPU, for example, ESBMC might not be able to complete the verification process for a several programs, resulting in \textit{``ERROR: Timed out''}.
We were able to run the verification process for $106,139$ programs. The $5861$ difference is because approximately $1$\% of the samples were not compilable by Clang while ESBMC encountered crashes on the rest; as such, ~$5$\% of samples were not classified.

In addition to the reported findings, we found several instances of \textit{``CWE-242: Use of Inherently Dangerous Function''}. Although ESBMC correctly reports several related functions as vulnerable, the reported line number of the vulnerability is often misleading. For instance, when the \texttt{gets()} function is invoked--which is declared in \texttt{io.c}--ESBMC marks a line number in \texttt{io.c} as the place of the vulnerability. This would be misleading for machine learning applications aiming to detect or fix vulnerabilities in the source code; therefore, we excluded such reports from the two csv files.

\section{Conclusions}
\label{sec:conclusion}

This paper shows that GPT-3.5-turbo notoriously introduces vulnerabilities when generating C code.
The broad range of programming scenarios was instrumental in unveiling a variety of coding strategies and evaluating how GPT-3.5 manages specific tasks. This provided an opportunity to pinpoint situations where it might utilize questionable techniques that could potentially introduce a vulnerability.
To facilitate a wide range of programming scenarios we created a zero-shot prompting technique and used GPT-3.5-turbo to generate C code. These programs constitute the FormAI dataset, containing $112,000$ independent compilable C programs.
We used the ESBMC bounded model checker to produce formally verifiable labels for bugs and vulnerabilities. In our experiment, we allocated a verification time of 30 seconds to each program, with the unwinding parameter set to $1$. In total 197800 vulnerable functions were detected by ESBMC. Some programs contain multiple different errors. The labeling is provided in a .csv file which includes: Filename, Vulnerability type, Function name, Line number, and Error type. In addition, we provide an additional .csv file containing the C code as a separate column.
Next, we connected the identified vulnerabilities to CWE identifiers. 
The FormAI dataset is a valuable resource for benchmarking vulnerability detection tools, or to train machine learning algorithms to possess the capabilities of the ESBMC module. The FormAI dataset proves to be a valuable instrument for fuzzing different applications, as we have demonstrated by identifying multiple bugs in the ESBMC and CBMC modules, as well as the Clang compiler.



@article{chakraborty_deep_2022,
	title = {Deep {Learning} {Based} {Vulnerability} {Detection}: {Are} {We} {There} {Yet}?},
	volume = {48},
	issn = {1939-3520},
	shorttitle = {Deep {Learning} {Based} {Vulnerability} {Detection}},
	doi = {10.1109/TSE.2021.3087402},
	abstract = {Automated detection of software vulnerabilities is a fundamental problem in software security. Existing program analysis techniques either suffer from high false positives or false negatives. Recent progress in Deep Learning (DL) has resulted in a surge of interest in applying DL for automated vulnerability detection. Several recent studies have demonstrated promising results achieving an accuracy of up to 95 percent at detecting vulnerabilities. In this paper, we ask, “how well do the state-of-the-art DL-based techniques perform in a real-world vulnerability prediction scenario?” To our surprise, we find that their performance drops by more than 50 percent. A systematic investigation of what causes such precipitous performance drop reveals that existing DL-based vulnerability prediction approaches suffer from challenges with the training data (e.g., data duplication, unrealistic distribution of vulnerable classes, etc.) and with the model choices (e.g., simple token-based models). As a result, these approaches often do not learn features related to the actual cause of the vulnerabilities. Instead, they learn unrelated artifacts from the dataset (e.g., specific variable/function names, etc.). Leveraging these empirical findings, we demonstrate how a more principled approach to data collection and model design, based on realistic settings of vulnerability prediction, can lead to better solutions. The resulting tools perform significantly better than the studied baseline—up to 33.57 percent boost in precision and 128.38 percent boost in recall compared to the best performing model in the literature. Overall, this paper elucidates existing DL-based vulnerability prediction systems’ potential issues and draws a roadmap for future DL-based vulnerability prediction research.},
	number = {9},
	journal = {IEEE Transactions on Software Engineering},
	author = {Chakraborty, Saikat and Krishna, Rahul and Ding, Yangruibo and Ray, Baishakhi},
	month = sep,
	year = {2022},
	note = {Conference Name: IEEE Transactions on Software Engineering},
	keywords = {Data models, Neural networks, Predictive models, Security, Testing, Training, Training data, Vulnerability, deep learning based vulnerability detection, graph neural network based vulnerability detection, real world vulnerabilities},
	pages = {3280--3296},
}

@article{kim_draper_2018,
	title = {Draper {VDISC} {Dataset} - {Vulnerability} {Detection} in {Source} {Code}},
	url = {https://osf.io/d45bw/},
	abstract = {The dataset consists of the source code of 1.27 million functions mined from open source software, labelled by static analysis for potential vulnerabilities. For more details on the dataset and benchmark results, see https://arxiv.org/abs/1807.04320 
    Hosted on the Open Science Framework},
	language = {en},
	urldate = {2023-06-27},
	author = {Kim, Louis and Russell, Rebecca},
	month = nov,
	year = {2018},
	note = {Publisher: OSF},
}

@article{chen_diversevul_2023,
	title = {{DiverseVul}: {A} {New} {Vulnerable} {Source} {Code} {Dataset} for {Deep} {Learning} {Based} {Vulnerability} {Detection}},
	shorttitle = {{DiverseVul}},
	url = {http://arxiv.org/abs/2304.00409},
	abstract = {We propose and release a new vulnerable source code dataset. We curate the dataset by crawling security issue websites, extracting vulnerability-fixing commits and source codes from the corresponding projects. Our new dataset contains 150 CWEs, 26,635 vulnerable functions, and 352,606 non-vulnerable functions extracted from 7,861 commits. Our dataset covers 305 more projects than all previous datasets combined. We show that increasing the diversity and volume of training data improves the performance of deep learning models for vulnerability detection.},
	language = {en},
	urldate = {2023-06-27},
	publisher = {arXiv},
	author = {Chen, Yizheng and Ding, Zhoujie and Chen, Xinyun and Wagner, David},
	month = apr,
	year = {2023},
	note = {arXiv:2304.00409 [cs]},
	keywords = {Computer Science - Artificial Intelligence, Computer Science - Cryptography and Security, Computer Science - Machine Learning, Computer Science - Software Engineering},
}
@online{zhoudataset_devign_2019,
	title = {Devign},
	url = {https://sites.google.com/view/devign},
	abstract = {Abstract
Vulnerabilities are the root cause of cyber threats. Thus vulnerability identification is crucial to protect the software systems from attacks. However, vulnerability discovery is a challenging and tedious process, and also requires specialized security expertise. In this paper,  we},
	language = {en-US},
 author = {Zhou, Yaqin and Liu, Shangqing and Siow, Jingkai and Du, Xiaoning and Liu, Yang},
	month = sep,
	year = {2019},
	urldate = {2023-06-27},
}


@article{zhou_devign_2019,
	title = {Devign: {Effective} {Vulnerability} {Identification} by {Learning} {Comprehensive} {Program} {Semantics} via {Graph} {Neural} {Networks}},
	shorttitle = {Devign},
	url = {http://arxiv.org/abs/1909.03496},
	abstract = {Vulnerability identiﬁcation is crucial to protect the software systems from attacks for cyber security. It is especially important to localize the vulnerable functions among the source code to facilitate the ﬁx. However, it is a challenging and tedious process, and also requires specialized security expertise. Inspired by the work on manually-deﬁned patterns of vulnerabilities from various code representation graphs and the recent advance on graph neural networks, we propose Devign, a general graph neural network based model for graph-level classiﬁcation through learning on a rich set of code semantic representations. It includes a novel Conv module to efﬁciently extract useful features in the learned rich node representations for graph-level classiﬁcation. The model is trained over manually labeled datasets built on 4 diversiﬁed large-scale open-source C projects that incorporate high complexity and variety of real source code instead of synthesis code used in previous works. The results of the extensive evaluation on the datasets demonstrate that Devign outperforms the state of the arts signiﬁcantly with an average of 10.51\% higher accuracy and 8.68\% F1 score, increases averagely 4.66\% accuracy and 6.37\% F1 by the Conv module.},
	language = {en},
	urldate = {2023-06-27},
	publisher = {arXiv},
	author = {Zhou, Yaqin and Liu, Shangqing and Siow, Jingkai and Du, Xiaoning and Liu, Yang},
	month = sep,
	year = {2019},
	note = {arXiv:1909.03496 [cs, stat]},
	keywords = {Computer Science - Cryptography and Security, Computer Science - Machine Learning, Computer Science - Software Engineering, Statistics - Machine Learning},
}

@article{black_software_2018,
	title = {A {Software} {Assurance} {Reference} {Dataset}: {Thousands} of {Programs} {With} {Known} {Bugs}},
	volume = {123},
	issn = {1044-677X},
	shorttitle = {A {Software} {Assurance} {Reference} {Dataset}},
	url = {https://www.ncbi.nlm.nih.gov/pmc/articles/PMC7339570/},
	doi = {10.6028/jres.123.005},
	urldate = {2023-06-27},
	journal = {Journal of Research of the National Institute of Standards and Technology},
	author = {Black, Paul E.},
	month = apr,
	year = {2018},
	pmid = {34877127},
	pmcid = {PMC7339570},
	pages = {1--3},
}

@inproceedings{fan_cc_2020,
	address = {New York, NY, USA},
	series = {{MSR} '20},
	title = {A {C}/{C}++ {Code} {Vulnerability} {Dataset} with {Code} {Changes} and {CVE} {Summaries}},
	isbn = {978-1-4503-7517-7},
	url = {https://doi.org/10.1145/3379597.3387501},
	doi = {10.1145/3379597.3387501},
	abstract = {We collected a large C/C++ code vulnerability dataset from open-source Github projects, namely Big-Vul. We crawled the public Common Vulnerabilities and Exposures (CVE) database and CVE-related source code repositories. Specifically, we collected the descriptive information of the vulnerabilities from the CVE database, e.g., CVE IDs, CVE severity scores, and CVE summaries. With the CVE information and its related published Github code repository links, we downloaded all of the code repositories and extracted vulnerability related code changes. In total, Big-Vul contains 3,754 code vulnerabilities spanning 91 different vulnerability types. All these code vulnerabilities are extracted from 348 Github projects. All information is stored in the CSV format. We linked the code changes with the CVE descriptive information. Thus, our Big-Vul can be used for various research topics, e.g., detecting and fixing vulnerabilities, analyzing the vulnerability related code changes. Big-Vul is publicly available on Github.},
	urldate = {2023-06-27},
	booktitle = {Proceedings of the 17th {International} {Conference} on {Mining} {Software} {Repositories}},
	publisher = {Association for Computing Machinery},
	author = {Fan, Jiahao and Li, Yi and Wang, Shaohua and Nguyen, Tien N.},
	month = sep,
	year = {2020},
	keywords = {C/C++ Code, Code Changes, Common Vulnerabilities and Exposures},
	pages = {508--512},
}

@article{shumailov_curse_2023,
	title = {The {Curse} of {Recursion}: {Training} on {Generated} {Data} {Makes} {Models} {Forget}},
	shorttitle = {The {Curse} of {Recursion}},
	url = {http://arxiv.org/abs/2305.17493},
	abstract = {Stable Diffusion revolutionised image creation from descriptive text. GPT-2, GPT-3(.5) and GPT-4 demonstrated astonishing performance across a variety of language tasks. ChatGPT introduced such language models to the general public. It is now clear that large language models (LLMs) are here to stay, and will bring about drastic change in the whole ecosystem of online text and images. In this paper we consider what the future might hold. What will happen to GPT-\{n\} once LLMs contribute much of the language found online? We find that use of model-generated content in training causes irreversible defects in the resulting models, where tails of the original content distribution disappear. We refer to this effect as model collapse1 and show that it can occur in Variational Autoencoders, Gaussian Mixture Models and LLMs. We build theoretical intuition behind the phenomenon and portray its ubiquity amongst all learned generative models. We demonstrate that it has to be taken seriously if we are to sustain the benefits of training from large-scale data scraped from the web. Indeed, the value of data collected about genuine human interactions with systems will be increasingly valuable in the presence of content generated by LLMs in data crawled from the Internet.},
	language = {en},
	urldate = {2023-06-27},
	publisher = {arXiv},
	author = {Shumailov, Ilia and Shumaylov, Zakhar and Zhao, Yiren and Gal, Yarin and Papernot, Nicolas and Anderson, Ross},
	month = may,
	year = {2023},
	note = {arXiv:2305.17493 [cs]},
	keywords = {Computer Science - Artificial Intelligence, Computer Science - Computation and Language, Computer Science - Computer Vision and Pattern Recognition, Computer Science - Cryptography and Security, Computer Science - Machine Learning},
}
@inproceedings{russell_automated_2018,
	title = {Automated {Vulnerability} {Detection} in {Source} {Code} {Using} {Deep} {Representation} {Learning}},
	doi = {10.1109/ICMLA.2018.00120},
	abstract = {Increasing numbers of software vulnerabilities are discovered every year whether they are reported publicly or discovered internally in proprietary code. These vulnerabilities can pose serious risk of exploit and result in system compromise, information leaks, or denial of service. We leveraged the wealth of C and C++ open-source code available to develop a largescale function-level vulnerability detection system using machine learning. To supplement existing labeled vulnerability datasets, we compiled a vast dataset of millions of open-source functions and labeled it with carefully-selected findings from three different static analyzers that indicate potential exploits. Using these datasets, we developed a fast and scalable vulnerability detection tool based on deep feature representation learning that directly interprets lexed source code. We evaluated our tool on code from both real software packages and the NIST SATE IV benchmark dataset. Our results demonstrate that deep feature representation learning on source code is a promising approach for automated software vulnerability detection.},
	booktitle = {2018 17th {IEEE} {International} {Conference} on {Machine} {Learning} and {Applications} ({ICMLA})},
	author = {Russell, Rebecca and Kim, Louis and Hamilton, Lei and Lazovich, Tomo and Harer, Jacob and Ozdemir, Onur and Ellingwood, Paul and McConley, Marc},
	month = dec,
	year = {2018},
	keywords = {artificial neural networks, computer security, data mining, Feature extraction, machine learning, Machine learning, Open source software, Security, Tools, Training},
	pages = {757--762},
	file = {IEEE Xplore Abstract Record:C\:\\Users\\ifiadmin\\Zotero\\storage\\WXFE7SCH\\8614145.html:text/html;IEEE Xplore Full Text PDF:C\:\\Users\\ifiadmin\\Zotero\\storage\\98IXBGMI\\Russell et al. - 2018 - Automated Vulnerability Detection in Source Code U.pdf:application/pdf},
}

@inproceedings{hutchinson_towards_2021,
	address = {New York, NY, USA},
	series = {{FAccT} '21},
	title = {Towards {Accountability} for {Machine} {Learning} {Datasets}: {Practices} from {Software} {Engineering} and {Infrastructure}},
	isbn = {978-1-4503-8309-7},
	shorttitle = {Towards {Accountability} for {Machine} {Learning} {Datasets}},
	url = {https://dl.acm.org/doi/10.1145/3442188.3445918},
	doi = {10.1145/3442188.3445918},
	abstract = {Datasets that power machine learning are often used, shared, and reused with little visibility into the processes of deliberation that led to their creation. As artificial intelligence systems are increasingly used in high-stakes tasks, system development and deployment practices must be adapted to address the very real consequences of how model development data is constructed and used in practice. This includes greater transparency about data, and accountability for decisions made when developing it. In this paper, we introduce a rigorous framework for dataset development transparency that supports decision-making and accountability. The framework uses the cyclical, infrastructural and engineering nature of dataset development to draw on best practices from the software development lifecycle. Each stage of the data development lifecycle yields documents that facilitate improved communication and decision-making, as well as drawing attention to the value and necessity of careful data work. The proposed framework makes visible the often overlooked work and decisions that go into dataset creation, a critical step in closing the accountability gap in artificial intelligence and a critical/necessary resource aligned with recent work on auditing processes.},
	urldate = {2023-06-26},
	booktitle = {Proceedings of the 2021 {ACM} {Conference} on {Fairness}, {Accountability}, and {Transparency}},
	publisher = {Association for Computing Machinery},
	author = {Hutchinson, Ben and Smart, Andrew and Hanna, Alex and Denton, Emily and Greer, Christina and Kjartansson, Oddur and Barnes, Parker and Mitchell, Margaret},
	month = mar,
	year = {2021},
	keywords = {datasets, machine learning, requirements engineering},
	pages = {560--575},
	file = {Full Text PDF:C\:\\Users\\ifiadmin\\Zotero\\storage\\Y8W2ILET\\Hutchinson et al. - 2021 - Towards Accountability for Machine Learning Datase.pdf:application/pdf},
}

@inproceedings{picard_ensuring_2020,
	title = {Ensuring {Dataset} {Quality} for {Machine} {Learning} {Certification}},
	doi = {10.1109/ISSREW51248.2020.00085},
	abstract = {In this paper, we address the problem of dataset quality in the context of Machine Learning (ML)-based critical systems. We briefly analyse the applicability of some existing standards dealing with data and show that the specificities of the ML context are neither properly captured nor taken into account. As a first answer to this concerning situation, we propose a dataset specification and verification process, and apply it on a signal recognition system from the railway domain. In addition, we also give a list of recommendations for the collection and management of datasets. This work is one step towards the dataset engineering process that will be required for ML to be used on safety critical systems.},
	booktitle = {2020 {IEEE} {International} {Symposium} on {Software} {Reliability} {Engineering} {Workshops} ({ISSREW})},
	author = {Picard, S. and Chapdelaine, C. and Cappi, C. and Gardes, L. and Jenn, E. and Lefevre, B. and Soumarmon, T.},
	month = oct,
	year = {2020},
	keywords = {Data integrity, Data processing, Databases, Machine learning, Rail transportation, Safety, Standards, certification process, datasets, machine learning},
	pages = {275--282},
}

@online{oliynyk_top_2019,
	title = {Top 10 {Best} {Embedded} {Systems} {Programming} {Languages}},
	url = {https://www.geeksforgeeks.org/top-10-best-embedded-systems-programming-languages/},
	abstract = {A Computer Science portal for geeks. It contains well written, well thought and well explained computer science and programming articles, quizzes and practice/competitive programming/company interview Questions.},
	language = {en-us},
	urldate = {2023-06-26},
	journal = {GeeksforGeeks},
	month = may,
	year = {2019},
        author= {Oliynyk, Kostiantyn},
	note = {Section: GBlog},
}

@online{risto_embedded_2022,
	title = {Embedded {Software} {Programming} {Languages}: {Pros}, {Cons}, and {Comparisons} of {Popular} {Languages}},
	shorttitle = {Embedded {Software} {Programming} {Languages}},
	url = {https://www.qt.io/embedded-development-talk/embedded-software-programming-languages-pros-cons-and-comparisons-of-popular-languages},
	abstract = {Learn about programming languages for embedded systems, pros \& cons of popular languages, download a comparison table, and expert advice.},
	language = {en},
        author={Avila,Risto},
        year={2022},
	urldate = {2023-06-26},
}

@online{manuel_llms_2023,
	title = {{LLMs} will fundamentally change software engineering},
	url = {https://dev.to/wesen/llms-will-fundamentally-change-software-engineering-3oj8},
	abstract = {(X-posted from here  Heard about Large Language Models like ChatGPT4, Bing, GPT3? I'm sure you...},
	language = {en},
	urldate = {2023-06-24},
	journal = {DEV Community},
	month = mar,
	year = {2023},
}
@inproceedings{gulwani_dimensions_2010,
	address = {New York, NY, USA},
	series = {{PPDP} '10},
	title = {Dimensions in program synthesis},
	isbn = {978-1-4503-0132-9},
	url = {https://doi.org/10.1145/1836089.1836091},
	doi = {10.1145/1836089.1836091},
	abstract = {Program Synthesis, which is the task of discovering programs that realize user intent, can be useful in several scenarios: enabling people with no programming background to develop utility programs, helping regular programmers automatically discover tricky/mundane details, program understanding, discovery of new algorithms, and even teaching. This paper describes three key dimensions in program synthesis: expression of user intent, space of programs over which to search, and the search technique. These concepts are illustrated by brief description of various program synthesis projects that target synthesis of a wide variety of programs such as standard undergraduate textbook algorithms e.g., sorting, dynamic programming), program inverses(e.g., decoders, deserializers), bitvector manipulation routines, deobfuscated programs, graph algorithms, text-manipulating routines, mutual exclusion algorithms, etc.},
	urldate = {2023-06-24},
	booktitle = {Proceedings of the 12th international {ACM} {SIGPLAN} symposium on {Principles} and practice of declarative programming},
	publisher = {Association for Computing Machinery},
	author = {Gulwani, Sumit},
	month = jul,
	year = {2010},
	keywords = {belief propagation, deductive synthesis, genetic programming, inductive synthesis, machine learning, probabilistic inference, programming by demonstration, programming by examples, sat solving, smt solving},
	pages = {13--24},
}

@online{somoye_is_2023,
	title = {Is {ChatGPT} free and unlimited? {In} short - yes},
	shorttitle = {Is {ChatGPT} free and unlimited?},
	url = {https://www.pcguide.com/apps/chat-gpt-free/},
	abstract = {Looking to use the famous online AI chatbot? Find out if ChatGPT is free in our article here, including what the name actually means.},
	language = {en-US},
	urldate = {2023-06-26},
	journal = {PC Guide},
	author = {Somoye, Funmi Looi},
	month = jun,
	year = {2023},
}

@article{gulwani_program_2017,
	title = {Program {Synthesis}},
	volume = {4},
	issn = {2325-1107, 2325-1131},
	url = {https://www.nowpublishers.com/article/Details/PGL-010},
	doi = {10.1561/2500000010},
	abstract = {Program Synthesis},
	language = {English},
	number = {1-2},
	urldate = {2023-06-24},
	journal = {Foundations and Trends® in Programming Languages},
	author = {Gulwani, Sumit and Polozov, Oleksandr and Singh, Rishabh},
	month = jul,
	year = {2017},
	note = {Publisher: Now Publishers, Inc.},
	pages = {1--119},
}

@article{wei_chain--thought_2023,
	title = {Chain-of-{Thought} {Prompting} {Elicits} {Reasoning} in {Large} {Language} {Models}},
	url = {http://arxiv.org/abs/2201.11903},
	doi = {10.48550/arXiv.2201.11903},
	abstract = {We explore how generating a chain of thought -- a series of intermediate reasoning steps -- significantly improves the ability of large language models to perform complex reasoning. In particular, we show how such reasoning abilities emerge naturally in sufficiently large language models via a simple method called chain of thought prompting, where a few chain of thought demonstrations are provided as exemplars in prompting. Experiments on three large language models show that chain of thought prompting improves performance on a range of arithmetic, commonsense, and symbolic reasoning tasks. The empirical gains can be striking. For instance, prompting a 540B-parameter language model with just eight chain of thought exemplars achieves state of the art accuracy on the GSM8K benchmark of math word problems, surpassing even finetuned GPT-3 with a verifier.},
	urldate = {2023-06-24},
	publisher = {arXiv},
	author = {Wei, Jason and Wang, Xuezhi and Schuurmans, Dale and Bosma, Maarten and Ichter, Brian and Xia, Fei and Chi, Ed and Le, Quoc and Zhou, Denny},
	month = jan,
	year = {2023},
	note = {arXiv:2201.11903 [cs]},
	keywords = {Computer Science - Artificial Intelligence, Computer Science - Computation and Language},
}

@article{white_prompt_2023,
	title = {A {Prompt} {Pattern} {Catalog} to {Enhance} {Prompt} {Engineering} with {ChatGPT}},
	url = {http://arxiv.org/abs/2302.11382},
	doi = {10.48550/arXiv.2302.11382},
	abstract = {Prompt engineering is an increasingly important skill set needed to converse effectively with large language models (LLMs), such as ChatGPT. Prompts are instructions given to an LLM to enforce rules, automate processes, and ensure specific qualities (and quantities) of generated output. Prompts are also a form of programming that can customize@misc{wei_chain--thought_2023,
	title = {Chain-of-{Thought} {Prompting} {Elicits} {Reasoning} in {Large} {Language} {Models}},
	url = {http://arxiv.org/abs/2201.11903},
	doi = {10.48550/arXiv.2201.11903},
	abstract = {We explore how generating a chain of thought -- a series of intermediate reasoning steps -- significantly improves the ability of large language models to perform complex reasoning. In particular, we show how such reasoning abilities emerge naturally in sufficiently large language models via a simple method called chain of thought prompting, where a few chain of thought demonstrations are provided as exemplars in prompting. Experiments on three large language models show that chain of thought prompting improves performance on a range of arithmetic, commonsense, and symbolic reasoning tasks. The empirical gains can be striking. For instance, prompting a 540B-parameter language model with just eight chain of thought exemplars achieves state of the art accuracy on the GSM8K benchmark of math word problems, surpassing even finetuned GPT-3 with a verifier.},
	urldate = {2023-06-24},
	publisher = {arXiv},
	author = {Wei, Jason and Wang, Xuezhi and Schuurmans, Dale and Bosma, Maarten and Ichter, Brian and Xia, Fei and Chi, Ed and Le, Quoc and Zhou, Denny},
	month = jan,
	year = {2023},
	note = {arXiv:2201.11903 [cs]},
	keywords = {Computer Science - Artificial Intelligence, Computer Science - Computation and Language},
} the outputs and interactions with an LLM. This paper describes a catalog of prompt engineering techniques presented in pattern form that have been applied to solve common problems when conversing with LLMs. Prompt patterns are a knowledge transfer method analogous to software patterns since they provide reusable solutions to common problems faced in a particular context, i.e., output generation and interaction when working with LLMs. This paper provides the following contributions to research on prompt engineering that apply LLMs to automate software development tasks. First, it provides a framework for documenting patterns for structuring prompts to solve a range of problems so that they can be adapted to different domains. Second, it presents a catalog of patterns that have been applied successfully to improve the outputs of LLM conversations. Third, it explains how prompts can be built from multiple patterns and illustrates prompt patterns that benefit from combination with other prompt patterns.},
	urldate = {2023-06-24},
	publisher = {arXiv},
	author = {White, Jules and Fu, Quchen and Hays, Sam and Sandborn, Michael and Olea, Carlos and Gilbert, Henry and Elnashar, Ashraf and Spencer-Smith, Jesse and Schmidt, Douglas C.},
	month = feb,
	year = {2023},
	note = {arXiv:2302.11382 [cs]},
	keywords = {Computer Science - Artificial Intelligence, Computer Science - Software Engineering},
}

@article{xing_prompt_2023,
	title = {Prompt {Sapper}: {LLM}-{Empowered} {Software} {Engineering} {Infrastructure} for {AI}-{Native} {Services}},
	shorttitle = {Prompt {Sapper}},
	url = {http://arxiv.org/abs/2306.02230},
	doi = {10.48550/arXiv.2306.02230},
	abstract = {Foundation models, such as GPT-4, DALL-E have brought unprecedented AI "operating system" effect and new forms of human-AI interaction, sparking a wave of innovation in AI-native services, where natural language prompts serve as executable "code" directly (prompt as executable code), eliminating the need for programming language as an intermediary and opening up the door to personal AI. Prompt Sapper has emerged in response, committed to support the development of AI-native services by AI chain engineering. It creates a large language model (LLM) empowered software engineering infrastructure for authoring AI chains through human-AI collaborative intelligence, unleashing the AI innovation potential of every individual, and forging a future where everyone can be a master of AI innovation. This article will introduce the R{\textbackslash}\&D motivation behind Prompt Sapper, along with its corresponding AI chain engineering methodology and technical practices.},
	urldate = {2023-06-24},
	publisher = {arXiv},
	author = {Xing, Zhenchang and Huang, Qing and Cheng, Yu and Zhu, Liming and Lu, Qinghua and Xu, Xiwei},
	month = jun,
	year = {2023},
	note = {arXiv:2306.02230 [cs]},
	keywords = {Computer Science - Software Engineering},
}
@article{wallace_software_1989,
	title = {Software verification and validation: an overview},
	volume = {6},
	issn = {0740-7459},
	shorttitle = {Software verification and validation},
	url = {http://ieeexplore.ieee.org/document/28119/},
	doi = {10.1109/52.28119},
	language = {en},
	number = {3},
	urldate = {2023-06-22},
	journal = {IEEE Software},
	author = {Wallace, D.R. and Fujii, R.U.},
	month = may,
	year = {1989},
	pages = {10--17},
}

@article{dsilva_survey_2008,
	title = {A {Survey} of {Automated} {Techniques} for {Formal} {Software} {Verification}},
	volume = {27},
	issn = {1937-4151},
	doi = {10.1109/TCAD.2008.923410},
	abstract = {The quality and the correctness of software are often the greatest concern in electronic systems. Formal verification tools can provide a guarantee that a design is free of specific flaws. This paper surveys algorithms that perform automatic static analysis of software to detect programming errors or prove their absence. The three techniques considered are static analysis with abstract domains, model checking, and bounded model checking. A short tutorial on these techniques is provided, highlighting their differences when applied to practical problems. This paper also surveys tools implementing these techniques and describes their merits and shortcomings.},
	number = {7},
	journal = {IEEE Transactions on Computer-Aided Design of Integrated Circuits and Systems},
	author = {D'Silva, Vijay and Kroening, Daniel and Weissenbacher, Georg},
	month = jul,
	year = {2008},
	note = {Conference Name: IEEE Transactions on Computer-Aided Design of Integrated Circuits and Systems},
	keywords = {Algorithm design and analysis, Automatic programming, Automatic testing, Bounded model checking (BMC), Formal verification, Hardware, Performance analysis, Software algorithms, Software performance, Software quality, Software systems, model checking, predicate abstraction, software verification, static analysis},
	pages = {1165--1178},
}

@online{openai_platform_2023,
	title = {{OpenAI} {Platform}},
	url = {https://platform.openai.com},
	abstract = {Explore developer resources, tutorials, API docs, and dynamic examples to get the most out of OpenAI's platform.},
	language = {en},
	urldate = {2023-06-22},
	author = {{OpenAI}},
}

@inproceedings{CordeiroF11,
 author       = {Lucas C. Cordeiro and
                 Bernd Fischer},
 editor       = {Richard N. Taylor and
                 Harald C. Gall and
                 Nenad Medvidovic},
 title        = {Verifying multi-threaded software using smt-based context-bounded
                 model checking},
 booktitle    = {Proceedings of the 33rd International Conference on Software Engineering,
                 {ICSE} 2011, Waikiki, Honolulu , HI, USA, May 21-28, 2011},
 pages        = {331--340},
 publisher    = {{ACM}},
 year         = {2011},
 doi          = {10.1145/1985793.1985839}
}

@inproceedings{Sheera00,
 author = "Sheeran, Mary and Singh, Satnam and St{\aa}lmarck, Gunnar",
 title = "Checking Safety Properties Using Induction And A {SAT}-Solver",
 booktitle = "Formal Methods In Computer-Aided Design",
 pages = "108--125",
 year = "2000",
 series = "{LNCS}",
 volume = "1954"
}

@inproceedings{Kinductor,
 author = {Donaldson, Alastair and Haller, Leopold and Kroening, Daniel and R{\"{u}}mmer, Philipp},
 title = "Software Verification Using $k$-Induction",
 booktitle = "Static Analysis Symposium",
 pages = "351--368",
 year = "2011",
 series = "{LNCS}",
 volume = "6887"
}

@article{GadelhaIC17,
  author       = {Mikhail Y. R. Gadelha and
                 Hussama Ibrahim Ismail and
                  Lucas C. Cordeiro},
  title        = {Handling loops in bounded model checking of {C} programs via k-induction},
  journal      = {Int. J. Softw. Tools Technol. Transf.},
  volume       = {19},
  number       = {1},
  pages        = {97--114},
  year         = {2017},
  doi          = {10.1007/s10009-015-0407-9}
}
@online{trend_micro_devops_resource_center_security_2023,
	title = {Security {Vulnerabilities} of {ChatGPT}-{Generated} {Code}},
	url = {https://www.trendmicro.com/en_no/devops/23/e/chatgpt-security-vulnerabilities.html},
	abstract = {ChatGPT is a Large Language Model (LLM) based on the GPT-3.5 architecture that OpenAI built and trained - learn the security vulnerabilities of ChatGPT-generated code.},
	language = {en-NO},
	urldate = {2023-06-22},
	journal = {Trend Micro},
	author = {{Trend Micro DevOps Resource Center}},
	month = may,
	year = {2023},
	note = {Section: expert perspective},
}

@online{umawing_chatgpt_2023,
	title = {{ChatGPT} writes insecure code},
	url = {https://www.malwarebytes.com/blog/news/2023/04/chatgpt-creates-not-so-secure-code-study-finds},
	abstract = {Researchers have found that ChatGPT, OpenAI's popular chatbot, is prone to generating insecure code.},
	language = {en},
	urldate = {2023-06-22},
	journal = {Malwarebytes},
	author = {Umawing, Jovi},
	month = apr,
	year = {2023},
}
@article{chavez_chat_2023,
	title = {Chat {Generative} {Pre}-trained {Transformer}: why we should embrace this technology},
	volume = {228},
	issn = {0002-9378},
	shorttitle = {Chat {Generative} {Pre}-trained {Transformer}},
	url = {https://www.sciencedirect.com/science/article/pii/S0002937823001552},
	doi = {10.1016/j.ajog.2023.03.010},
	abstract = {With the advent of artificial intelligence that not only can learn from us but also can communicate with us in plain language, humans are embarking on a brave new future. The interaction between humans and artificial intelligence has never been so widespread. Chat Generative Pre-trained Transformer is an artificial intelligence resource that has potential uses in the practice of medicine. As clinicians, we have the opportunity to help guide and develop new ways to use this powerful tool. Optimal use of any tool requires a certain level of comfort. This is best achieved by appreciating its power and limitations. Being part of the process is crucial in maximizing its use in our field. This clinical opinion demonstrates the potential uses of Chat Generative Pre-trained Transformer for obstetrician-gynecologists and encourages readers to serve as the driving force behind this resource.},
	language = {en},
	number = {6},
	urldate = {2023-06-22},
	journal = {American Journal of Obstetrics and Gynecology},
	author = {Chavez, Martin R. and Butler, Thomas S. and Rekawek, Patricia and Heo, Hye and Kinzler, Wendy L.},
	month = jun,
	year = {2023},
	keywords = {Chat Generative Pre-trained Transformer, OpenAI, artificial intelligence, future of medicine, large language models, plain conversational language interfaces},
	pages = {706--711},
}

@inproceedings{ross_programmers_2023,
	address = {New York, NY, USA},
	series = {{IUI} '23},
	title = {The {Programmer}’s {Assistant}: {Conversational} {Interaction} with a {Large} {Language} {Model} for {Software} {Development}},
	isbn = {9798400701061},
	shorttitle = {The {Programmer}’s {Assistant}},
	url = {https://dl.acm.org/doi/10.1145/3581641.3584037},
	doi = {10.1145/3581641.3584037},
	abstract = {Large language models (LLMs) have recently been applied in software engineering to perform tasks such as translating code between programming languages, generating code from natural language, and autocompleting code as it is being written. When used within development tools, these systems typically treat each model invocation independently from all previous invocations, and only a specific limited functionality is exposed within the user interface. This approach to user interaction misses an opportunity for users to more deeply engage with the model by having the context of their previous interactions, as well as the context of their code, inform the model’s responses. We developed a prototype system – the Programmer’s Assistant – in order to explore the utility of conversational interactions grounded in code, as well as software engineers’ receptiveness to the idea of conversing with, rather than invoking, a code-fluent LLM. Through an evaluation with 42 participants with varied levels of programming experience, we found that our system was capable of conducting extended, multi-turn discussions, and that it enabled additional knowledge and capabilities beyond code generation to emerge from the LLM. Despite skeptical initial expectations for conversational programming assistance, participants were impressed by the breadth of the assistant’s capabilities, the quality of its responses, and its potential for improving their productivity. Our work demonstrates the unique potential of conversational interactions with LLMs for co-creative processes like software development.},
	urldate = {2023-06-22},
	booktitle = {Proceedings of the 28th {International} {Conference} on {Intelligent} {User} {Interfaces}},
	publisher = {Association for Computing Machinery},
	author = {Ross, Steven I. and Martinez, Fernando and Houde, Stephanie and Muller, Michael and Weisz, Justin D.},
	month = mar,
	year = {2023},
	keywords = {code-fluent large language models, conversational interaction, foundation models, human-centered AI},
	pages = {491--514},
}

@article{bui_codetf_2023,
	title = {{CodeTF}: {One}-stop {Transformer} {Library} for {State}-of-the-art {Code} {LLM}},
	shorttitle = {{CodeTF}},
	url = {http://arxiv.org/abs/2306.00029},
	abstract = {Code intelligence plays a key role in transforming modern software engineering. Recently, deep learning-based models, especially Transformer-based large language models (LLMs), have demonstrated remarkable potential in tackling these tasks by leveraging massive open-source code data and programming language features. However, the development and deployment of such models often require expertise in both machine learning and software engineering, creating a barrier for the model adoption. In this paper, we present CodeTF, an open-source Transformer-based library for state-of-the-art Code LLMs and code intelligence. Following the principles of modular design and extensible framework, we design CodeTF with a unified interface to enable rapid access and development across different types of models, datasets and tasks. Our library supports a collection of pretrained Code LLM models and popular code benchmarks, including a standardized interface to train and serve code LLMs efficiently, and data features such as language-specific parsers and utility functions for extracting code attributes. In this paper, we describe the design principles, the architecture, key modules and components, and compare with other related library tools. Finally, we hope CodeTF is able to bridge the gap between machine learning/generative AI and software engineering, providing a comprehensive open-source solution for developers, researchers, and practitioners.},
	language = {en},
	urldate = {2023-06-22},
	publisher = {arXiv},
	author = {Bui, Nghi D. Q. and Le, Hung and Wang, Yue and Li, Junnan and Gotmare, Akhilesh Deepak and Hoi, Steven C. H.},
	month = may,
	year = {2023},
	note = {arXiv:2306.00029 [cs]},
	keywords = {Computer Science - Artificial Intelligence, Computer Science - Software Engineering},
}

@article{sandoval_lost_2023,
	title = {Lost at {C}: {A} {User} {Study} on the {Security} {Implications} of {Large} {Language} {Model} {Code} {Assistants}},
	shorttitle = {Lost at {C}},
	url = {http://arxiv.org/abs/2208.09727},
	abstract = {Large Language Models (LLMs) such as OpenAI Codex are increasingly being used as AI-based coding assistants. Understanding the impact of these tools on developers’ code is paramount, especially as recent work showed that LLMs may suggest cybersecurity vulnerabilities. We conduct a securitydriven user study (N=58) to assess code written by student programmers when assisted by LLMs. Given the potential severity of low-level bugs as well as their relative frequency in real-world projects, we tasked participants with implementing a singly-linked ‘shopping list’ structure in C. Our results indicate that the security impact in this setting (low-level C with pointer and array manipulations) is small: AI-assisted users produce critical security bugs at a rate no greater than 10\% more than the control, indicating the use of LLMs does not introduce new security risks.},
	language = {en},
	urldate = {2023-06-10},
	publisher = {arXiv},
	author = {Sandoval, Gustavo and Pearce, Hammond and Nys, Teo and Karri, Ramesh and Garg, Siddharth and Dolan-Gavitt, Brendan},
	month = feb,
	year = {2023},
	note = {arXiv:2208.09727 [cs]},
	keywords = {Computer Science - Cryptography and Security},
}

@article{cai_large_2023,
	title = {Large {Language} {Models} as {Tool} {Makers}},
	url = {http://arxiv.org/abs/2305.17126},
	doi = {10.48550/arXiv.2305.17126},
	abstract = {Recent research shows the potential of enhancing the problem-solving ability of large language models (LLMs) through the use of external tools. However, prior work along this line depends on the availability of existing tools. In this work, we take an initial step towards removing this dependency by proposing a closed-loop framework, referred to as LLMs As Tool Makers (LATM), where LLMs create their own reusable tools for problem-solving. Our approach consists of two key phases: 1) tool making: an LLM acts as the tool maker that crafts tools for given tasks, where a tool is implemented as a Python utility function. 2) tool using: an LLM acts as the tool user, which applies the tool built by the tool maker for problem-solving. The tool user can be either the same or a different LLM from the tool maker. Tool-making enables an LLM to continually generate tools that can be applied to different requests so that future requests can call the corresponding APIs when beneficial for solving the tasks. Furthermore, the division of labor among LLMs for tool-making and tool-using phases introduces the opportunity to achieve cost effectiveness without degrading the quality of generated tools and problem solutions. For example, recognizing that tool-making demands more sophisticated capabilities than tool-using, we can apply a powerful yet resource-intensive model as the tool maker, and a lightweight while cost-effective model as the tool user. We validate the effectiveness of our approach across a variety of complex reasoning tasks, including Big-Bench tasks. With GPT-4 as the tool maker and GPT-3.5 as the tool user, LATM can achieve performance that is on par with using GPT-4 for both tool making and tool using, while the inference cost is significantly reduced.},
	urldate = {2023-06-10},
	publisher = {arXiv},
	author = {Cai, Tianle and Wang, Xuezhi and Ma, Tengyu and Chen, Xinyun and Zhou, Denny},
	month = may,
	year = {2023},
	note = {arXiv:2305.17126 [cs, stat]},
	keywords = {Computer Science - Artificial Intelligence, Computer Science - Computation and Language, Computer Science - Machine Learning, Statistics - Machine Learning},
}

@article{pearce_asleep_2021,
	title = {Asleep at the {Keyboard}? {Assessing} the {Security} of {GitHub} {Copilot}'s {Code} {Contributions}},
	shorttitle = {Asleep at the {Keyboard}?},
	url = {http://arxiv.org/abs/2108.09293},
	doi = {10.48550/arXiv.2108.09293},
	abstract = {There is burgeoning interest in designing AI-based systems to assist humans in designing computing systems, including tools that automatically generate computer code. The most notable of these comes in the form of the first self-described `AI pair programmer', GitHub Copilot, a language model trained over open-source GitHub code. However, code often contains bugs - and so, given the vast quantity of unvetted code that Copilot has processed, it is certain that the language model will have learned from exploitable, buggy code. This raises concerns on the security of Copilot's code contributions. In this work, we systematically investigate the prevalence and conditions that can cause GitHub Copilot to recommend insecure code. To perform this analysis we prompt Copilot to generate code in scenarios relevant to high-risk CWEs (e.g. those from MITRE's "Top 25" list). We explore Copilot's performance on three distinct code generation axes -- examining how it performs given diversity of weaknesses, diversity of prompts, and diversity of domains. In total, we produce 89 different scenarios for Copilot to complete, producing 1,689 programs. Of these, we found approximately 40\% to be vulnerable.},
	urldate = {2023-06-10},
	publisher = {arXiv},
	author = {Pearce, Hammond and Ahmad, Baleegh and Tan, Benjamin and Dolan-Gavitt, Brendan and Karri, Ramesh},
	month = dec,
	year = {2021},
	note = {arXiv:2108.09293 [cs]},
	keywords = {Computer Science - Artificial Intelligence, Computer Science - Cryptography and Security},
}

@article{pearce_examining_2022,
	title = {Examining {Zero}-{Shot} {Vulnerability} {Repair} with {Large} {Language} {Models}},
	url = {http://arxiv.org/abs/2112.02125},
	abstract = {Human developers can produce code with cybersecurity bugs. Can emerging ‘smart’ code completion tools help repair those bugs? In this work, we examine the use of large language models (LLMs) for code (such as OpenAI’s Codex and AI21’s Jurassic J-1) for zero-shot vulnerability repair. We investigate challenges in the design of prompts that coax LLMs into generating repaired versions of insecure code. This is difﬁcult due to the numerous ways to phrase key information—both semantically and syntactically—with natural languages. We perform a large scale study of ﬁve commercially available, blackbox, “off-the-shelf” LLMs, as well as an open-source model and our own locally-trained model, on a mix of synthetic, handcrafted, and real-world security bug scenarios. Our experiments demonstrate that while the approach has promise (the LLMs could collectively repair 100\% of our synthetically generated and hand-crafted scenarios), a qualitative evaluation of the model’s performance over a corpus of historical real-world examples highlights challenges in generating functionally correct code.},
	language = {en},
	urldate = {2023-06-10},
	publisher = {arXiv},
	author = {Pearce, Hammond and Tan, Benjamin and Ahmad, Baleegh and Karri, Ramesh and Dolan-Gavitt, Brendan},
	month = aug,
	year = {2022},
	note = {arXiv:2112.02125 [cs]},
	keywords = {Computer Science - Artificial Intelligence, Computer Science - Cryptography and Security},
}


@artile{kocon_chatgpt_2023,
	title = {{ChatGPT}: {Jack} of all trades, master of none},
	shorttitle = {{ChatGPT}},
	url = {http://arxiv.org/abs/2302.10724},
	abstract = {OpenAI has released the Chat Generative Pre-trained Transformer (ChatGPT) and revolutionized the approach in artificial intelligence to human-model interaction. The first contact with the chatbot reveals its ability to provide detailed and precise answers in various areas. Several publications on ChatGPT evaluation test its effectiveness on well-known natural language processing (NLP) tasks. However, the existing studies are mostly non-automated and tested on a very limited scale. In this work, we examined ChatGPT’s capabilities on 25 diverse analytical NLP tasks, most of them subjective even to humans, such as sentiment analysis, emotion recognition, offensiveness, and stance detection. In contrast, the other tasks require more objective reasoning like word sense disambiguation, linguistic acceptability, and question answering. We also evaluated GPT-4 model on five selected subsets of NLP tasks. We automated ChatGPT and GPT-4 prompting process and analyzed more than 49k responses. Our comparison of its results with available State-of-the-Art (SOTA) solutions showed that the average loss in quality of the ChatGPT model was about 25\% for zero-shot and few-shot evaluation. For GPT-4 model, a loss for semantic tasks is significantly lower than for ChatGPT. We showed that the more difficult the task (lower SOTA performance), the higher the ChatGPT loss. It especially refers to pragmatic NLP problems like emotion recognition. We also tested the ability to personalize ChatGPT responses for selected subjective tasks via Random Contextual Few-Shot Personalization, and we obtained significantly better user-based predictions. Additional qualitative analysis revealed a ChatGPT bias, most likely due to the rules imposed on human trainers by OpenAI. Our results provide the basis for a fundamental discussion of whether the high quality of recent predictive NLP models can indicate a tool’s usefulness to society and how the learning and validation procedures for such systems should be established.},
	language = {en},
	urldate = {2023-06-10},
	publisher = {arXiv},
	author = {Kocon, Jan and Cichecki, Igor and Kaszyca, Oliwier and Kochanek, Mateusz and Szydło, Dominika and Baran, Joanna and Bielaniewicz, Julita and Gruza, Marcin and Janz, Arkadiusz and Kanclerz, Kamil and Kocon, Anna and Koptyra, Bartlomiej and Mieleszczenko-Kowszewicz, Wiktoria and Milkowski, Piotr and Oleksy, Marcin and Piasecki, Maciej and Radlinski, Lukasz and Wojtasik, Konrad and Wozniak, Stanislaw and Kazienko, Przemyslaw},
	month = jun,
	year = {2023},
	note = {arXiv:2302.10724 [cs]},
	keywords = {Computer Science - Artificial Intelligence, Computer Science - Computation and Language, Computer Science - Computers and Society, Computer Science - Machine Learning},
}

@article{ma_scope_2023,
	title = {The {Scope} of {ChatGPT} in {Software} {Engineering}: {A} {Thorough} {Investigation}},
	shorttitle = {The {Scope} of {ChatGPT} in {Software} {Engineering}},
	url = {http://arxiv.org/abs/2305.12138},
	doi = {10.48550/arXiv.2305.12138},
	abstract = {ChatGPT demonstrates immense potential to transform software engineering (SE) by exhibiting outstanding performance in tasks such as code and document generation. However, the high reliability and risk control requirements of SE make the lack of interpretability for ChatGPT a concern. To address this issue, we carried out a study evaluating ChatGPT's capabilities and limitations in SE. We broke down the abilities needed for AI models to tackle SE tasks into three categories: 1) syntax understanding, 2) static behavior understanding, and 3) dynamic behavior understanding. Our investigation focused on ChatGPT's ability to comprehend code syntax and semantic structures, including abstract syntax trees (AST), control flow graphs (CFG), and call graphs (CG). We assessed ChatGPT's performance on cross-language tasks involving C, Java, Python, and Solidity. Our findings revealed that while ChatGPT excels at understanding code syntax (AST), it struggles with comprehending code semantics, particularly dynamic semantics. We conclude that ChatGPT possesses capabilities akin to an Abstract Syntax Tree (AST) parser, demonstrating initial competencies in static code analysis. Additionally, our study highlights that ChatGPT is susceptible to hallucination when interpreting code semantic structures and fabricating non-existent facts. These results underscore the need to explore methods for verifying the correctness of ChatGPT's outputs to ensure its dependability in SE. More importantly, our study provide an iniital answer why the generated codes from LLMs are usually synatx correct but vulnerabale.},
	urldate = {2023-06-10},
	publisher = {arXiv},
	author = {Ma, Wei and Liu, Shangqing and Wang, Wenhan and Hu, Qiang and Liu, Ye and Zhang, Cen and Nie, Liming and Liu, Yang},
	month = may,
	year = {2023},
	note = {arXiv:2305.12138 [cs]},
	keywords = {Computer Science - Artificial Intelligence, Computer Science - Software Engineering},
}

@article{liu_is_2023,
	title = {Is {Your} {Code} {Generated} by {ChatGPT} {Really} {Correct}? {Rigorous} {Evaluation} of {Large} {Language} {Models} for {Code} {Generation}},
	shorttitle = {Is {Your} {Code} {Generated} by {ChatGPT} {Really} {Correct}?},
	url = {http://arxiv.org/abs/2305.01210},
	doi = {10.48550/arXiv.2305.01210},
	abstract = {Program synthesis has been long studied with recent approaches focused on directly using the power of Large Language Models (LLMs) to generate code according to user intent written in natural language. Code evaluation datasets, containing curated synthesis problems with input/output test-cases, are used to measure the performance of various LLMs on code synthesis. However, test-cases in these datasets can be limited in both quantity and quality for fully assessing the functional correctness of the generated code. Such limitation in the existing benchmarks begs the following question: In the era of LLMs, is the code generated really correct? To answer this, we propose EvalPlus -- a code synthesis benchmarking framework to rigorously evaluate the functional correctness of LLM-synthesized code. In short, EvalPlus takes in the base evaluation dataset and uses an automatic input generation step to produce and diversify large amounts of new test inputs using both LLM-based and mutation-based input generators to further validate the synthesized code. We extend the popular HUMANEVAL benchmark and build HUMANEVAL+ with 81x additionally generated tests. Our extensive evaluation across 14 popular LLMs demonstrates that HUMANEVAL+ is able to catch significant amounts of previously undetected wrong code synthesized by LLMs, reducing the pass@k by 15.1\% on average! Moreover, we even found several incorrect ground-truth implementations in HUMANEVAL. Our work not only indicates that prior popular code synthesis evaluation results do not accurately reflect the true performance of LLMs for code synthesis but also opens up a new direction to improve programming benchmarks through automated test input generation.},
	urldate = {2023-06-10},
	publisher = {arXiv},
	author = {Liu, Jiawei and Xia, Chunqiu Steven and Wang, Yuyao and Zhang, Lingming},
	month = may,
	year = {2023},
	note = {arXiv:2305.01210 [cs]},
	keywords = {Computer Science - Computation and Language, Computer Science - Machine Learning, Computer Science - Software Engineering},
}

@article{cordeiro_smt-based_2012,
	title = {{SMT}-{Based} {Bounded} {Model} {Checking} for {Embedded} {ANSI}-{C} {Software}},
	volume = {38},
	issn = {1939-3520},
	doi = {10.1109/TSE.2011.59},
	abstract = {Propositional bounded model checking has been applied successfully to verify embedded software, but remains limited by increasing propositional formula sizes and the loss of high-level information during the translation preventing potential optimizations to reduce the state space to be explored. These limitations can be overcome by encoding high-level information in theories richer than propositional logic and using SMT solvers for the generated verification conditions. Here, we propose the application of different background theories and SMT solvers to the verification of embedded software written in ANSI-C in order to improve scalability and precision in a completely automatic way. We have modified and extended the encodings from previous SMT-based bounded model checkers to provide more accurate support for variables of finite bit width, bit-vector operations, arrays, structures, unions, and pointers. We have integrated the CVC3, Boolector, and Z3 solvers with the CBMC front-end and evaluated them using both standard software model checking benchmarks and typical embedded software applications from telecommunications, control systems, and medical devices. The experiments show that our ESBMC model checker can analyze larger problems than existing tools and substantially reduce the verification time.},
	number = {4},
	journal = {IEEE Transactions on Software Engineering},
	author = {Cordeiro, Lucas and Fischer, Bernd and Marques-Silva, Joao},
	month = jul,
	year = {2012},
	note = {Conference Name: IEEE Transactions on Software Engineering},
	keywords = {Electronic mail, Embedded software, Encoding, Optimization, Safety, Software engineering, Space exploration, formal methods, model checking, verification},
	pages = {957--974},
}


@online{openai_openai_2023,
	title = {{OpenAI} {API}},
	url = {https://platform.openai.com},
	abstract = {An API for accessing new AI models developed by OpenAI},
	language = {en},
	urldate = {2023-06-05},
}

@online{gadelha_esbmc_2023,
	title = {{ESBMC}: 5.0: {An} {Industrial}-{Strength} {Model} {Checker}},
	shorttitle = {{ESBMC}},
	url = {https://github.com/esbmc/esbmc},
	abstract = {The efficient SMT-based context-bounded model checker (ESBMC)},
	urldate = {2023-06-01},
	author = {Gadelha, Mikhail R. and Monteiro, Felipe R. and Morse, Jeremy and Cordeiro, Lucas C. and Fischer, Bernd and Nicole, Denis A.},
	month = jun,
	year = {2023},
	note = {original-date: 2015-06-20T19:35:34Z},
}

@online{secdev_software_2023,
	title = {Software {Development} {Life} {Cycle} ({SDLC})},
	url = {https://snyk.io/learn/sdlc-software-development-life-cycle/},
	abstract = {Learn more about SDLC (Software Development Life Cycle), SDLC phases and methodologies to help keep everyone on the same page and working towards a common goal.},
	language = {en-US},
	urldate = {2023-06-01},
	journal = {Snyk},
}

@article{yao_tree_2023,
	title = {Tree of {Thoughts}: {Deliberate} {Problem} {Solving} with {Large} {Language} {Models}},
	shorttitle = {Tree of {Thoughts}},
	url = {http://arxiv.org/abs/2305.10601},
	abstract = {Language models are increasingly being deployed for general problem solving across a wide range of tasks, but are still conﬁned to token-level, left-to-right decision-making processes during inference. This means they can fall short in tasks that require exploration, strategic lookahead, or where initial decisions play a pivotal role. To surmount these challenges, we introduce a new framework for language model inference, “Tree of Thoughts” (ToT), which generalizes over the popular “Chain of Thought” approach to prompting language models, and enables exploration over coherent units of text (“thoughts”) that serve as intermediate steps toward problem solving. ToT allows LMs to perform deliberate decision making by considering multiple different reasoning paths and self-evaluating choices to decide the next course of action, as well as looking ahead or backtracking when necessary to make global choices. Our experiments show that ToT signiﬁcantly enhances language models’ problem-solving abilities on three novel tasks requiring non-trivial planning or search: Game of 24, Creative Writing, and Mini Crosswords. For instance, in Game of 24, while GPT-4 with chain-of-thought prompting only solved 4\% of tasks, our method achieved a success rate of 74\%. Code repo with all prompts: https://github.com/ysymyth/tree-of-thought-llm.},
	language = {en},
	urldate = {2023-05-29},
	publisher = {arXiv},
	author = {Yao, Shunyu and Yu, Dian and Zhao, Jeffrey and Shafran, Izhak and Griffiths, Thomas L. and Cao, Yuan and Narasimhan, Karthik},
	month = may,
	year = {2023},
	note = {arXiv:2305.10601 [cs]},
	keywords = {Computer Science - Artificial Intelligence, Computer Science - Computation and Language, Computer Science - Machine Learning},
}

@article{charalambous_new_2023,
	title = {A {New} {Era} in {Software} {Security}: {Towards} {Self}-{Healing} {Software} via {Large} {Language} {Models} and {Formal} {Verification}},
	shorttitle = {A {New} {Era} in {Software} {Security}},
	url = {http://arxiv.org/abs/2305.14752},
	doi = {10.48550/arXiv.2305.14752},
	abstract = {In this paper we present a novel solution that combines the capabilities of Large Language Models (LLMs) with Formal Verification strategies to verify and automatically repair software vulnerabilities. Initially, we employ Bounded Model Checking (BMC) to locate the software vulnerability and derive a counterexample. The counterexample provides evidence that the system behaves incorrectly or contains a vulnerability. The counterexample that has been detected, along with the source code, are provided to the LLM engine. Our approach involves establishing a specialized prompt language for conducting code debugging and generation to understand the vulnerability's root cause and repair the code. Finally, we use BMC to verify the corrected version of the code generated by the LLM. As a proof of concept, we create ESBMC-AI based on the Efficient SMT-based Context-Bounded Model Checker (ESBMC) and a pre-trained Transformer model, specifically gpt-3.5-turbo, to detect and fix errors in C programs. Our experimentation involved generating a dataset comprising 1000 C code samples, each consisting of 20 to 50 lines of code. Notably, our proposed method achieved an impressive success rate of up to 80\% in repairing vulnerable code encompassing buffer overflow and pointer dereference failures. We assert that this automated approach can effectively incorporate into the software development lifecycle's continuous integration and deployment (CI/CD) process.},
	urldate = {2023-05-31},
	publisher = {arXiv},
	author = {Charalambous, Yiannis and Tihanyi, Norbert and Jain, Ridhi and Sun, Youcheng and Ferrag, Mohamed Amine and Cordeiro, Lucas C.},
	month = may,
	year = {2023},
	note = {arXiv:2305.14752 [cs]},
	keywords = {Computer Science - Artificial Intelligence, Computer Science - Formal Languages and Automata Theory, Computer Science - Machine Learning, Computer Science - Software Engineering},
}

@techreport{openai_gpt-4_2023,
	title = {{GPT}-4 {Technical} {Report}},
	url = {http://arxiv.org/abs/2303.08774},
	abstract = {We report the development of GPT-4, a large-scale, multimodal model which can accept image and text inputs and produce text outputs. While less capable than humans in many real-world scenarios, GPT-4 exhibits human-level performance on various professional and academic benchmarks, including passing a simulated bar exam with a score around the top 10\% of test takers. GPT-4 is a Transformerbased model pre-trained to predict the next token in a document. The post-training alignment process results in improved performance on measures of factuality and adherence to desired behavior. A core component of this project was developing infrastructure and optimization methods that behave predictably across a wide range of scales. This allowed us to accurately predict some aspects of GPT-4’s performance based on models trained with no more than 1/1,000th the compute of GPT-4.},
	language = {en},
	urldate = {2023-05-29},
	publisher = {arXiv},
	author = {OpenAI},
	month = mar,
	year = {2023},
	note = {arXiv:2303.08774 [cs]},
	keywords = {Computer Science - Artificial Intelligence, Computer Science - Computation and Language},
}

@article{rey1996testing,
  title={Testing software for characteristics other than correctness: Safety, failure tolerance, and security},
  author={rey Voas, Je},
  year={1996},
  publisher={Citeseer}
}

@article{perry_users_2022,
	title = {Do {Users} {Write} {More} {Insecure} {Code} with {AI} {Assistants}?},
	url = {http://arxiv.org/abs/2211.03622},
	abstract = {We conduct the ﬁrst large-scale user study examining how users interact with an AI Code assistant to solve a variety of security related tasks across different programming languages. Overall, we ﬁnd that participants who had access to an AI assistant based on OpenAI’s codex-davinci-002 model wrote signiﬁcantly less secure code than those without access. Additionally, participants with access to an AI assistant were more likely to believe they wrote secure code than those without access to the AI assistant. Furthermore, we ﬁnd that participants who trusted the AI less and engaged more with the language and format of their prompts (e.g. re-phrasing, adjusting temperature) provided code with fewer security vulnerabilities. Finally, in order to better inform the design of future AI-based Code assistants, we provide an in-depth analysis of participants’ language and interaction behavior, as well as release our user interface as an instrument to conduct similar studies in the future.},
	language = {en},
	urldate = {2023-05-30},
	publisher = {arXiv},
	author = {Perry, Neil and Srivastava, Megha and Kumar, Deepak and Boneh, Dan},
	month = dec,
	year = {2022},
	note = {arXiv:2211.03622 [cs]},
	keywords = {Computer Science - Cryptography and Security},
}

@online{codetasks_papers_2023,
	title = {Papers with {Code} - {Computer} {Code}},
	url = {https://paperswithcode.com/area/computer-code},
	abstract = {Browse 52 tasks • 81 datasets • 71},
	language = {en},
	urldate = {2023-05-30},
}

@article{grahn_analysis_2021,
	title = {An {Analysis} of {C}/{C}++ {Datasets} for {Machine} {Learning}-{Assisted} {Software} {Vulnerability} {Detection}},
	abstract = {As machine learning-assisted vulnerability detection research matures, it is critical to understand the datasets being used by existing papers. In this paper, we explore 7 C/C++ datasets and evaluate their suitability for machine learning-assisted vulnerability detection. We also present a new dataset, named Wild C, containing over 10.3 million individual opensource C/C++ files – a sufficiently large sample to be reasonably considered representative of typical C/C++ code. To facilitate comparison, we tokenize all of the datasets and perform the analysis at this level. We make three primary contributions. First, while all the datasets differ from our Wild C dataset, some do so to a greater degree. This includes divergence in file lengths and token usage frequency. Additionally, none of the datasets contain the entirety of the C/C++ vocabulary. These missing tokens account for up to 11\% of all token usage. Second, we find all the datasets contain duplication with some containing a significant amount. In the Juliet dataset, we describe augmentations of test cases making the dataset susceptible to data leakage. This augmentation occurs with such frequency that a random 80/20 split has roughly 58\% overlap of the test with the training data. Finally, we collect and process a large dataset of C code, named Wild C. This dataset is designed to serve as a representative sample of all C/C++ code and is the basis for our analyses.},
	language = {en},
	author = {Grahn, Daniel and Zhang, Junjie},
	year = {2021},
}

@article{vaswani_attention_2017,
	title = {Attention {Is} {All} {You} {Need}},
	url = {http://arxiv.org/abs/1706.03762},
	doi = {10.48550/arXiv.1706.03762},
	abstract = {The dominant sequence transduction models are based on complex recurrent or convolutional neural networks in an encoder-decoder configuration. The best performing models also connect the encoder and decoder through an attention mechanism. We propose a new simple network architecture, the Transformer, based solely on attention mechanisms, dispensing with recurrence and convolutions entirely. Experiments on two machine translation tasks show these models to be superior in quality while being more parallelizable and requiring significantly less time to train. Our model achieves 28.4 BLEU on the WMT 2014 English-to-German translation task, improving over the existing best results, including ensembles by over 2 BLEU. On the WMT 2014 English-to-French translation task, our model establishes a new single-model state-of-the-art BLEU score of 41.8 after training for 3.5 days on eight GPUs, a small fraction of the training costs of the best models from the literature. We show that the Transformer generalizes well to other tasks by applying it successfully to English constituency parsing both with large and limited training data.},
	urldate = {2023-05-29},
	publisher = {arXiv},
	author = {Vaswani, Ashish and Shazeer, Noam and Parmar, Niki and Uszkoreit, Jakob and Jones, Llion and Gomez, Aidan N. and Kaiser, Lukasz and Polosukhin, Illia},
	month = dec,
	year = {2017},
	note = {arXiv:1706.03762 [cs]},
	keywords = {Computer Science - Computation and Language, Computer Science - Machine Learning},
}

@article{shinn_reflexion_2023,
	title = {Reflexion: {Language} {Agents} with {Verbal} {Reinforcement} {Learning}},
	shorttitle = {Reflexion},
	url = {http://arxiv.org/abs/2303.11366},
	abstract = {Large language models (LLMs) have been increasingly used to interact with external environments (e.g., games, compilers, APIs) as goal-driven agents. However, it remains challenging for these language agents to quickly and efﬁciently learn from trial-and-error as traditional reinforcement learning methods require extensive training samples and expensive model ﬁne-tuning. We propose Reﬂexion, a novel framework to reinforce language agents not by updating weights, but instead through linguistic feedback. Concretely, Reﬂexion agents verbally reﬂect on task feedback signals, then maintain their own reﬂective text in an episodic memory buffer to induce better decision-making in subsequent trials. Reﬂexion is ﬂexible enough to incorporate various types (scalar values or free-form language) and sources (external or internally simulated) of feedback signals, and obtains signiﬁcant improvements over a baseline agent across diverse tasks (sequential decision-making, coding, language reasoning). For example, Reﬂexion achieves a 91\% pass@1 accuracy on the HumanEval coding benchmark, surpassing the previous state-of-the-art GPT-4 that achieves 80\%. We also conduct ablation and analysis studies using different feedback signals, feedback incorporation methods, and agent types, and provide insights into how they affect performance. We release all code and demos at https://github.com/noahshinn024/reflexion.},
	language = {en},
	urldate = {2023-05-29},
	publisher = {arXiv},
	author = {Shinn, Noah and Cassano, Federico and Labash, Beck and Gopinath, Ashwin and Narasimhan, Karthik and Yao, Shunyu},
	month = may,
	year = {2023},
	note = {arXiv:2303.11366 [cs]},
	keywords = {Computer Science - Artificial Intelligence, Computer Science - Computation and Language, Computer Science - Machine Learning},
}

@online{humaneval_papers_2023,
	title = {Papers with {Code} - {HumanEval} {Benchmark} ({Code} {Generation})},
	url = {https://paperswithcode.com/sota/code-generation-on-humaneval},
	abstract = {The current state-of-the-art on HumanEval is Reflexion (GPT-4). See a full comparison of 30 papers with code.},
	language = {en},
	urldate = {2023-05-29},
}

@article{anil_palm_2023,
	title = {{PaLM} 2 {Technical} {Report}},
	url = {http://arxiv.org/abs/2305.10403},
	abstract = {We introduce PaLM 2, a new state-of-the-art language model that has better multilingual and reasoning capabilities and is more compute-efﬁcient than its predecessor PaLM. PaLM 2 is a Transformer-based model trained using a mixture of objectives. Through extensive evaluations on English and multilingual language, and reasoning tasks, we demonstrate that PaLM 2 has signiﬁcantly improved quality on downstream tasks across different model sizes, while simultaneously exhibiting faster and more efﬁcient inference compared to PaLM. This improved efﬁciency enables broader deployment while also allowing the model to respond faster, for a more natural pace of interaction. PaLM 2 demonstrates robust reasoning capabilities exempliﬁed by large improvements over PaLM on BIG-Bench and other reasoning tasks. PaLM 2 exhibits stable performance on a suite of responsible AI evaluations, and enables inference-time control over toxicity without additional overhead or impact on other capabilities. Overall, PaLM 2 achieves state-of-the-art performance across a diverse set of tasks and capabilities.},
	language = {en},
	urldate = {2023-05-29},
	publisher = {arXiv},
	author = {Anil, Rohan and Dai, Andrew M. and Firat, Orhan and Johnson, Melvin and Lepikhin, Dmitry and Passos, Alexandre and Shakeri, Siamak and Taropa, Emanuel and Bailey, Paige and Chen, Zhifeng and Chu, Eric and Clark, Jonathan H. and Shafey, Laurent El and Huang, Yanping and Meier-Hellstern, Kathy and Mishra, Gaurav and Moreira, Erica and Omernick, Mark and Robinson, Kevin and Ruder, Sebastian and Tay, Yi and Xiao, Kefan and Xu, Yuanzhong and Zhang, Yujing and Abrego, Gustavo Hernandez and Ahn, Junwhan and Austin, Jacob and Barham, Paul and Botha, Jan and Bradbury, James and Brahma, Siddhartha and Brooks, Kevin and Catasta, Michele and Cheng, Yong and Cherry, Colin and Choquette-Choo, Christopher A. and Chowdhery, Aakanksha and Crepy, Clément and Dave, Shachi and Dehghani, Mostafa and Dev, Sunipa and Devlin, Jacob and Díaz, Mark and Du, Nan and Dyer, Ethan and Feinberg, Vlad and Feng, Fangxiaoyu and Fienber, Vlad and Freitag, Markus and Garcia, Xavier and Gehrmann, Sebastian and Gonzalez, Lucas and Gur-Ari, Guy and Hand, Steven and Hashemi, Hadi and Hou, Le and Howland, Joshua and Hu, Andrea and Hui, Jeffrey and Hurwitz, Jeremy and Isard, Michael and Ittycheriah, Abe and Jagielski, Matthew and Jia, Wenhao and Kenealy, Kathleen and Krikun, Maxim and Kudugunta, Sneha and Lan, Chang and Lee, Katherine and Lee, Benjamin and Li, Eric and Li, Music and Li, Wei and Li, YaGuang and Li, Jian and Lim, Hyeontaek and Lin, Hanzhao and Liu, Zhongtao and Liu, Frederick and Maggioni, Marcello and Mahendru, Aroma and Maynez, Joshua and Misra, Vedant and Moussalem, Maysam and Nado, Zachary and Nham, John and Ni, Eric and Nystrom, Andrew and Parrish, Alicia and Pellat, Marie and Polacek, Martin and Polozov, Alex and Pope, Reiner and Qiao, Siyuan and Reif, Emily and Richter, Bryan and Riley, Parker and Ros, Alex Castro and Roy, Aurko and Saeta, Brennan and Samuel, Rajkumar and Shelby, Renee and Slone, Ambrose and Smilkov, Daniel and So, David R. and Sohn, Daniel and Tokumine, Simon and Valter, Dasha and Vasudevan, Vijay and Vodrahalli, Kiran and Wang, Xuezhi and Wang, Pidong and Wang, Zirui and Wang, Tao and Wieting, John and Wu, Yuhuai and Xu, Kelvin and Xu, Yunhan and Xue, Linting and Yin, Pengcheng and Yu, Jiahui and Zhang, Qiao and Zheng, Steven and Zheng, Ce and Zhou, Weikang and Zhou, Denny and Petrov, Slav and Wu, Yonghui},
	month = may,
	year = {2023},
	note = {arXiv:2305.10403 [cs]},
	keywords = {Computer Science - Artificial Intelligence, Computer Science - Computation and Language},
}

@inproceedings{tan_codeflaws_2017,
	title = {Codeflaws: a programming competition benchmark for evaluating automated program repair tools},
	shorttitle = {Codeflaws},
	doi = {10.1109/ICSE-C.2017.76},
	abstract = {Several automated program repair techniques have been proposed to reduce the time and effort spent in bug-fixing. While these repair tools are designed to be generic such that they could address many software faults, different repair tools may fix certain types of faults more effectively than other tools. Therefore, it is important to compare more objectively the effectiveness of different repair tools on various fault types. However, existing benchmarks on automated program repairs do not allow thorough investigation of the relationship between fault types and the effectiveness of repair tools. We present Codeflaws, a set of 3902 defects from 7436 programs automatically classified across 39 defect classes (we refer to different types of fault as defect classes derived from the syntactic differences between a buggy program and a patched program).},
	booktitle = {2017 {IEEE}/{ACM} 39th {International} {Conference} on {Software} {Engineering} {Companion} ({ICSE}-{C})},
	author = {Tan, Shin Hwei and Yi, Jooyong and {Yulis} and Mechtaev, Sergey and Roychoudhury, Abhik},
	month = may,
	year = {2017},
	keywords = {Benchmark testing, Conferences, Maintenance engineering, Programming, Software, Software engineering, Tools, automated program repair, benchmark, defect classes, empirical evaluation},
	pages = {180--182},
}

@article{orvalho_c-pack_2022,
	title = {C-{Pack} of {IPAs}: {A} {C90} {Program} {Benchmark} of {Introductory} {Programming} {Assignments}},
	shorttitle = {C-{Pack} of {IPAs}},
	url = {http://arxiv.org/abs/2206.08768},
	doi = {10.48550/arXiv.2206.08768},
	abstract = {Due to the vast number of students enrolled in Massive Open Online Courses (MOOCs), there has been an increasing number of automated program repair techniques focused on introductory programming assignments (IPAs). Such techniques take advantage of previous correct student implementations in order to provide automated, comprehensive, and personalized feedback to students. This paper presents C-Pack-IPAs, a publicly available benchmark of students' programs submitted for 25 different IPAs. C-Pack-IPAs contains semantically correct, semantically incorrect, and syntactically incorrect programs plus a test suite for each IPA. Hence, C-Pack-IPAs can be used to help evaluate the development of novel semantic, as well as syntactic, automated program repair frameworks, focused on providing feedback to novice programmers.},
	urldate = {2023-05-28},
	publisher = {arXiv},
	author = {Orvalho, Pedro and Janota, Mikoláš and Manquinho, Vasco},
	month = jun,
	year = {2022},
	note = {arXiv:2206.08768 [cs]},
	keywords = {Computer Science - Artificial Intelligence, Computer Science - Computers and Society, Computer Science - Programming Languages, Computer Science - Software Engineering},
}

@article{gupta_deepfix_2017,
	title = {{DeepFix}: {Fixing} {Common} {C} {Language} {Errors} by {Deep} {Learning}},
	volume = {31},
	copyright = {Copyright (c)},
	issn = {2374-3468},
	shorttitle = {{DeepFix}},
	url = {https://ojs.aaai.org/index.php/AAAI/article/view/10742},
	doi = {10.1609/aaai.v31i1.10742},
	abstract = {The problem of automatically fixing programming errors is a very active research topic in software engineering. This is a challenging problem as fixing even a single error may require analysis of the entire program. In practice, a number of errors arise due to programmer's inexperience with the programming language or lack of attention to detail. We call these common programming errors. These are analogous to grammatical errors in natural languages. Compilers detect such errors, but their error messages are usually inaccurate. In this work, we present an end-to-end solution, called DeepFix, that can fix multiple such errors in a program without relying on any external tool to locate or fix them. At the heart of DeepFix is a multi-layered sequence-to-sequence neural network with attention which is trained to predict erroneous program locations along with the required correct statements. On a set of 6971 erroneous C programs written by students for 93 programming tasks, DeepFix could fix 1881 (27\%) programs completely and 1338 (19\%) programs partially.},
	language = {en},
	number = {1},
	urldate = {2023-05-28},
	journal = {Proceedings of the AAAI Conference on Artificial Intelligence},
	author = {Gupta, Rahul and Pal, Soham and Kanade, Aditya and Shevade, Shirish},
	month = feb,
	year = {2017},
	note = {Number: 1},
	keywords = {programming education},
}

@article{liu_deepfuzz_2019,
	title = {{DeepFuzz}: {Automatic} {Generation} of {Syntax} {Valid} {C} {Programs} for {Fuzz} {Testing}},
	volume = {33},
	copyright = {Copyright (c) 2019 Association for the Advancement of Artificial Intelligence},
	issn = {2374-3468},
	shorttitle = {{DeepFuzz}},
	url = {https://ojs.aaai.org/index.php/AAAI/article/view/3895},
	doi = {10.1609/aaai.v33i01.33011044},
	abstract = {Compilers are among the most fundamental programming tools for building software. However, production compilers remain buggy. Fuzz testing is often leveraged with newlygenerated, or mutated inputs in order to find new bugs or security vulnerabilities. In this paper, we propose a grammarbased fuzzing tool called DEEPFUZZ. Based on a generative Sequence-to-Sequence model, DEEPFUZZ automatically and continuously generates well-formed C programs. We use this set of new C programs to fuzz off-the-shelf C compilers, e.g., GCC and Clang/LLVM. We present a detailed case study to analyze the success rate and coverage improvement of the generated C programs for fuzz testing. We analyze the performance of DEEPFUZZ with three types of sampling methods as well as three types of generation strategies. Consequently, DEEPFUZZ improved the testing efficacy in regards to the line, function, and branch coverage. In our preliminary study, we found and reported 8 bugs of GCC, all of which are actively being addressed by developers.},
	language = {en},
	number = {01},
	urldate = {2023-05-28},
	journal = {Proceedings of the AAAI Conference on Artificial Intelligence},
	author = {Liu, Xiao and Li, Xiaoting and Prajapati, Rupesh and Wu, Dinghao},
	month = jul,
	year = {2019},
	note = {Number: 01},
	pages = {1044--1051},
}

@article{le_goues_manybugs_2015,
	title = {The {ManyBugs} and {IntroClass} {Benchmarks} for {Automated} {Repair} of {C} {Programs}},
	volume = {41},
	issn = {1939-3520},
	doi = {10.1109/TSE.2015.2454513},
	abstract = {The field of automated software repair lacks a set of common benchmark problems. Although benchmark sets are used widely throughout computer science, existing benchmarks are not easily adapted to the problem of automatic defect repair, which has several special requirements. Most important of these is the need for benchmark programs with reproducible, important defects and a deterministic method for assessing if those defects have been repaired. This article details the need for a new set of benchmarks, outlines requirements, and then presents two datasets, ManyBugs and IntroClass, consisting between them of 1,183 defects in 15 C programs. Each dataset is designed to support the comparative evaluation of automatic repair algorithms asking a variety of experimental questions. The datasets have empirically defined guarantees of reproducibility and benchmark quality, and each study object is categorized to facilitate qualitative evaluation and comparisons by category of bug or program. The article presents baseline experimental results on both datasets for three existing repair methods, GenProg, AE, and TrpAutoRepair, to reduce the burden on researchers who adopt these datasets for their own comparative evaluations.},
	number = {12},
	journal = {IEEE Transactions on Software Engineering},
	author = {Le Goues, Claire and Holtschulte, Neal and Smith, Edward K. and Brun, Yuriy and Devanbu, Premkumar and Forrest, Stephanie and Weimer, Westley},
	month = dec,
	year = {2015},
	note = {Conference Name: IEEE Transactions on Software Engineering},
	keywords = {Automated program repair, Benchmark testing, Computer bugs, Electronic mail, IntroClass, Maintenance engineering, ManyBugs, Software systems, benchmark, reproducibility, subject defect},
	pages = {1236--1256},
}

@online{noauthor_codeflaws_2023,
	title = {Codeflaws},
	copyright = {MIT},
	url = {https://github.com/codeflaws/codeflaws},
	abstract = {This repository the benchmark with 3902 defects extracted from C programs in Codeforces (http://codeforces.com/)},
	urldate = {2023-05-28},
	publisher = {codeflaws},
	month = apr,
	year = {2023},
}

@online{noauthor_introclass_2023,
	title = {{IntroClass} {ReadMe}},
	copyright = {BSD-3-Clause},
	url = {https://github.com/ProgramRepair/IntroClass},
	abstract = {The IntroClass benchmark.  http://repairbenchmarks.cs.umass.edu/},
	urldate = {2023-05-28},
	publisher = {ProgramRepair},
	month = may,
	year = {2023},
}

@article{jr_juliet_2012,
	title = {The {Juliet} 1.1 {C}/{C}++ and {Java} {Test} {Suite}},
	volume = {45},
	url = {https://www.nist.gov/publications/juliet-11-cc-and-java-test-suite},
	abstract = {The Juliet Test Suite 1.1 is a collection of over 81,000 synthetic C/C++ and Java programs with known flaws.},
	language = {en},
	number = {10},
	urldate = {2023-05-28},
	journal = {NIST},
	author = {Jr, Frederick E. Boland and Black, Paul E.},
	month = oct,
	year = {2012},
	note = {Last Modified: 2021-10-12T11:10-04:00
Publisher: Frederick E. Boland Jr., Paul E. Black},
	pages = {88--90},
}


@article{hampton_ransomware_2018,
	title = {Ransomware behavioural analysis on windows platforms},
	volume = {40},
	issn = {22142126},
	url = {https://linkinghub.elsevier.com/retrieve/pii/S2214212617306506},
	doi = {10.1016/j.jisa.2018.02.008},
	language = {en},
	urldate = {2023-05-25},
	journal = {Journal of Information Security and Applications},
	author = {Hampton, Nikolai and Baig, Zubair and Zeadally, Sherali},
	month = jun,
	year = {2018},
	pages = {44--51},
}

@article{atapattu_approach_2021,
	title = {An {Approach} to {Detect} {Fileless} {Malware} that {Maintains} {Persistence} in {Windows} {Environment}},
	abstract = {The rapid enhancement of the Internet in the past few years has increasingly impacted the general public’s work and life. As a drawback, this enhancement has also led to a major increase in malicious software on the internet causing great security threats to the consumers of the internet. Currently, a new type of malware class called Fileless malware has come into action causing more destructive damages. As the name Fileless suggests, these types of malware programs are not files or executables, but a malicious activity that runs entirely in the memory, leaving the slightest evidence on the targeted host machine. Microsoft Windows is one of the most widely used operating systems both in personal desktop computers and enterprise computer systems and is highly targeted by Fileless malware. This paper provides an approach to detect fileless malware that maintains persistence in the Windows environment using Fileless malware behavioural data and deep learningbased classification models.},
	language = {en},
	journal = {Sri Lanka},
	author = {Atapattu, Malmi},
	year = {2021},
}

@inproceedings{barr-smith_survivalism_2021,
	address = {San Francisco, CA, USA},
	title = {Survivalism: {Systematic} {Analysis} of {Windows} {Malware} {Living}-{Off}-{The}-{Land}},
	isbn = {978-1-72818-934-5},
	shorttitle = {Survivalism},
	url = {https://ieeexplore.ieee.org/document/9519480/},
	doi = {10.1109/SP40001.2021.00047},
	abstract = {As malware detection algorithms and methods become more sophisticated, malware authors adopt equally sophisticated evasion mechanisms to defeat them. Anecdotal evidence claims Living-Off-The-Land (LotL) techniques are one of the major evasion techniques used in many malware attacks. These techniques leverage binaries already present in the system to conduct malicious actions. We present the ﬁrst large-scale systematic investigation of the use of these techniques by malware on Windows systems.},
	language = {en},
	urldate = {2023-05-25},
	booktitle = {2021 {IEEE} {Symposium} on {Security} and {Privacy} ({SP})},
	publisher = {IEEE},
	author = {Barr-Smith, Frederick and Ugarte-Pedrero, Xabier and Graziano, Mariano and Spolaor, Riccardo and Martinovic, Ivan},
	month = may,
	year = {2021},
	pages = {1557--1574},
}

@inproceedings{hendler_amsi-based_2020,
	address = {Taipei Taiwan},
	title = {{AMSI}-{Based} {Detection} of {Malicious} {PowerShell} {Code} {Using} {Contextual} {Embeddings}},
	isbn = {978-1-4503-6750-9},
	url = {https://dl.acm.org/doi/10.1145/3320269.3384742},
	doi = {10.1145/3320269.3384742},
	abstract = {PowerShell is a command-line shell, supporting a scripting language. It is widely used in organizations for configuration management and task automation but is also increasingly used for launching cyber attacks against organizations, mainly because it is pre-installed on Windows machines and exposes strong functionality that may be leveraged by attackers. This makes the problem of detecting malicious PowerShell code both urgent and challenging. Microsoft’s Antimalware Scan Interface (AMSI), built into Windows 10, allows defending systems to scan all the code passed to scripting engines such as PowerShell prior to its execution. In this work, we conduct the first study of malicious PowerShell code detection using the information made available by AMSI.},
	language = {en},
	urldate = {2023-05-25},
	booktitle = {Proceedings of the 15th {ACM} {Asia} {Conference} on {Computer} and {Communications} {Security}},
	publisher = {ACM},
	author = {Hendler, Danny and Kels, Shay and Rubin, Amir},
	month = oct,
	year = {2020},
	pages = {679--693},
}

@article{xu_new_2016,
	title = {A {New} {Paradigm} of {Software} {Service} {Engineering} in the {Era} of {Big} {Data} and {Big} {Service}},
	url = {http://arxiv.org/abs/1608.08342},
	doi = {10.48550/arXiv.1608.08342},
	abstract = {Servitization is one of the most significant trends that reshapes the information world and society in recent years. The requirement of collecting,storing, processing, and sharing of the Big Data has led to massive software resources being developed and made accessible as web-based services to facilitate such process. These services that handle the Big Data come from various domains and heterogeneous networks, and converge into a huge complicated service network (or ecosystem), called the Big Service.The key issue facing the big data and big service ecosystem is how to optimally configure and operate the related service resources to serve the specific requirements of possible applications, i.e., how to reuse the existing service resources effectively and efficiently to develop the new applications or software services, to meet the massive individualized requirements of end-users.Based on analyzing the big service ecosystem, we present in this paper a new paradigm for software service engineering, RE2SEP (Requirement-Engineering Two-Phase of Service Engineering Paradigm), which includes three components: service-oriented requirement engineering, domain-oriented service engineering, and software service development approach. RE2SEP enables the rapid design and implementation of service solutions to match the requirement propositions of massive individualized customers in the Big Service ecosystem. A case study on people's mobility service in a smart city environment is given to demonstrate the application of RE2SEP.RE2SEP can potentially revolutionize the traditional life-cycle oriented software engineering, leading to a new approach to software service engineering.},
	urldate = {2023-05-25},
	publisher = {arXiv},
	author = {Xu, Xiaofei and Motta, Gianmario and Wang, Xianzhi and Tu, Zhiying and Xu, Hanchuan},
	month = aug,
	year = {2016},
	note = {arXiv:1608.08342 [cs]},
	keywords = {Computer Science - Software Engineering},
}

@article{li_multi-step_2023,
	title = {Multi-step {Jailbreaking} {Privacy} {Attacks} on {ChatGPT}},
	url = {http://arxiv.org/abs/2304.05197},
	doi = {10.48550/arXiv.2304.05197},
	abstract = {With the rapid progress of large language models (LLMs), many downstream NLP tasks can be well solved given good prompts. Though model developers and researchers work hard on dialog safety to avoid generating harmful content from LLMs, it is still challenging to steer AI-generated content (AIGC) for the human good. As powerful LLMs are devouring existing text data from various domains (e.g., GPT-3 is trained on 45TB texts), it is natural to doubt whether the private information is included in the training data and what privacy threats can these LLMs and their downstream applications bring. In this paper, we study the privacy threats from OpenAI's model APIs and New Bing enhanced by ChatGPT and show that application-integrated LLMs may cause more severe privacy threats ever than before. To this end, we conduct extensive experiments to support our claims and discuss LLMs' privacy implications.},
	urldate = {2023-05-02},
	publisher = {arXiv},
	author = {Li, Haoran and Guo, Dadi and Fan, Wei and Xu, Mingshi and Song, Yangqiu},
	month = apr,
	year = {2023},
	note = {arXiv:2304.05197 [cs]},
	keywords = {Computer Science - Computation and Language, Computer Science - Cryptography and Security},
}

@online{jennifer_malicious_2021,
        author = {Jennifer, Fernick},
	title = {On the malicious use of large language models like {GPT}-3},
	url = {https://research.nccgroup.com/2021/12/31/on-the-malicious-use-of-large-language-models-like-gpt-3/},
	abstract = {(Or, “Can large language models generate exploits?”) While attacking machine learning systems is a hot topic for which attacks have begun to be demonstrated, I believe that there are a number of en…},
	language = {en-GB},
	urldate = {2023-04-28},
	journal = {NCC Group Research Blog},
	month = dec,
	year = {2021},
}





@article{vaswani2017attention,
  title={Attention is all you need},
  author={Vaswani, Ashish and Shazeer, Noam and Parmar, Niki and Uszkoreit, Jakob and Jones, Llion and Gomez, Aidan N and Kaiser, {\L}ukasz and Polosukhin, Illia},
  journal={Advances in neural information processing systems},
  volume={30},
  year={2017}
}

@misc{refs:openai:1,
  author = {{OpenAI}},
  year = {n.d.},
  title = {Chat Plugins},
  note = {Accessed Apr 21, 2023.  \url{https://platform.openai.com/docs/plugins/introduction}}
}

@misc{gpt35,
  title = {GPT-3.5},
  author = {{OpenAI}},
  year = {2022},
  note = {Accessed May 17, 2023.  \url{https://platform.openai.com/docs/models/gpt-3-5}}
}

@misc{gpt4,
  author = {{OpenAI}},
  year = {2023},
  title = {GPT-4},
  note = {Accessed May 17, 2023.  \url{https://openai.com/research/gpt-4}}
}

@article{rajasekharan2023reliable,
  title={Reliable Natural Language Understanding with Large Language Models and Answer Set Programming},
  author={Rajasekharan, Abhiramon and Zeng, Yankai and Padalkar, Parth and Gupta, Gopal},
  journal={arXiv preprint arXiv:2302.03780},
  year={2023}
}

@article{ge2023openagi,
  title={OpenAGI: When LLM Meets Domain Experts},
  author={Ge, Yingqiang and Hua, Wenyue and Ji, Jianchao and Tan, Juntao and Xu, Shuyuan and Zhang, Yongfeng},
  journal={arXiv preprint arXiv:2304.04370},
  year={2023}
}


@article{t5,
  title={Exploring the limits of transfer learning with a unified text-to-text transformer},
  author={Raffel, Colin and Shazeer, Noam and Roberts, Adam and Lee, Katherine and Narang, Sharan and Matena, Michael and Zhou, Yanqi and Li, Wei and Liu, Peter J},
  journal={The Journal of Machine Learning Research},
  volume={21},
  number={1},
  pages={5485--5551},
  year={2020},
  publisher={JMLRORG}
}

@article{devlin2018bert,
  title={Bert: Pre-training of deep bidirectional transformers for language understanding},
  author={Devlin, Jacob and Chang, Ming-Wei and Lee, Kenton and Toutanova, Kristina},
  journal={arXiv preprint arXiv:1810.04805},
  year={2018}
}

@inproceedings{GadelhaSC0N19,
  author       = {Mikhail Y. R. Gadelha and
                  Enrico Steffinlongo and
                  Lucas C. Cordeiro and
                  Bernd Fischer and
                  Denis A. Nicole},
  editor       = {Joanne M. Atlee and
                  Tevfik Bultan and
                  Jon Whittle},
  title        = {SMT-based refutation of spurious bug reports in the clang static analyzer},
  booktitle    = {Proceedings of the 41st International Conference on Software Engineering:
                  Companion Proceedings, {ICSE} 2019, Montreal, QC, Canada, May 25-31,
                  2019},
  pages        = {11--14},
  publisher    = {{IEEE} / {ACM}},
  year         = {2019},
  doi          = {10.1109/ICSE-Companion.2019.00026}
}

@article{MonteiroGC22,
  author       = {Felipe R. Monteiro and
                  Mikhail R. Gadelha and
                  Lucas C. Cordeiro},
  title        = {Model checking {C++} programs},
  journal      = {Softw. Test. Verification Reliab.},
  volume       = {32},
  number       = {1},
  year         = {2022},
  doi          = {10.1002/stvr.1793}
}

@inproceedings{SongMFC22,
  author       = {Kunjian Song and
                  Nedas Matulevicius and
                  Eddie B. de Lima Filho and
                  Lucas C. Cordeiro},
  title        = {ESBMC-Solidity: An SMT-Based Model Checker for Solidity Smart Contracts},
  booktitle    = {44th {IEEE/ACM} International Conference on Software Engineering ({ICSE})},
  pages        = {65--69},
  publisher    = {{ACM/IEEE}},
  year         = {2022},
  doi          = {10.1145/3510454.3516855}
}

@inproceedings{CordeiroKS19,
  author       = {Lucas C. Cordeiro and
                  Daniel Kroening and
                  Peter Schrammel},
  title        = {{JBMC:} Bounded Model Checking for Java Bytecode - (Competition Contribution)},
  booktitle    = {Tools and Algorithms for the Construction and Analysis of Systems ({TACAS})},
  series       = {LNCS},
  volume       = {11429},
  pages        = {219--223},
  publisher    = {Springer},
  year         = {2019},
  doi          = {10.1007/978-3-030-17502-3\_17}
}

@article{AlvesCF17,
  author       = {Erickson H. da S. Alves and
                  Lucas C. Cordeiro and
                  Eddie Batista de Lima Filho},
  title        = {A method to localize faults in concurrent {C} programs},
  journal      = {J. Syst. Softw.},
  volume       = {132},
  pages        = {336--352},
  year         = {2017},
  doi          = {10.1016/j.jss.2017.03.010}
}

@book{Aho:2006:CPT:1177220,
 author = "Aho, Alfred V. and Lam, Monica S. and Sethi, Ravi and Ullman, Jeffrey D.",
 title = "Compilers: Principles, Techniques, And Tools",
 year = "2006",
 edition = "2nd",
 publisher = "Addison-Wesley Longman Publishing Co., Inc."
}

@inproceedings{AldughaimAGFC23,
  author       = {Mohannad Aldughaim and
                  Kaled M. Alshmrany and
                  Mikhail R. Gadelha and
                  Rosiane de Freitas and
                  Lucas C. Cordeiro},
  title        = {FuSeBMC{\_}IA: Interval Analysis and Methods for Test Case Generation
                  - (Competition Contribution)},
  booktitle    = {26th International
                  Conference Fundamental Approaches to Software Engineering ({FASE})},
  series       = {LNCS},
  volume       = {13991},
  pages        = {324--329},
  publisher    = {Springer},
  year         = {2023},
  doi          = {10.1007/978-3-031-30826-0\_18}
}

@article{CordeiroFM12,
  author       = {Lucas C. Cordeiro and
                  Bernd Fischer and
                  Jo{\~{a}}o Marques{-}Silva},
  title        = {SMT-Based Bounded Model Checking for Embedded {ANSI-C} Software},
  journal      = {{IEEE} Trans. Software Eng.},
  volume       = {38},
  number       = {4},
  pages        = {957--974},
  year         = {2012},
  doi          = {10.1109/TSE.2011.59}
}

@misc{gpt,
  title={Improving language understanding by generative pre-training},
  author={Radford, Alec and Narasimhan, Karthik and Salimans, Tim and Sutskever, Ilya and others},
  year={2018},
  publisher={OpenAI}
}

@article{austin2021program,
  title={Program synthesis with large language models},
  author={Austin, Jacob and Odena, Augustus and Nye, Maxwell and Bosma, Maarten and Michalewski, Henryk and Dohan, David and Jiang, Ellen and Cai, Carrie and Terry, Michael and Le, Quoc and others},
  journal={arXiv preprint arXiv:2108.07732},
  year={2021}
}

@article{sobania2023analysis,
  title={An analysis of the automatic bug fixing performance of chatgpt},
  author={Sobania, Dominik and Briesch, Martin and Hanna, Carol and Petke, Justyna},
  journal={arXiv preprint arXiv:2301.08653},
  year={2023}
}

@article{imani2023mathprompter,
  title={Mathprompter: Mathematical reasoning using large language models},
  author={Imani, Shima and Du, Liang and Shrivastava, Harsh},
  journal={arXiv preprint arXiv:2303.05398},
  year={2023}
}

@article{cao2023prompt,
  title={A study on Prompt Design, Advantages and Limitations of ChatGPT for Deep Learning Program Repair},
  author={Cao, Jialun and Li, Meiziniu and Wen, Ming and Cheung, Shing-chi},
  journal={arXiv:2304.08191},
  year={2023}
}

@misc{deng2023fuzzgpt,
      title={Large Language Models are Edge-Case Fuzzers: Testing Deep Learning Libraries via FuzzGPT}, 
      author={Yinlin Deng and Chunqiu Steven Xia and Chenyuan Yang and Shizhuo Dylan Zhang and Shujing Yang and Lingming Zhang},
      year={2023},
      eprint={2304.02014},
      archivePrefix={arXiv},
      primaryClass={cs.SE}
}
@article{getafix,
  title={Getafix: Learning to fix bugs automatically},
  author={Bader, Johannes and Scott, Andrew and Pradel, Michael and Chandra, Satish},
  journal={Proceedings of the ACM on Programming Languages},
  volume={3},
  number={OOPSLA},
  pages={1--27},
  year={2019},
  publisher={ACM New York, NY, USA}
}

@inproceedings{johnson2013don,
  title={Why don't software developers use static analysis tools to find bugs?},
  author={Johnson, Brittany and Song, Yoonki and Murphy-Hill, Emerson and Bowdidge, Robert},
  booktitle={2013 35th International Conference on Software Engineering (ICSE)},
  pages={672--681},
  year={2013},
  organization={IEEE}
}

@article{khoury_how_2023,
	title = {How {Secure} is {Code} {Generated} by {ChatGPT}?},
	url = {http://arxiv.org/abs/2304.09655},
	abstract = {In recent years, large language models have been responsible for great advances in the field of artificial intelligence (AI). ChatGPT in particular, an AI chatbot developed and recently released by OpenAI, has taken the field to the next level. The conversational model is able not only to process human-like text, but also to translate natural language into code. However, the safety of programs generated by ChatGPT should not be overlooked. In this paper, we perform an experiment to address this issue. Specifically, we ask ChatGPT to generate a number of program and evaluate the security of the resulting source code. We further investigate whether ChatGPT can be prodded to improve the security by appropriate prompts, and discuss the ethical aspects of using AI to generate code. Results suggest that ChatGPT is aware of potential vulnerabilities, but nonetheless often generates source code that are not robust to certain attacks.},
	language = {en},
	urldate = {2023-05-30},
	publisher = {arXiv},
	author = {Khoury, Raphaël and Avila, Anderson R. and Brunelle, Jacob and Camara, Baba Mamadou},
	month = apr,
	year = {2023},
	note = {arXiv:2304.09655 [cs]},
	keywords = {Computer Science - Cryptography and Security},
}

@article{burow2017control,
  title={Control-flow integrity: Precision, security, and performance},
  author={Burow, Nathan and Carr, Scott A and Nash, Joseph and Larsen, Per and Franz, Michael and Brunthaler, Stefan and Payer, Mathias},
  journal={ACM Computing Surveys (CSUR)},
  volume={50},
  number={1},
  pages={1--33},
  year={2017},
  publisher={ACM New York, NY, USA}
}

@inproceedings{hovemeyer2007finding,
  title={Finding more null pointer bugs, but not too many},
  author={Hovemeyer, David and Pugh, William},
  booktitle={Proceedings of the 7th ACM SIGPLAN-SIGSOFT workshop on Program analysis for software tools and engineering},
  pages={9--14},
  year={2007}
}

@inproceedings{gadelha2018esbmc,
  title={ESBMC 5.0: an industrial-strength C model checker},
  author={Gadelha, Mikhail R and Monteiro, Felipe R and Morse, Jeremy and Cordeiro, Lucas C and Fischer, Bernd and Nicole, Denis A},
  booktitle={Proceedings of the 33rd ACM/IEEE International Conference on Automated Software Engineering},
  pages={888--891},
  year={2018}
}

@inproceedings{lajko2022fine,
  title={Fine-tuning GPT-2 to patch programs, is it worth it?},
  author={Lajk{\'o}, M{\'a}rk and Horv{\'a}th, D{\'a}niel and Csuvik, Viktor and Vid{\'a}cs, L{\'a}szl{\'o}},
  booktitle={Computational Science and Its Applications--ICCSA 2022 Workshops: Malaga, Spain, July 4--7, 2022, Proceedings, Part IV},
  pages={79--91},
  year={2022},
  organization={Springer}
}
@inproceedings{findbugs,
  title={Evaluating and tuning a static analysis to find null pointer bugs},
  author={Hovemeyer, David and Spacco, Jaime and Pugh, William},
  booktitle={Proceedings of the 6th ACM SIGPLAN-SIGSOFT workshop on Program analysis for software tools and engineering},
  pages={13--19},
  year={2005}
}

@inproceedings{gupta2017deepfix,
  title={Deepfix: Fixing common c language errors by deep learning},
  author={Gupta, Rahul and Pal, Soham and Kanade, Aditya and Shevade, Shirish},
  booktitle={Proceedings of the aaai conference on artificial intelligence},
  volume={31},
  number={1},
  year={2017}
}

@article{prenner2021automatic,
  title={Automatic Program Repair with OpenAI's Codex: Evaluating QuixBugs},
  author={Prenner, Julian Aron and Robbes, Romain},
  journal={arXiv preprint arXiv:2111.03922},
  year={2021}
}

@article{fan2022improving,
  title={Improving automatically generated code from Codex via Automated Program Repair},
  author={Fan, Zhiyu and Gao, Xiang and Roychoudhury, Abhik and Tan, Shin Hwei},
  journal={arXiv preprint arXiv:2205.10583},
  year={2022}
}

@article{trummer2022codexdb,
  title={CodexDB: Synthesizing code for query processing from natural language instructions using GPT-3 Codex},
  author={Trummer, Immanuel},
  journal={Proceedings of the VLDB Endowment},
  volume={15},
  number={11},
  pages={2921--2928},
  year={2022},
  publisher={VLDB Endowment}
}



@article{chen2021evaluating,
  title={Evaluating Large Language Models Trained on Code},
  author={Chen, Mark and Tworek, Jerry and Jun, Heewoo and Yuan, Qiming and Pinto, Henrique Ponde de Oliveira and Kaplan, Jared and Edwards, Harri and Burda, Yuri and Joseph, Nicholas and Brockman, Greg and Ray, Alex and Puri, Raul and Krueger, Gretchen and Petrov, Michael and Khlaaf, Heidy and Sastry, Girish and Mishkin, Pamela and Chan, Brooke and Gray, Scott and Ryder, Nick and Pavlov, Mikhail and Power, Alethea and Kaiser, Lukasz and Bavarian, Mohammad and Winter, Clemens and Tillet, Philippe and Such, Felipe Petroski and Cummings, Dave and Plappert, Matthias and Chantzis, Fotios and Barnes, Elizabeth and Herbert-Voss, Ariel and Guss, William Hebgen and Nichol, Alex and Paino, Alex and Tezak, Nikolas and Tang, Jie and Babuschkin, Igor and Balaji, Suchir and Jain, Shantanu and Saunders, William and Hesse, Christopher and Carr, Andrew N. and Leike, Jan and Achiam, Josh and Misra, Vedant and Morikawa, Evan and Radford, Alec and Knight, Matthew and Brundage, Miles and Murati, Mira and Mayer, Katie and Welinder, Peter and McGrew, Bob and Amodei, Dario and McCandlish, Sam and Sutskever, Ilya and Zaremba, Wojciech},
  journal={arXiv preprint arXiv:2107.03374},
  year={2021},
  url={http://arxiv.org/abs/2107.03374}
}

@inproceedings{li2020dlfix,
  title={Dlfix: Context-based code transformation learning for automated program repair},
  author={Li, Yi and Wang, Shaohua and Nguyen, Tien N},
  booktitle={Proceedings of the ACM/IEEE 42nd International Conference on Software Engineering},
  pages={602--614},
  year={2020}
}

@inproceedings{lutellier2020coconut,
  title={Coconut: combining context-aware neural translation models using ensemble for program repair},
  author={Lutellier, Thibaud and Pham, Hung Viet and Pang, Lawrence and Li, Yitong and Wei, Moshi and Tan, Lin},
  booktitle={Proceedings of the 29th ACM SIGSOFT international symposium on software testing and analysis},
  pages={101--114},
  year={2020}
}
@article{bhayat2021towards,
  title={Towards a Hybrid Approach to Protect Against Memory Safety Vulnerabilities},
  author={Bhayat, Ahmed and Cordeiro, Lucas and Reger, Giles and Shmarov, Fedor and Korovin, Konstantin and Melham, Tom and Alshamrany, Kaled and A Mustafa, Mustafa and Olivier, Pierre},
  year={2021},
  publisher={TechRxiv}
}
@inproceedings{alshmrany2022fusebmc,
  title={FuSeBMC v4: Smart Seed Generation for Hybrid Fuzzing: (Competition Contribution)},
  author={Alshmrany, Kaled M and Aldughaim, Mohannad and Bhayat, Ahmed and Cordeiro, Lucas C},
  booktitle={Fundamental Approaches to Software Engineering: 25th International Conference, FASE 2022, Held as Part of the European Joint Conferences on Theory and Practice of Software, ETAPS 2022, Munich, Germany, April 2--7, 2022, Proceedings},
  pages={336--340},
  year={2022},
  organization={Springer International Publishing Cham}
}
@inproceedings{song2022esbmc,
  title={ESBMC-solidity: an SMT-based model checker for solidity smart contracts},
  author={Song, Kunjian and Matulevicius, Nedas and de Lima Filho, Eddie B and Cordeiro, Lucas C},
  booktitle={Proceedings of the ACM/IEEE 44th International Conference on Software Engineering: Companion Proceedings},
  pages={65--69},
  year={2022}
}
@article{aljaafari2022combining,
  title={Combining BMC and Fuzzing Techniques for Finding Software Vulnerabilities in Concurrent Programs},
  author={Aljaafari, Fatimah K and Menezes, Rafael and Manino, Edoardo and Shmarov, Fedor and Mustafa, Mustafa A and Cordeiro, Lucas C},
  journal={IEEE Access},
  volume={10},
  pages={121365--121384},
  year={2022},
  publisher={IEEE}
}
@article{chen2019sequencer,
  title={Sequencer: Sequence-to-sequence learning for end-to-end program repair},
  author={Chen, Zimin and Kommrusch, Steve and Tufano, Michele and Pouchet, Louis-No{\"e}l and Poshyvanyk, Denys and Monperrus, Martin},
  journal={IEEE Transactions on Software Engineering},
  volume={47},
  number={9},
  pages={1943--1959},
  year={2019},
  publisher={IEEE}
}

@inproceedings{zhu2021syntaxnmt,
  title={A syntax-guided edit decoder for neural program repair},
  author={Zhu, Qihao and Sun, Zeyu and Xiao, Yuan-an and Zhang, Wenjie and Yuan, Kang and Xiong, Yingfei and Zhang, Lu},
  booktitle={Proceedings of the 29th ACM Joint Meeting on European Software Engineering Conference and Symposium on the Foundations of Software Engineering},
  pages={341--353},
  year={2021}
}
@inproceedings{promptprogramming,
  title={Prompt programming for large language models: Beyond the few-shot paradigm},
  author={Reynolds, Laria and McDonell, Kyle},
  booktitle={Extended Abstracts of the 2021 CHI Conference on Human Factors in Computing Systems},
  pages={1--7},
  year={2021}
}

@inproceedings{alpharepair,
  title={Less training, more repairing please: revisiting automated program repair via zero-shot learning},
  author={Xia, Chunqiu Steven and Zhang, Lingming},
  booktitle={Proceedings of the 30th ACM Joint European Software Engineering Conference and Symposium on the Foundations of Software Engineering},
  pages={959--971},
  year={2022}
}

@inproceedings{pearce2022examining,
  title={Examining Zero-Shot Vulnerability Repair with Large Language Models},
  author={Pearce, Hammond and Tan, Benjamin and Ahmad, Baleegh and Karri, Ramesh and Dolan-Gavitt, Brendan},
  booktitle={2023 IEEE Symposium on Security and Privacy (SP)},
  pages={1--18},
  year={2022},
  organization={IEEE Computer Society}
}

@inproceedings{lajko2022towards,
  title={Towards JavaScript program repair with generative pre-trained transformer (GPT-2)},
  author={Lajk{\'o}, M{\'a}rk and Csuvik, Viktor and Vid{\'a}cs, L{\'a}szl{\'o}},
  booktitle={Proceedings of the Third International Workshop on Automated Program Repair},
  pages={61--68},
  year={2022}
}

@article{sutskever2014nmt,
  title={Sequence to sequence learning with neural networks},
  author={Sutskever, Ilya and Vinyals, Oriol and Le, Quoc V},
  journal={Advances in neural information processing systems},
  volume={27},
  year={2014}
}

@inproceedings{sadowski2014developers,
  title={How developers use data race detection tools},
  author={Sadowski, Caitlin and Yi, Jaeheon},
  booktitle={Proceedings of the 5th Workshop on Evaluation and Usability of Programming Languages and Tools},
  pages={43--51},
  year={2014}
}

@INPROCEEDINGS{cure,
  author={Jiang, Nan and Lutellier, Thibaud and Tan, Lin},
  booktitle={2021 IEEE/ACM 43rd International Conference on Software Engineering (ICSE)}, 
  title={CURE: Code-Aware Neural Machine Translation for Automatic Program Repair}, 
  year={2021},
  volume={},
  number={},
  pages={1161-1173},
  doi={10.1109/ICSE43902.2021.00107}}


@inproceedings{li2022dear,
  title={Dear: A novel deep learning-based approach for automated program repair},
  author={Li, Yi and Wang, Shaohua and Nguyen, Tien N},
  booktitle={Proceedings of the 44th International Conference on Software Engineering},
  pages={511--523},
  year={2022}
}

@article{zhang2010conmem,
  title={ConMem: detecting severe concurrency bugs through an effect-oriented approach},
  author={Zhang, Wei and Sun, Chong and Lu, Shan},
  journal={ACM Sigplan Notices},
  volume={45},
  number={3},
  pages={179--192},
  year={2010},
  publisher={ACM New York, NY, USA}
}

@inproceedings{zaman2011security,
  title={Security versus performance bugs: a case study on firefox},
  author={Zaman, Shahed and Adams, Bram and Hassan, Ahmed E},
  booktitle={Proceedings of the 8th working conference on mining software repositories},
  pages={93--102},
  year={2011}
}

@inproceedings{bajwa2015unintentional,
  title={Unintentional bugs to vulnerability mapping in android applications},
  author={Bajwa, Garima and Fazeen, Mohamed and Dantu, Ram and Tanpure, Sonal},
  booktitle={2015 IEEE International Conference on Intelligence and Security Informatics (ISI)},
  pages={176--178},
  year={2015},
  organization={IEEE}
}

@book{dustin1999automated,
  title={Automated software testing: Introduction, management, and performance: Introduction, management, and performance},
  author={Dustin, Elfriede and Rashka, Jeff and Paul, John},
  year={1999},
  publisher={Addison-Wesley Professional}
}

@article{godefroid2008automating,
  title={Automating software testing using program analysis},
  author={Godefroid, Patrice and de Halleux, Peli and Nori, Aditya V and Rajamani, Sriram K and Schulte, Wolfram and Tillmann, Nikolai and Levin, Michael Y},
  journal={IEEE software},
  volume={25},
  number={5},
  pages={30--37},
  year={2008},
  publisher={IEEE}
}

@article{goues2019automatedrepair,
  title={Automated program repair},
  author={Goues, Claire Le and Pradel, Michael and Roychoudhury, Abhik},
  journal={Communications of the ACM},
  volume={62},
  number={12},
  pages={56--65},
  year={2019},
  publisher={ACM New York, NY, USA}
}

@article{evalplus,
  title={Is Your Code Generated by ChatGPT Really Correct? Rigorous Evaluation of Large Language Models for Code Generation},
  author={Liu, Jiawei and Xia, Chunqiu Steven and Wang, Yuyao and Zhang, Lingming},
  journal={arXiv preprint arXiv:2305.01210},
  year={2023}
}


@article{compcode,
  title={Compilable neural code generation with compiler feedback},
  author={Wang, Xin and Wang, Yasheng and Wan, Yao and Mi, Fei and Li, Yitong and Zhou, Pingyi and Liu, Jin and Wu, Hao and Jiang, Xin and Liu, Qun},
  journal={arXiv preprint arXiv:2203.05132},
  year={2022}
}

@article{tian2023assistant,
  title={Is ChatGPT the Ultimate Programming Assistant--How far is it?},
  author={Tian, Haoye and Lu, Weiqi and Li, Tsz On and Tang, Xunzhu and Cheung, Shing-Chi and Klein, Jacques and Bissyand{\'e}, Tegawend{\'e} F},
  journal={arXiv preprint arXiv:2304.11938},
  year={2023}
} 
@article{brown2020language,
  title={Language models are few-shot learners},
  author={Brown, Tom and Mann, Benjamin and Ryder, Nick and Subbiah, Melanie and Kaplan, Jared D and Dhariwal, Prafulla and Neelakantan, Arvind and Shyam, Pranav and Sastry, Girish and Askell, Amanda and others},
  journal={Advances in neural information processing systems},
  volume={33},
  pages={1877--1901},
  year={2020}
}


@inproceedings{csmith,
  title={Finding and understanding bugs in C compilers},
  author={Yang, Xuejun and Chen, Yang and Eide, Eric and Regehr, John},
  booktitle={Proceedings of the 32nd ACM SIGPLAN conference on Programming language design and implementation},
  pages={283--294},
  year={2011}
}

@misc{strasser2023pitfalls,
      title={On pitfalls (and advantages) of sophisticated large language models}, 
      author={Anna Strasser},
      year={2023},
      eprint={2303.17511},
      archivePrefix={arXiv},
      primaryClass={cs.CY}
}

@article{austin2021program2,
  title={Program synthesis with large language models},
  author={Austin, Jacob and Odena, Augustus and Nye, Maxwell and Bosma, Maarten and Michalewski, Henryk and Dohan, David and Jiang, Ellen and Cai, Carrie and Terry, Michael and Le, Quoc and others},
  journal={arXiv preprint arXiv:2108.07732},
  year={2021}
}

@inproceedings{jain2023code,
  title={A Code Centric Evaluation of C/C++ Vulnerability Datasets for Deep Learning Based Vulnerability Detection Techniques},
  author={Jain, Ridhi and Gervasoni, Nicole and Ndhlovu, Mthandazo and Rawat, Sanjay},
  booktitle={Proceedings of the 16th Innovations in Software Engineering Conference},
  pages={1--10},
  year={2023}
}

@inproceedings{white2016deep,
  title={Deep learning code fragments for code clone detection},
  author={White, Martin and Tufano, Michele and Vendome, Christopher and Poshyvanyk, Denys},
  booktitle={Proceedings of the 31st IEEE/ACM international conference on automated software engineering},
  pages={87--98},
  year={2016}
}

@inproceedings{tu2022remgen,
  title={Remgen: Remanufacturing a Random Program Generator for Compiler Testing},
  author={Tu, Haoxin and Jiang, He and Li, Xiaochen and Ren, Zhilei and Zhou, Zhide and Jiang, Lingxiao},
  booktitle={2022 IEEE 33rd International Symposium on Software Reliability Engineering (ISSRE)},
  pages={529--540},
  year={2022},
  organization={IEEE}
}

@inproceedings{da2021anghabench,
  title={Anghabench: A suite with one million compilable c benchmarks for code-size reduction},
  author={Da Silva, Anderson Faustino and Kind, Bruno Conde and de Souza Magalh{\~a}es, Jos{\'e} Wesley and Rocha, Jer{\^o}nimo Nunes and Guimaraes, Breno Campos Ferreira and Pereira, Fernando Magno Quin{\~a}o},
  booktitle={2021 IEEE/ACM International Symposium on Code Generation and Optimization (CGO)},
  pages={378--390},
  year={2021},
  organization={IEEE}
}


@inproceedings{barany2018ldrgen,
  title={Liveness-driven random program generation},
  author={Barany, Gerg{\"o}},
  booktitle={Logic-Based Program Synthesis and Transformation: 27th International Symposium, LOPSTR 2017, Namur, Belgium, October 10-12, 2017, Revised Selected Papers 27},
  pages={112--127},
  year={2018},
  organization={Springer}
}

@inproceedings{randprog1,
  title={Volatiles are miscompiled, and what to do about it},
  author={Eide, Eric and Regehr, John},
  booktitle={Proceedings of the 8th ACM international conference on Embedded software},
  pages={255--264},
  year={2008}
}

@inproceedings{zhao2018deepsim,
  title={Deepsim: deep learning code functional similarity},
  author={Zhao, Gang and Huang, Jeff},
  booktitle={Proceedings of the 2018 26th ACM Joint Meeting on European Software Engineering Conference and Symposium on the Foundations of Software Engineering},
  pages={141--151},
  year={2018}
}

@inproceedings{sarsa2022automatic,
  title={Automatic generation of programming exercises and code explanations using large language models},
  author={Sarsa, Sami and Denny, Paul and Hellas, Arto and Leinonen, Juho},
  booktitle={Proceedings of the 2022 ACM Conference on International Computing Education Research-Volume 1},
  pages={27--43},
  year={2022}
}

@inproceedings{jain2022jigsaw,
  title={Jigsaw: Large language models meet program synthesis},
  author={Jain, Naman and Vaidyanath, Skanda and Iyer, Arun and Natarajan, Nagarajan and Parthasarathy, Suresh and Rajamani, Sriram and Sharma, Rahul},
  booktitle={Proceedings of the 44th International Conference on Software Engineering},
  pages={1219--1231},
  year={2022}
}

@article{lampinen2022can,
  title={Can language models learn from explanations in context?},
  author={Lampinen, Andrew K and Dasgupta, Ishita and Chan, Stephanie CY and Matthewson, Kory and Tessler, Michael Henry and Creswell, Antonia and McClelland, James L and Wang, Jane X and Hill, Felix},
  journal={arXiv preprint arXiv:2204.02329},
  year={2022}
}

@inproceedings{chakraborty2022natgen,
  title={NatGen: generative pre-training by “naturalizing” source code},
  author={Chakraborty, Saikat and Ahmed, Toufique and Ding, Yangruibo and Devanbu, Premkumar T and Ray, Baishakhi},
  booktitle={Proceedings of the 30th ACM Joint European Software Engineering Conference and Symposium on the Foundations of Software Engineering},
  pages={18--30},
  year={2022}
}

@article{dakhel2023github,
  title={Github copilot ai pair programmer: Asset or liability?},
  author={Dakhel, Arghavan Moradi and Majdinasab, Vahid and Nikanjam, Amin and Khomh, Foutse and Desmarais, Michel C and Jiang, Zhen Ming},
  journal={Journal of Systems and Software},
  pages={111734},
  year={2023},
  publisher={Elsevier}
}

@article{llmAPRuntrust,
  title={Automated Repair of Programs from Large Language Models},
  author={Fan, Zhiyu and Gao, Xiang and Roychoudhury, Abhik and Tan, Shin Hwei},
  journal={arXiv preprint arXiv:2205.10583},
  year={2022}
}

@inproceedings{ahmed2022few,
  title={Few-shot training LLMs for project-specific code-summarization},
  author={Ahmed, Toufique and Devanbu, Premkumar},
  booktitle={37th IEEE/ACM International Conference on Automated Software Engineering},
  pages={1--5},
  year={2022}
}
@inproceedings{deeprepair,
  title={Sorting and transforming program repair ingredients via deep learning code similarities},
  author={White, Martin and Tufano, Michele and Martinez, Matias and Monperrus, Martin and Poshyvanyk, Denys},
  booktitle={2019 IEEE 26th International Conference on Software Analysis, Evolution and Reengineering (SANER)},
  pages={479--490},
  year={2019},
  organization={IEEE}
}

@article{jin2023inferfix,
  title={InferFix: End-to-End Program Repair with LLMs},
  author={Jin, Matthew and Shahriar, Syed and Tufano, Michele and Shi, Xin and Lu, Shuai and Sundaresan, Neel and Svyatkovskiy, Alexey},
  journal={arXiv preprint arXiv:2303.07263},
  year={2023}
}

@article{li2023finding,
  title={Finding Failure-Inducing Test Cases with ChatGPT},
  author={Li, Tsz-On and Zong, Wenxi and Wang, Yibo and Tian, Haoye and Wang, Ying and Cheung, Shing-Chi},
  journal={arXiv preprint arXiv:2304.11686},
  year={2023}
}

@article{gazzola2019automatic,
  title={Automatic Software Repair: A Survey},
  author={Gazzola, Luca and Micucci, Daniela and Mariani, Leonardo},
  journal={IEEE Transactions on Software Engineering},
  volume={45},
  number={01},
  pages={34--67},
  year={2019},
  publisher={IEEE Computer Society}
}

@article{xia2023conversational,
  title={Conversational automated program repair},
  author={Xia, Chunqiu Steven and Zhang, Lingming},
  journal={arXiv preprint arXiv:2301.13246},
  year={2023}
}

@inproceedings{yang2020applying,
  title={Applying deep learning algorithm to automatic bug localization and repair},
  author={Yang, Geunseok and Min, Kyeongsic and Lee, Byungjeong},
  booktitle={Proceedings of the 35th Annual ACM symposium on applied computing},
  pages={1634--1641},
  year={2020}
}

@InProceedings{SVCOMP2023,
author="Beyer, Dirk",
editor="Sankaranarayanan, Sriram
and Sharygina, Natasha",
title="Competition on Software Verification and Witness Validation: SV-COMP 2023",
booktitle="Tools and Algorithms for the Construction and Analysis of Systems",
year="2023",
publisher="Springer Nature Switzerland",
address="Cham",
pages="495--522",
abstract="The 12th edition of the Competition on Software Verification (SV-COMP 2023) is again the largest overview of tools for software verification, evaluating 52 verification systems from 34 teams from 10 countries. Besides providing an overview of the state of the art in automatic software verification, the goal of the competition is to establish standards, provide a platform for exchange to developers of such tools, educate PhD students on reproducibility approaches and benchmarking, and provide computing resources to developers that do not have access to compute clusters. The competition consisted of 23 805 verification tasks for C programs and 586 verification tasks for Java programs. The specifications include reachability, memory safety, overflows, and termination. This year, the competition introduced a new competition track on witness validation, where validators for verification witnesses are evaluated with respect to their quality.",
isbn="978-3-031-30820-8"
}


@inproceedings{lin2017quixbugs,
  title={QuixBugs: A multi-lingual program repair benchmark set based on the Quixey Challenge},
  author={Lin, Derrick and Koppel, James and Chen, Angela and Solar-Lezama, Armando},
  booktitle={Proceedings Companion of the 2017 ACM SIGPLAN international conference on systems, programming, languages, and applications: software for humanity},
  pages={55--56},
  year={2017}
}

@article{ma2023scope,
  title={The Scope of ChatGPT in Software Engineering: A Thorough Investigation},
  author={Ma, Wei and Liu, Shangqing and Wang, Wenhan and Hu, Qiang and Liu, Ye and Zhang, Cen and Nie, Liming and Liu, Yang},
  journal={arXiv preprint arXiv:2305.12138},
  year={2023}
}

\begin{thebibliography}{47}


\ifx \showCODEN    \undefined \def \showCODEN     #1{\unskip}     \fi
\ifx \showDOI      \undefined \def \showDOI       #1{#1}\fi
\ifx \showISBNx    \undefined \def \showISBNx     #1{\unskip}     \fi
\ifx \showISBNxiii \undefined \def \showISBNxiii  #1{\unskip}     \fi
\ifx \showISSN     \undefined \def \showISSN      #1{\unskip}     \fi
\ifx \showLCCN     \undefined \def \showLCCN      #1{\unskip}     \fi
\ifx \shownote     \undefined \def \shownote      #1{#1}          \fi
\ifx \showarticletitle \undefined \def \showarticletitle #1{#1}   \fi
\ifx \showURL      \undefined \def \showURL       {\relax}        \fi
\providecommand\bibfield[2]{#2}
\providecommand\bibinfo[2]{#2}
\providecommand\natexlab[1]{#1}
\providecommand\showeprint[2][]{arXiv:#2}

\bibitem[Aho et~al\mbox{.}(2006)]%
        {Aho:2006:CPT:1177220}
\bibfield{author}{\bibinfo{person}{Alfred~V. Aho}, \bibinfo{person}{Monica~S.
  Lam}, \bibinfo{person}{Ravi Sethi}, {and} \bibinfo{person}{Jeffrey~D.
  Ullman}.} \bibinfo{year}{2006}\natexlab{}.
\newblock \bibinfo{booktitle}{\emph{Compilers: Principles, Techniques, And
  Tools} (\bibinfo{edition}{2nd} ed.)}.
\newblock \bibinfo{publisher}{Addison-Wesley Longman Publishing Co., Inc.}
\newblock


\bibitem[Avila(2022)]%
        {risto_embedded_2022}
\bibfield{author}{\bibinfo{person}{Risto Avila}.}
  \bibinfo{year}{2022}\natexlab{}.
\newblock \bibinfo{booktitle}{\emph{Embedded {Software} {Programming}
  {Languages}: {Pros}, {Cons}, and {Comparisons} of {Popular} {Languages}}}.
\newblock
\urldef\tempurl%
\url{https://www.qt.io/embedded-development-talk/embedded-software-programming-languages-pros-cons-and-comparisons-of-popular-languages}
\showURL{%
\tempurl}


\bibitem[Beyer(2023)]%
        {SVCOMP2023}
\bibfield{author}{\bibinfo{person}{Dirk Beyer}.}
  \bibinfo{year}{2023}\natexlab{}.
\newblock \showarticletitle{Competition on Software Verification and Witness
  Validation: SV-COMP 2023}. In \bibinfo{booktitle}{\emph{Tools and Algorithms
  for the Construction and Analysis of Systems}},
  \bibfield{editor}{\bibinfo{person}{Sriram Sankaranarayanan} {and}
  \bibinfo{person}{Natasha Sharygina}} (Eds.). \bibinfo{publisher}{Springer
  Nature Switzerland}, \bibinfo{address}{Cham}, \bibinfo{pages}{495--522}.
\newblock
\showISBNx{978-3-031-30820-8}


\bibitem[Black(2018)]%
        {black_software_2018}
\bibfield{author}{\bibinfo{person}{Paul~E. Black}.}
  \bibinfo{year}{2018}\natexlab{}.
\newblock \showarticletitle{A {Software} {Assurance} {Reference} {Dataset}:
  {Thousands} of {Programs} {With} {Known} {Bugs}}.
\newblock \bibinfo{journal}{\emph{Journal of Research of the National Institute
  of Standards and Technology}}  \bibinfo{volume}{123} (\bibinfo{date}{April}
  \bibinfo{year}{2018}), \bibinfo{pages}{1--3}.
\newblock
\showISSN{1044-677X}
\urldef\tempurl%
\url{https://doi.org/10.6028/jres.123.005}
\showDOI{\tempurl}


\bibitem[Bui et~al\mbox{.}(2023)]%
        {bui_codetf_2023}
\bibfield{author}{\bibinfo{person}{Nghi D.~Q. Bui}, \bibinfo{person}{Hung Le},
  \bibinfo{person}{Yue Wang}, \bibinfo{person}{Junnan Li},
  \bibinfo{person}{Akhilesh~Deepak Gotmare}, {and} \bibinfo{person}{Steven
  C.~H. Hoi}.} \bibinfo{year}{2023}\natexlab{}.
\newblock \showarticletitle{{CodeTF}: {One}-stop {Transformer} {Library} for
  {State}-of-the-art {Code} {LLM}}.
\newblock  (\bibinfo{date}{May} \bibinfo{year}{2023}).
\newblock
\urldef\tempurl%
\url{http://arxiv.org/abs/2306.00029}
\showURL{%
\tempurl}
\newblock
\shownote{arXiv:2306.00029 [cs]}.


\bibitem[Chakraborty et~al\mbox{.}(2022)]%
        {chakraborty_deep_2022}
\bibfield{author}{\bibinfo{person}{Saikat Chakraborty}, \bibinfo{person}{Rahul
  Krishna}, \bibinfo{person}{Yangruibo Ding}, {and} \bibinfo{person}{Baishakhi
  Ray}.} \bibinfo{year}{2022}\natexlab{}.
\newblock \showarticletitle{Deep {Learning} {Based} {Vulnerability}
  {Detection}: {Are} {We} {There} {Yet}?}
\newblock \bibinfo{journal}{\emph{IEEE Transactions on Software Engineering}}
  \bibinfo{volume}{48}, \bibinfo{number}{9} (\bibinfo{date}{Sept.}
  \bibinfo{year}{2022}), \bibinfo{pages}{3280--3296}.
\newblock
\showISSN{1939-3520}
\urldef\tempurl%
\url{https://doi.org/10.1109/TSE.2021.3087402}
\showDOI{\tempurl}
\newblock
\shownote{Conference Name: IEEE Transactions on Software Engineering}.


\bibitem[Charalambous et~al\mbox{.}(2023)]%
        {charalambous_new_2023}
\bibfield{author}{\bibinfo{person}{Yiannis Charalambous},
  \bibinfo{person}{Norbert Tihanyi}, \bibinfo{person}{Ridhi Jain},
  \bibinfo{person}{Youcheng Sun}, \bibinfo{person}{Mohamed~Amine Ferrag}, {and}
  \bibinfo{person}{Lucas~C. Cordeiro}.} \bibinfo{year}{2023}\natexlab{}.
\newblock \showarticletitle{A {New} {Era} in {Software} {Security}: {Towards}
  {Self}-{Healing} {Software} via {Large} {Language} {Models} and {Formal}
  {Verification}}.
\newblock  (\bibinfo{date}{May} \bibinfo{year}{2023}).
\newblock
\urldef\tempurl%
\url{https://doi.org/10.48550/arXiv.2305.14752}
\showDOI{\tempurl}
\newblock
\shownote{arXiv:2305.14752 [cs]}.


\bibitem[Chavez et~al\mbox{.}(2023)]%
        {chavez_chat_2023}
\bibfield{author}{\bibinfo{person}{Martin~R. Chavez},
  \bibinfo{person}{Thomas~S. Butler}, \bibinfo{person}{Patricia Rekawek},
  \bibinfo{person}{Hye Heo}, {and} \bibinfo{person}{Wendy~L. Kinzler}.}
  \bibinfo{year}{2023}\natexlab{}.
\newblock \showarticletitle{Chat {Generative} {Pre}-trained {Transformer}: why
  we should embrace this technology}.
\newblock \bibinfo{journal}{\emph{American Journal of Obstetrics and
  Gynecology}} \bibinfo{volume}{228}, \bibinfo{number}{6} (\bibinfo{date}{June}
  \bibinfo{year}{2023}), \bibinfo{pages}{706--711}.
\newblock
\showISSN{0002-9378}
\urldef\tempurl%
\url{https://doi.org/10.1016/j.ajog.2023.03.010}
\showDOI{\tempurl}


\bibitem[Chen et~al\mbox{.}(2023)]%
        {chen_diversevul_2023}
\bibfield{author}{\bibinfo{person}{Yizheng Chen}, \bibinfo{person}{Zhoujie
  Ding}, \bibinfo{person}{Xinyun Chen}, {and} \bibinfo{person}{David Wagner}.}
  \bibinfo{year}{2023}\natexlab{}.
\newblock \showarticletitle{{DiverseVul}: {A} {New} {Vulnerable} {Source}
  {Code} {Dataset} for {Deep} {Learning} {Based} {Vulnerability} {Detection}}.
\newblock  (\bibinfo{date}{April} \bibinfo{year}{2023}).
\newblock
\urldef\tempurl%
\url{http://arxiv.org/abs/2304.00409}
\showURL{%
\tempurl}
\newblock
\shownote{arXiv:2304.00409 [cs]}.


\bibitem[Cordeiro et~al\mbox{.}(2012)]%
        {cordeiro_smt-based_2012}
\bibfield{author}{\bibinfo{person}{Lucas Cordeiro}, \bibinfo{person}{Bernd
  Fischer}, {and} \bibinfo{person}{Joao Marques-Silva}.}
  \bibinfo{year}{2012}\natexlab{}.
\newblock \showarticletitle{{SMT}-{Based} {Bounded} {Model} {Checking} for
  {Embedded} {ANSI}-{C} {Software}}.
\newblock \bibinfo{journal}{\emph{IEEE Transactions on Software Engineering}}
  \bibinfo{volume}{38}, \bibinfo{number}{4} (\bibinfo{date}{July}
  \bibinfo{year}{2012}), \bibinfo{pages}{957--974}.
\newblock
\showISSN{1939-3520}
\urldef\tempurl%
\url{https://doi.org/10.1109/TSE.2011.59}
\showDOI{\tempurl}
\newblock
\shownote{Conference Name: IEEE Transactions on Software Engineering}.


\bibitem[D'Silva et~al\mbox{.}(2008)]%
        {dsilva_survey_2008}
\bibfield{author}{\bibinfo{person}{Vijay D'Silva}, \bibinfo{person}{Daniel
  Kroening}, {and} \bibinfo{person}{Georg Weissenbacher}.}
  \bibinfo{year}{2008}\natexlab{}.
\newblock \showarticletitle{A {Survey} of {Automated} {Techniques} for {Formal}
  {Software} {Verification}}.
\newblock \bibinfo{journal}{\emph{IEEE Transactions on Computer-Aided Design of
  Integrated Circuits and Systems}} \bibinfo{volume}{27}, \bibinfo{number}{7}
  (\bibinfo{date}{July} \bibinfo{year}{2008}), \bibinfo{pages}{1165--1178}.
\newblock
\showISSN{1937-4151}
\urldef\tempurl%
\url{https://doi.org/10.1109/TCAD.2008.923410}
\showDOI{\tempurl}
\newblock
\shownote{Conference Name: IEEE Transactions on Computer-Aided Design of
  Integrated Circuits and Systems}.


\bibitem[Fan et~al\mbox{.}(2020)]%
        {fan_cc_2020}
\bibfield{author}{\bibinfo{person}{Jiahao Fan}, \bibinfo{person}{Yi Li},
  \bibinfo{person}{Shaohua Wang}, {and} \bibinfo{person}{Tien~N. Nguyen}.}
  \bibinfo{year}{2020}\natexlab{}.
\newblock \showarticletitle{A {C}/{C}++ {Code} {Vulnerability} {Dataset} with
  {Code} {Changes} and {CVE} {Summaries}}. In
  \bibinfo{booktitle}{\emph{Proceedings of the 17th {International}
  {Conference} on {Mining} {Software} {Repositories}}}
  \emph{(\bibinfo{series}{{MSR} '20})}. \bibinfo{publisher}{Association for
  Computing Machinery}, \bibinfo{address}{New York, NY, USA},
  \bibinfo{pages}{508--512}.
\newblock
\showISBNx{978-1-4503-7517-7}
\urldef\tempurl%
\url{https://doi.org/10.1145/3379597.3387501}
\showDOI{\tempurl}


\bibitem[Gadelha et~al\mbox{.}(2018)]%
        {gadelha2018esbmc}
\bibfield{author}{\bibinfo{person}{Mikhail~R Gadelha},
  \bibinfo{person}{Felipe~R Monteiro}, \bibinfo{person}{Jeremy Morse},
  \bibinfo{person}{Lucas~C Cordeiro}, \bibinfo{person}{Bernd Fischer}, {and}
  \bibinfo{person}{Denis~A Nicole}.} \bibinfo{year}{2018}\natexlab{}.
\newblock \showarticletitle{ESBMC 5.0: an industrial-strength C model checker}.
  In \bibinfo{booktitle}{\emph{Proceedings of the 33rd ACM/IEEE International
  Conference on Automated Software Engineering}}. \bibinfo{pages}{888--891}.
\newblock


\bibitem[Gadelha et~al\mbox{.}(2023)]%
        {gadelha_esbmc_2023}
\bibfield{author}{\bibinfo{person}{Mikhail~R. Gadelha},
  \bibinfo{person}{Felipe~R. Monteiro}, \bibinfo{person}{Jeremy Morse},
  \bibinfo{person}{Lucas~C. Cordeiro}, \bibinfo{person}{Bernd Fischer}, {and}
  \bibinfo{person}{Denis~A. Nicole}.} \bibinfo{year}{2023}\natexlab{}.
\newblock \bibinfo{booktitle}{\emph{{ESBMC}: 5.0: {An} {Industrial}-{Strength}
  {Model} {Checker}}}.
\newblock
\urldef\tempurl%
\url{https://github.com/esbmc/esbmc}
\showURL{%
\tempurl}
\newblock
\shownote{original-date: 2015-06-20T19:35:34Z}.


\bibitem[Gadelha et~al\mbox{.}(2019)]%
        {GadelhaSC0N19}
\bibfield{author}{\bibinfo{person}{Mikhail Y.~R. Gadelha},
  \bibinfo{person}{Enrico Steffinlongo}, \bibinfo{person}{Lucas~C. Cordeiro},
  \bibinfo{person}{Bernd Fischer}, {and} \bibinfo{person}{Denis~A. Nicole}.}
  \bibinfo{year}{2019}\natexlab{}.
\newblock \showarticletitle{SMT-based refutation of spurious bug reports in the
  clang static analyzer}. In \bibinfo{booktitle}{\emph{Proceedings of the 41st
  International Conference on Software Engineering: Companion Proceedings,
  {ICSE} 2019, Montreal, QC, Canada, May 25-31, 2019}},
  \bibfield{editor}{\bibinfo{person}{Joanne~M. Atlee}, \bibinfo{person}{Tevfik
  Bultan}, {and} \bibinfo{person}{Jon Whittle}} (Eds.).
  \bibinfo{publisher}{{IEEE} / {ACM}}, \bibinfo{pages}{11--14}.
\newblock
\urldef\tempurl%
\url{https://doi.org/10.1109/ICSE-Companion.2019.00026}
\showDOI{\tempurl}


\bibitem[Hutchinson et~al\mbox{.}(2021)]%
        {hutchinson_towards_2021}
\bibfield{author}{\bibinfo{person}{Ben Hutchinson}, \bibinfo{person}{Andrew
  Smart}, \bibinfo{person}{Alex Hanna}, \bibinfo{person}{Emily Denton},
  \bibinfo{person}{Christina Greer}, \bibinfo{person}{Oddur Kjartansson},
  \bibinfo{person}{Parker Barnes}, {and} \bibinfo{person}{Margaret Mitchell}.}
  \bibinfo{year}{2021}\natexlab{}.
\newblock \showarticletitle{Towards {Accountability} for {Machine} {Learning}
  {Datasets}: {Practices} from {Software} {Engineering} and {Infrastructure}}.
  In \bibinfo{booktitle}{\emph{Proceedings of the 2021 {ACM} {Conference} on
  {Fairness}, {Accountability}, and {Transparency}}}
  \emph{(\bibinfo{series}{{FAccT} '21})}. \bibinfo{publisher}{Association for
  Computing Machinery}, \bibinfo{address}{New York, NY, USA},
  \bibinfo{pages}{560--575}.
\newblock
\showISBNx{978-1-4503-8309-7}
\urldef\tempurl%
\url{https://doi.org/10.1145/3442188.3445918}
\showDOI{\tempurl}


\bibitem[Imani et~al\mbox{.}(2023)]%
        {imani2023mathprompter}
\bibfield{author}{\bibinfo{person}{Shima Imani}, \bibinfo{person}{Liang Du},
  {and} \bibinfo{person}{Harsh Shrivastava}.} \bibinfo{year}{2023}\natexlab{}.
\newblock \showarticletitle{Mathprompter: Mathematical reasoning using large
  language models}.
\newblock \bibinfo{journal}{\emph{arXiv preprint arXiv:2303.05398}}
  (\bibinfo{year}{2023}).
\newblock


\bibitem[Jain et~al\mbox{.}(2023)]%
        {jain2023code}
\bibfield{author}{\bibinfo{person}{Ridhi Jain}, \bibinfo{person}{Nicole
  Gervasoni}, \bibinfo{person}{Mthandazo Ndhlovu}, {and}
  \bibinfo{person}{Sanjay Rawat}.} \bibinfo{year}{2023}\natexlab{}.
\newblock \showarticletitle{A Code Centric Evaluation of C/C++ Vulnerability
  Datasets for Deep Learning Based Vulnerability Detection Techniques}. In
  \bibinfo{booktitle}{\emph{Proceedings of the 16th Innovations in Software
  Engineering Conference}}. \bibinfo{pages}{1--10}.
\newblock


\bibitem[Jr and Black(2012)]%
        {jr_juliet_2012}
\bibfield{author}{\bibinfo{person}{Frederick E.~Boland Jr} {and}
  \bibinfo{person}{Paul~E. Black}.} \bibinfo{year}{2012}\natexlab{}.
\newblock \showarticletitle{The {Juliet} 1.1 {C}/{C}++ and {Java} {Test}
  {Suite}}.
\newblock \bibinfo{journal}{\emph{NIST}} \bibinfo{volume}{45},
  \bibinfo{number}{10} (\bibinfo{date}{Oct.} \bibinfo{year}{2012}),
  \bibinfo{pages}{88--90}.
\newblock
\urldef\tempurl%
\url{https://www.nist.gov/publications/juliet-11-cc-and-java-test-suite}
\showURL{%
\tempurl}
\newblock
\shownote{Last Modified: 2021-10-12T11:10-04:00 Publisher: Frederick E. Boland
  Jr., Paul E. Black}.


\bibitem[Khoury et~al\mbox{.}(2023)]%
        {khoury_how_2023}
\bibfield{author}{\bibinfo{person}{Raphaël Khoury},
  \bibinfo{person}{Anderson~R. Avila}, \bibinfo{person}{Jacob Brunelle}, {and}
  \bibinfo{person}{Baba~Mamadou Camara}.} \bibinfo{year}{2023}\natexlab{}.
\newblock \showarticletitle{How {Secure} is {Code} {Generated} by {ChatGPT}?}
\newblock  (\bibinfo{date}{April} \bibinfo{year}{2023}).
\newblock
\urldef\tempurl%
\url{http://arxiv.org/abs/2304.09655}
\showURL{%
\tempurl}
\newblock
\shownote{arXiv:2304.09655 [cs]}.


\bibitem[Kim and Russell(2018)]%
        {kim_draper_2018}
\bibfield{author}{\bibinfo{person}{Louis Kim} {and} \bibinfo{person}{Rebecca
  Russell}.} \bibinfo{year}{2018}\natexlab{}.
\newblock \showarticletitle{Draper {VDISC} {Dataset} - {Vulnerability}
  {Detection} in {Source} {Code}}.
\newblock  (\bibinfo{date}{Nov.} \bibinfo{year}{2018}).
\newblock
\urldef\tempurl%
\url{https://osf.io/d45bw/}
\showURL{%
\tempurl}
\newblock
\shownote{Publisher: OSF}.


\bibitem[Liu et~al\mbox{.}(2023)]%
        {liu_is_2023}
\bibfield{author}{\bibinfo{person}{Jiawei Liu}, \bibinfo{person}{Chunqiu~Steven
  Xia}, \bibinfo{person}{Yuyao Wang}, {and} \bibinfo{person}{Lingming Zhang}.}
  \bibinfo{year}{2023}\natexlab{}.
\newblock \showarticletitle{Is {Your} {Code} {Generated} by {ChatGPT} {Really}
  {Correct}? {Rigorous} {Evaluation} of {Large} {Language} {Models} for {Code}
  {Generation}}.
\newblock  (\bibinfo{date}{May} \bibinfo{year}{2023}).
\newblock
\urldef\tempurl%
\url{https://doi.org/10.48550/arXiv.2305.01210}
\showDOI{\tempurl}
\newblock
\shownote{arXiv:2305.01210 [cs]}.


\bibitem[Ma et~al\mbox{.}(2023a)]%
        {ma2023scope}
\bibfield{author}{\bibinfo{person}{Wei Ma}, \bibinfo{person}{Shangqing Liu},
  \bibinfo{person}{Wenhan Wang}, \bibinfo{person}{Qiang Hu},
  \bibinfo{person}{Ye Liu}, \bibinfo{person}{Cen Zhang},
  \bibinfo{person}{Liming Nie}, {and} \bibinfo{person}{Yang Liu}.}
  \bibinfo{year}{2023}\natexlab{a}.
\newblock \showarticletitle{The Scope of ChatGPT in Software Engineering: A
  Thorough Investigation}.
\newblock \bibinfo{journal}{\emph{arXiv preprint arXiv:2305.12138}}
  (\bibinfo{year}{2023}).
\newblock


\bibitem[Ma et~al\mbox{.}(2023b)]%
        {ma_scope_2023}
\bibfield{author}{\bibinfo{person}{Wei Ma}, \bibinfo{person}{Shangqing Liu},
  \bibinfo{person}{Wenhan Wang}, \bibinfo{person}{Qiang Hu},
  \bibinfo{person}{Ye Liu}, \bibinfo{person}{Cen Zhang},
  \bibinfo{person}{Liming Nie}, {and} \bibinfo{person}{Yang Liu}.}
  \bibinfo{year}{2023}\natexlab{b}.
\newblock \showarticletitle{The {Scope} of {ChatGPT} in {Software}
  {Engineering}: {A} {Thorough} {Investigation}}.
\newblock  (\bibinfo{date}{May} \bibinfo{year}{2023}).
\newblock
\urldef\tempurl%
\url{https://doi.org/10.48550/arXiv.2305.12138}
\showDOI{\tempurl}
\newblock
\shownote{arXiv:2305.12138 [cs]}.


\bibitem[{OpenAI}(2022)]%
        {gpt35}
\bibfield{author}{\bibinfo{person}{{OpenAI}}.} \bibinfo{year}{2022}\natexlab{}.
\newblock \bibinfo{title}{GPT-3.5}.
\newblock
\newblock
\newblock
\shownote{Accessed May 17, 2023.
  \url{https://platform.openai.com/docs/models/gpt-3-5}}.


\bibitem[OpenAI(2023)]%
        {openai_gpt-4_2023}
\bibfield{author}{\bibinfo{person}{OpenAI}.} \bibinfo{year}{2023}\natexlab{}.
\newblock \bibinfo{booktitle}{\emph{{GPT}-4 {Technical} {Report}}}.
\newblock \bibinfo{type}{{T}echnical {R}eport}.
\newblock
\urldef\tempurl%
\url{http://arxiv.org/abs/2303.08774}
\showURL{%
\tempurl}
\newblock
\shownote{arXiv:2303.08774 [cs]}.


\bibitem[Pearce et~al\mbox{.}(2021)]%
        {pearce_asleep_2021}
\bibfield{author}{\bibinfo{person}{Hammond Pearce}, \bibinfo{person}{Baleegh
  Ahmad}, \bibinfo{person}{Benjamin Tan}, \bibinfo{person}{Brendan
  Dolan-Gavitt}, {and} \bibinfo{person}{Ramesh Karri}.}
  \bibinfo{year}{2021}\natexlab{}.
\newblock \showarticletitle{Asleep at the {Keyboard}? {Assessing} the
  {Security} of {GitHub} {Copilot}'s {Code} {Contributions}}.
\newblock  (\bibinfo{date}{Dec.} \bibinfo{year}{2021}).
\newblock
\urldef\tempurl%
\url{https://doi.org/10.48550/arXiv.2108.09293}
\showDOI{\tempurl}
\newblock
\shownote{arXiv:2108.09293 [cs]}.


\bibitem[Pearce et~al\mbox{.}(2022)]%
        {pearce_examining_2022}
\bibfield{author}{\bibinfo{person}{Hammond Pearce}, \bibinfo{person}{Benjamin
  Tan}, \bibinfo{person}{Baleegh Ahmad}, \bibinfo{person}{Ramesh Karri}, {and}
  \bibinfo{person}{Brendan Dolan-Gavitt}.} \bibinfo{year}{2022}\natexlab{}.
\newblock \showarticletitle{Examining {Zero}-{Shot} {Vulnerability} {Repair}
  with {Large} {Language} {Models}}.
\newblock  (\bibinfo{date}{Aug.} \bibinfo{year}{2022}).
\newblock
\urldef\tempurl%
\url{http://arxiv.org/abs/2112.02125}
\showURL{%
\tempurl}
\newblock
\shownote{arXiv:2112.02125 [cs]}.


\bibitem[Perry et~al\mbox{.}(2022)]%
        {perry_users_2022}
\bibfield{author}{\bibinfo{person}{Neil Perry}, \bibinfo{person}{Megha
  Srivastava}, \bibinfo{person}{Deepak Kumar}, {and} \bibinfo{person}{Dan
  Boneh}.} \bibinfo{year}{2022}\natexlab{}.
\newblock \showarticletitle{Do {Users} {Write} {More} {Insecure} {Code} with
  {AI} {Assistants}?}
\newblock  (\bibinfo{date}{Dec.} \bibinfo{year}{2022}).
\newblock
\urldef\tempurl%
\url{http://arxiv.org/abs/2211.03622}
\showURL{%
\tempurl}
\newblock
\shownote{arXiv:2211.03622 [cs]}.


\bibitem[Picard et~al\mbox{.}(2020)]%
        {picard_ensuring_2020}
\bibfield{author}{\bibinfo{person}{S. Picard}, \bibinfo{person}{C.
  Chapdelaine}, \bibinfo{person}{C. Cappi}, \bibinfo{person}{L. Gardes},
  \bibinfo{person}{E. Jenn}, \bibinfo{person}{B. Lefevre}, {and}
  \bibinfo{person}{T. Soumarmon}.} \bibinfo{year}{2020}\natexlab{}.
\newblock \showarticletitle{Ensuring {Dataset} {Quality} for {Machine}
  {Learning} {Certification}}. In \bibinfo{booktitle}{\emph{2020 {IEEE}
  {International} {Symposium} on {Software} {Reliability} {Engineering}
  {Workshops} ({ISSREW})}}. \bibinfo{pages}{275--282}.
\newblock
\urldef\tempurl%
\url{https://doi.org/10.1109/ISSREW51248.2020.00085}
\showDOI{\tempurl}


\bibitem[rey Voas(1996)]%
        {rey1996testing}
\bibfield{author}{\bibinfo{person}{Je rey Voas}.}
  \bibinfo{year}{1996}\natexlab{}.
\newblock \showarticletitle{Testing software for characteristics other than
  correctness: Safety, failure tolerance, and security}.
\newblock  (\bibinfo{year}{1996}).
\newblock


\bibitem[Ross et~al\mbox{.}(2023)]%
        {ross_programmers_2023}
\bibfield{author}{\bibinfo{person}{Steven~I. Ross}, \bibinfo{person}{Fernando
  Martinez}, \bibinfo{person}{Stephanie Houde}, \bibinfo{person}{Michael
  Muller}, {and} \bibinfo{person}{Justin~D. Weisz}.}
  \bibinfo{year}{2023}\natexlab{}.
\newblock \showarticletitle{The {Programmer}’s {Assistant}: {Conversational}
  {Interaction} with a {Large} {Language} {Model} for {Software}
  {Development}}. In \bibinfo{booktitle}{\emph{Proceedings of the 28th
  {International} {Conference} on {Intelligent} {User} {Interfaces}}}
  \emph{(\bibinfo{series}{{IUI} '23})}. \bibinfo{publisher}{Association for
  Computing Machinery}, \bibinfo{address}{New York, NY, USA},
  \bibinfo{pages}{491--514}.
\newblock
\showISBNx{9798400701061}
\urldef\tempurl%
\url{https://doi.org/10.1145/3581641.3584037}
\showDOI{\tempurl}


\bibitem[Russell et~al\mbox{.}(2018)]%
        {russell_automated_2018}
\bibfield{author}{\bibinfo{person}{Rebecca Russell}, \bibinfo{person}{Louis
  Kim}, \bibinfo{person}{Lei Hamilton}, \bibinfo{person}{Tomo Lazovich},
  \bibinfo{person}{Jacob Harer}, \bibinfo{person}{Onur Ozdemir},
  \bibinfo{person}{Paul Ellingwood}, {and} \bibinfo{person}{Marc McConley}.}
  \bibinfo{year}{2018}\natexlab{}.
\newblock \showarticletitle{Automated {Vulnerability} {Detection} in {Source}
  {Code} {Using} {Deep} {Representation} {Learning}}. In
  \bibinfo{booktitle}{\emph{2018 17th {IEEE} {International} {Conference} on
  {Machine} {Learning} and {Applications} ({ICMLA})}}.
  \bibinfo{pages}{757--762}.
\newblock
\urldef\tempurl%
\url{https://doi.org/10.1109/ICMLA.2018.00120}
\showDOI{\tempurl}


\bibitem[Sadowski and Yi(2014)]%
        {sadowski2014developers}
\bibfield{author}{\bibinfo{person}{Caitlin Sadowski} {and}
  \bibinfo{person}{Jaeheon Yi}.} \bibinfo{year}{2014}\natexlab{}.
\newblock \showarticletitle{How developers use data race detection tools}. In
  \bibinfo{booktitle}{\emph{Proceedings of the 5th Workshop on Evaluation and
  Usability of Programming Languages and Tools}}. \bibinfo{pages}{43--51}.
\newblock


\bibitem[Sandoval et~al\mbox{.}(2023)]%
        {sandoval_lost_2023}
\bibfield{author}{\bibinfo{person}{Gustavo Sandoval}, \bibinfo{person}{Hammond
  Pearce}, \bibinfo{person}{Teo Nys}, \bibinfo{person}{Ramesh Karri},
  \bibinfo{person}{Siddharth Garg}, {and} \bibinfo{person}{Brendan
  Dolan-Gavitt}.} \bibinfo{year}{2023}\natexlab{}.
\newblock \showarticletitle{Lost at {C}: {A} {User} {Study} on the {Security}
  {Implications} of {Large} {Language} {Model} {Code} {Assistants}}.
\newblock  (\bibinfo{date}{Feb.} \bibinfo{year}{2023}).
\newblock
\urldef\tempurl%
\url{http://arxiv.org/abs/2208.09727}
\showURL{%
\tempurl}
\newblock
\shownote{arXiv:2208.09727 [cs]}.


\bibitem[Shumailov et~al\mbox{.}(2023)]%
        {shumailov_curse_2023}
\bibfield{author}{\bibinfo{person}{Ilia Shumailov}, \bibinfo{person}{Zakhar
  Shumaylov}, \bibinfo{person}{Yiren Zhao}, \bibinfo{person}{Yarin Gal},
  \bibinfo{person}{Nicolas Papernot}, {and} \bibinfo{person}{Ross Anderson}.}
  \bibinfo{year}{2023}\natexlab{}.
\newblock \showarticletitle{The {Curse} of {Recursion}: {Training} on
  {Generated} {Data} {Makes} {Models} {Forget}}.
\newblock  (\bibinfo{date}{May} \bibinfo{year}{2023}).
\newblock
\urldef\tempurl%
\url{http://arxiv.org/abs/2305.17493}
\showURL{%
\tempurl}
\newblock
\shownote{arXiv:2305.17493 [cs]}.


\bibitem[Somoye(2023)]%
        {somoye_is_2023}
\bibfield{author}{\bibinfo{person}{Funmi~Looi Somoye}.}
  \bibinfo{year}{2023}\natexlab{}.
\newblock \bibinfo{booktitle}{\emph{Is {ChatGPT} free and unlimited? {In} short
  - yes}}.
\newblock
\urldef\tempurl%
\url{https://www.pcguide.com/apps/chat-gpt-free/}
\showURL{%
\tempurl}


\bibitem[Umawing(2023)]%
        {umawing_chatgpt_2023}
\bibfield{author}{\bibinfo{person}{Jovi Umawing}.}
  \bibinfo{year}{2023}\natexlab{}.
\newblock \bibinfo{booktitle}{\emph{{ChatGPT} writes insecure code}}.
\newblock
\urldef\tempurl%
\url{https://www.malwarebytes.com/blog/news/2023/04/chatgpt-creates-not-so-secure-code-study-finds}
\showURL{%
\tempurl}


\bibitem[Wallace and Fujii(1989)]%
        {wallace_software_1989}
\bibfield{author}{\bibinfo{person}{D.R. Wallace} {and} \bibinfo{person}{R.U.
  Fujii}.} \bibinfo{year}{1989}\natexlab{}.
\newblock \showarticletitle{Software verification and validation: an overview}.
\newblock \bibinfo{journal}{\emph{IEEE Software}} \bibinfo{volume}{6},
  \bibinfo{number}{3} (\bibinfo{date}{May} \bibinfo{year}{1989}),
  \bibinfo{pages}{10--17}.
\newblock
\showISSN{0740-7459}
\urldef\tempurl%
\url{https://doi.org/10.1109/52.28119}
\showDOI{\tempurl}


\bibitem[Wei et~al\mbox{.}(2023)]%
        {wei_chain--thought_2023}
\bibfield{author}{\bibinfo{person}{Jason Wei}, \bibinfo{person}{Xuezhi Wang},
  \bibinfo{person}{Dale Schuurmans}, \bibinfo{person}{Maarten Bosma},
  \bibinfo{person}{Brian Ichter}, \bibinfo{person}{Fei Xia},
  \bibinfo{person}{Ed Chi}, \bibinfo{person}{Quoc Le}, {and}
  \bibinfo{person}{Denny Zhou}.} \bibinfo{year}{2023}\natexlab{}.
\newblock \showarticletitle{Chain-of-{Thought} {Prompting} {Elicits}
  {Reasoning} in {Large} {Language} {Models}}.
\newblock  (\bibinfo{date}{Jan.} \bibinfo{year}{2023}).
\newblock
\urldef\tempurl%
\url{https://doi.org/10.48550/arXiv.2201.11903}
\showDOI{\tempurl}
\newblock
\shownote{arXiv:2201.11903 [cs]}.


\bibitem[White et~al\mbox{.}(2023)]%
        {white_prompt_2023}
\bibfield{author}{\bibinfo{person}{Jules White}, \bibinfo{person}{Quchen Fu},
  \bibinfo{person}{Sam Hays}, \bibinfo{person}{Michael Sandborn},
  \bibinfo{person}{Carlos Olea}, \bibinfo{person}{Henry Gilbert},
  \bibinfo{person}{Ashraf Elnashar}, \bibinfo{person}{Jesse Spencer-Smith},
  {and} \bibinfo{person}{Douglas~C. Schmidt}.} \bibinfo{year}{2023}\natexlab{}.
\newblock \showarticletitle{A {Prompt} {Pattern} {Catalog} to {Enhance}
  {Prompt} {Engineering} with {ChatGPT}}.
\newblock  (\bibinfo{date}{Feb.} \bibinfo{year}{2023}).
\newblock
\urldef\tempurl%
\url{https://doi.org/10.48550/arXiv.2302.11382}
\showDOI{\tempurl}
\newblock
\shownote{arXiv:2302.11382 [cs]}.


\bibitem[White et~al\mbox{.}(2016)]%
        {white2016deep}
\bibfield{author}{\bibinfo{person}{Martin White}, \bibinfo{person}{Michele
  Tufano}, \bibinfo{person}{Christopher Vendome}, {and} \bibinfo{person}{Denys
  Poshyvanyk}.} \bibinfo{year}{2016}\natexlab{}.
\newblock \showarticletitle{Deep learning code fragments for code clone
  detection}. In \bibinfo{booktitle}{\emph{Proceedings of the 31st IEEE/ACM
  international conference on automated software engineering}}.
  \bibinfo{pages}{87--98}.
\newblock


\bibitem[Xing et~al\mbox{.}(2023)]%
        {xing_prompt_2023}
\bibfield{author}{\bibinfo{person}{Zhenchang Xing}, \bibinfo{person}{Qing
  Huang}, \bibinfo{person}{Yu Cheng}, \bibinfo{person}{Liming Zhu},
  \bibinfo{person}{Qinghua Lu}, {and} \bibinfo{person}{Xiwei Xu}.}
  \bibinfo{year}{2023}\natexlab{}.
\newblock \showarticletitle{Prompt {Sapper}: {LLM}-{Empowered} {Software}
  {Engineering} {Infrastructure} for {AI}-{Native} {Services}}.
\newblock  (\bibinfo{date}{June} \bibinfo{year}{2023}).
\newblock
\urldef\tempurl%
\url{https://doi.org/10.48550/arXiv.2306.02230}
\showDOI{\tempurl}
\newblock
\shownote{arXiv:2306.02230 [cs]}.


\bibitem[Yao et~al\mbox{.}(2023)]%
        {yao_tree_2023}
\bibfield{author}{\bibinfo{person}{Shunyu Yao}, \bibinfo{person}{Dian Yu},
  \bibinfo{person}{Jeffrey Zhao}, \bibinfo{person}{Izhak Shafran},
  \bibinfo{person}{Thomas~L. Griffiths}, \bibinfo{person}{Yuan Cao}, {and}
  \bibinfo{person}{Karthik Narasimhan}.} \bibinfo{year}{2023}\natexlab{}.
\newblock \showarticletitle{Tree of {Thoughts}: {Deliberate} {Problem}
  {Solving} with {Large} {Language} {Models}}.
\newblock  (\bibinfo{date}{May} \bibinfo{year}{2023}).
\newblock
\urldef\tempurl%
\url{http://arxiv.org/abs/2305.10601}
\showURL{%
\tempurl}
\newblock
\shownote{arXiv:2305.10601 [cs]}.


\bibitem[Zhao and Huang(2018)]%
        {zhao2018deepsim}
\bibfield{author}{\bibinfo{person}{Gang Zhao} {and} \bibinfo{person}{Jeff
  Huang}.} \bibinfo{year}{2018}\natexlab{}.
\newblock \showarticletitle{Deepsim: deep learning code functional similarity}.
  In \bibinfo{booktitle}{\emph{Proceedings of the 2018 26th ACM Joint Meeting
  on European Software Engineering Conference and Symposium on the Foundations
  of Software Engineering}}. \bibinfo{pages}{141--151}.
\newblock


\bibitem[Zhou et~al\mbox{.}(2019a)]%
        {zhoudataset_devign_2019}
\bibfield{author}{\bibinfo{person}{Yaqin Zhou}, \bibinfo{person}{Shangqing
  Liu}, \bibinfo{person}{Jingkai Siow}, \bibinfo{person}{Xiaoning Du}, {and}
  \bibinfo{person}{Yang Liu}.} \bibinfo{year}{2019}\natexlab{a}.
\newblock \bibinfo{booktitle}{\emph{Devign}}.
\newblock
\urldef\tempurl%
\url{https://sites.google.com/view/devign}
\showURL{%
\tempurl}


\bibitem[Zhou et~al\mbox{.}(2019b)]%
        {zhou_devign_2019}
\bibfield{author}{\bibinfo{person}{Yaqin Zhou}, \bibinfo{person}{Shangqing
  Liu}, \bibinfo{person}{Jingkai Siow}, \bibinfo{person}{Xiaoning Du}, {and}
  \bibinfo{person}{Yang Liu}.} \bibinfo{year}{2019}\natexlab{b}.
\newblock \showarticletitle{Devign: {Effective} {Vulnerability}
  {Identification} by {Learning} {Comprehensive} {Program} {Semantics} via
  {Graph} {Neural} {Networks}}.
\newblock  (\bibinfo{date}{Sept.} \bibinfo{year}{2019}).
\newblock
\urldef\tempurl%
\url{http://arxiv.org/abs/1909.03496}
\showURL{%
\tempurl}
\newblock
\shownote{arXiv:1909.03496 [cs, stat]}.


\end{thebibliography}
\end{document}